\documentclass[]{article}

\usepackage{amsfonts}
\usepackage{amsmath}
\usepackage{amssymb}
\usepackage{amsthm}
\usepackage{caption}
\usepackage{graphicx}
\usepackage{float}
\usepackage{xcolor}
\usepackage{placeins}
\usepackage{url}
\usepackage{xurl}
\usepackage[colorlinks, allcolors=blue]{hyperref}

\title{Numerical investigation of the effect of macro control measures on epidemics transport  via a coupled PDE crowd flow - epidemics spreading dynamics model}

\author{A.I. Delis  \\
	School of Production Engineering \& Management, \\Technical University of Crete, \\
	University Campus, Chania, Greece, 73100 \\
	adelis@tuc.gr  \\
	\and 
	N. Bekiaris-Liberis \\
	School of Production Engineering \& Management,  \\Technical University of Crete, \\ 
	University Campus, Chania, Greece, 73100 \\
	bekiaris-liberis@ece.tuc.gr
	}

\begin{document}

\maketitle

\begin{abstract} 
This work aims to provide an approach to the macroscopic modeling and simulation of  pedestrian flow, coupled with contagion spreading, towards numerical investigation of the effect of certain, macro-control measures on epidemics transport dynamics.
To model the dynamics of the pedestrians, a second-order macroscopic model, coupled with an Eikonal equation, is used. This model is coupled with a macroscopic Susceptible-Exposed-Infected-Susceptible-Vaccinated (SEISV) contagion model, where the force-of-infection $\beta$ coefficient is modeled via a drift-diffusion equation, which is affected by the air-flow dynamics due to the ventilation. The air-flow dynamics are obtained assuming a potential flow that can imitate the existence of ventilation in the computational domain.  Numerical approximations are considered for the coupled model along with numerical tests and results. In particular, we investigate the effect of employment of different, epidemics transport control measures, which may be implemented through real-time manipulation of i) ventilation rate and direction, ii) maximum speed of pedestrians, and iii) average distances between pedestrians, and through iv) incorporation in the crowd of masked or vaccinated individuals.
Such simulations of disease spreading in a moving crowd can potentially provide valuable information about the risks of infection in relevant situations  and support the design of systematic intervention/control measures.
\end{abstract}


\maketitle

\section{Introduction} \label{Intro}
The ramifications of epidemics spreading in economy \cite{Anderson} and physical/mental health \cite{Delmastro}, highlighted the need of accurate prediction of epidemics spreading. One way towards this is through modeling the dynamics of epidemics spreading. Due to the complexity of detailed description of epidemics spreading when accounting for individual-to-individual interactions, see, for example, \cite{Willem}, macroscopic models may be utilized to describe epidemics spreading dynamics, particularly in cases of large crowds/spaces, see, for example, \cite{Vyn}. Even though such models capture epidemics spreading dynamics on a macroscopic scale, they may not capture the transporting behavior of epidemics, which is, in fact, the main, macroscopic phenomenon in epidemics spreading dynamics, emerging from people transport. Furthermore, there are at least four potential macroscopic  ways of control of epidemics spreading in closed spaces,  namely via manipulation of the air-flow  rate and direction, see, e.g., \cite{Hoss}, or via real-time manipulation of the maximum speed of and average distances between pedestrians (through recommendations or other means), see, for example, \cite{Kah, Wadoo10, Hans18, Beek2024},  or via incorporating in the crowd masked/vaccinated individuals. For these two reasons, in the present work, we perform numerical tests to study the epidemics transport dynamics, under the effect of various ventilation air-flow rates/directions, for various values of the maximum pedestrians' speed and of their average distances, and in the presence of masked/vaccinated individuals, utilizing a Partial Differential Equation (PDE) model that accounts for crowd flow and epidemics spreading dynamics, as well as for the effect of ventilation.

Macroscopic modeling of coupled crowd motion and epidemic spreading dynamics is relatively unexplored. Here we provide such a model that may be useful in situations of crowds occupying  confined spaces where a disease  may easily spread such as, e.g., measles, influenza, or COVID-19.
In particular, we utilize a PDE model  consisting of three components; a component that describes crowd flow dynamics in  two-dimensional (2D) spaces, a component that describes epidemics spreading, essentially, being a macroscopic version of a  SEISV-type model, and a component that describes the effect of aerosol dynamics to spreading, in particular, describing the effect of ventilation induced air-flow. The model we present here can be viewed as a modified version of the model considered in \cite{salam1}, in that we use a less complex crowd flow model (modified from \cite{Kah} and \cite{Maity24}) and use potential theory to compute the ventilation induced air-flow field,   besides also incorporating transport dynamics for  masked/vaccinated individuals. Papers \cite{salam2}, \cite{Kim}, and \cite{Agnelli2015} also present relevant PDE models\footnote{Although here we review only PDE macroscopic models, there exist microscopic models, which describe coupled pedestrians-epidemics transport dynamics, accounting for movements of individuals and individual-to-individual interactions, see, e.g., \cite{Hag}.}, consisting of a crowd flow dynamics component and an epidemics spreading component with a different type of infection coefficient. Our work can be also viewed as relevant to papers that consider macroscopic models of  people mobility (via describing the respective traffic flow dynamics) and epidemics spreading dynamics, such as, e.g., \cite{Hag, Satt, Arino, Tizz, Bert, Guan,Niazi}.  Although these works are relevant, they describe the coupled people flow - epidemics spreading process on a higher, macroscopic level (e.g., on the level of a country or a city), not considering the crowd movements in closed spaces. Therefore, here we employ an integrated model that considers the  interactions between pedestrian movements and disease transmission dynamics in closed spaces, enabling the assessment of how crowd dynamics influence disease spreading in relevant scenarios.

In the present paper, we perform numerical tests employing the proposed PDE model. In particular, we employ a second-order, accurate finite-volume-based numerical scheme to simultaneously solve on a 2D domain the three components of the model, namely, the crowd flow, epidemics spreading, and aerosol dynamics models; while employing a finite-difference scheme for numerically solving a 2D Laplace equation to obtain a simplified velocity field (based on potential theory), corresponding to the operation of the ventilation system. Implementing these numerical schemes, we categorize the tests performed in three different types as follows.  In the first, we consider people moving towards a single exit, together with accounting for variable ventilation rates, different air-flow directions, different maximum speed and pressure coefficients, as well as accounting for the presence of masked/vaccinated individuals. In the second, we perform similar tests, but in comparison with the first type of tests, we consider two exits. In the third, we account for a larger space and a larger number of total individuals moving towards an exit. 

Based on the results from the tests performed we compare the various potential macro control measures with respect to the resulting total number of exposed individuals, which is considered as an index of spreading degree. In general, our findings indicate that the total number of exposed individuals decreases with an increase in the ventilation rate, or when people move against the direction of air-flow, or when the maximum speed of individuals increases, or when the pressure coefficient increases (implying larger average distances between individuals), or when masked/vaccinated individuals are included in the crowd. However, the effect of increased pressure coefficient is less significant when the flow capacity on the exit is large; while when the flow capacity at the exit is low, further increasing the maximum speed may not be beneficial for spreading. The exact quantification of the total number of individuals depends on the actual magnitudes of ventilation rate, maximum speed, and pressure coefficients, as well as the number of masked/vaccinated individuals.
Essentially, our results illustrate the possibility to utilize the density (and speed) profile in time/space in real time (i.e., to not only employing an average, total number of people), to manipulate ventilation air-flow field in closed spaces towards balancing epidemics spreading suppression and energy, as well as to manipulate in a systematic manner, the average distances between individuals and their maximum speed (via, e.g., recommendations from speakers), towards epidemics spreading suppression.

The paper is organized as follows. We start in Section 2 presenting the coupled crowd flow-epidemics spreading dynamics model. In Section 3 we present the numerical schemes employed for solving the complete model. In Section 4 we present and interpret the simulation results obtained from the numerical implementation of the model. We provide concluding remarks  in Section 5.

\section{Coupled Modeling Equations}

\subsection{Crowd Flow Modeling}

The model presented here is  classified
as a nonlinear, hyperbolic PDE  system (with source terms present). 
Second-order models in 2D
 use three-coupled PDEs to describe crowd flow; the mass conservation
and two equations that resemble  the momentum
equations in fluid flow with a modification to the pressure term
to mimic crowd motion. In essence, we consider pedestrians as a  ``thinking fluid". The  model considered here is known for its isotropic nature, which is realistic
assuming
pedestrian motion is influenced from all directions. 
This model stems from the well-known Payne-Whitam (PW) traffic flow model, \cite{LW, Payne}, adapted to crowd dynamics \cite{Kah, Maity24}.
We present the model in conservation law form, which eases the application of proper numerical
methods (e.g., finite volume ones). In general, we consider  a 2D  connected domain 
$\Omega \subset \mathbb{R}^2$ corresponding to some walking facility, possibly equipped with
some entrances or exits. In this work, we do not consider the case of having some entrances, thus the boundary of the domain $\partial \Omega = \Gamma_0 \cup \Gamma_w$,  where outflow boundaries are denoted by $\Gamma_0$ and the
walls by $\Gamma_w$.

By denoting as ${\bf x}=(x,y) \in \Omega$ the spatial variables and  $t>0 $ the time, we define as $\rho(x,y,t)$  the pedestrian density (that has to  stay non-negative and bounded: $ 0 \le \rho({\bf x},t) \le \rho_{\max}$) and as $u({\bf x},t)$ and $v({\bf x},t)$ the $x-$component and $y-$component  of the velocity vector ${\bf v}$, respectively. The model equations,  along with their initial conditions, can be written as
\begin{eqnarray}
&&  \rho_t + \nabla\cdot \left(\rho{\bf v}\right) = 0,  \\
&&(\rho {\bf v})_t  + \nabla\cdot \left(\rho {\bf v} \otimes {\bf v} + P(\rho)\right) = \frac{1}{\tau}\rho \left(  V(\rho) \vec{\bf \mu} - {\bf v}\right); \\
&&\rho({\bf x},0) = \rho_0({\bf x}),  \\
&&{\bf v}({\bf x},0) = {\bf v}_0({\bf x}), 
\end{eqnarray}
where $P(\rho)$ is an internal pressure function.
 In the momentum equations  (2) the relaxation source term drives  ${\bf v}$ towards  the desired speed $V(\rho)\vec{\mu}$,
where $V(\rho)$ is an equilibrium speed-density relation, $\vec{\mu}$ is the desired direction vector, and $\tau$ is a relaxation time.

For the  speed-density relation we  implement  the following relation 
\begin{equation}
V(\rho) = u_{\max} \displaystyle{e^{-\alpha(\rho/\rho_{\max})^2}}, 
\end{equation}
where $\alpha > 0$ is a constant, $u_{\max}$ is the free-flow velocity, and $V(\rho_{\max}) \approx 0$ with $\rho_{\max}$ being the congestion density at which the motion is hardly
possible. The critical density, where the pedestrian flow is at a maximum, is given by $\rho_{cr}=\rho_{\max}/\sqrt{2\alpha}$. We note here that other equilibrium speed-density relations  can also be implemented. For the  internal pressure function we choose $P(\rho)=\rho C_0^2$, with $C_0^2$ being constant representing, an anticipation factor.

With regard to the  desired direction of motion $\vec{\mu}$,  under the assumptions that pedestrians route choice is based on  the shortest path to a destination and the
avoidance of high density regions, following \cite{Hudg, Goatin1, Goatin2, salam1, salam2},
we define a potential field $\Phi$ that describes the instantaneous travel cost to a destination with $\nabla \Phi = [\Phi_x, \Phi_y]^{\rm T}$, and set
     \begin{equation}
     \vec{\mu} = - \frac{\nabla \Phi}{\| \nabla \Phi \|}
     \end{equation}
      The potential $\Phi: \Omega \to \mathbb{R}$ is defined by the  Eikonal equation
      \begin{eqnarray}
      \| \nabla \Phi \| &=&\frac{1}{V(\rho)},  \quad  t>0 , \quad {\bf x}\; \in\; \Omega,  \\
      \Phi &=& 0,      \quad  {\bf x} \; \in \; \Gamma_o,  \quad\Phi = \infty,     \quad  {\bf x} \; \in  \; \Gamma_w.
      \end{eqnarray}
     In simulations we assign a finite very large value for $\Phi$ when ${\bf x} \; \in  \; \Gamma_w.$ 
     
      This  choice for $\vec{\mu}$   implies that pedestrians have a knowledge of the density distribution in the whole domain at each
time instant, and use this to estimate their travel time. Such a reaction might be triggered, for example, by a global view on the pedestrian crowd or by information supplied by a real-time evacuation assistant. 

Further,  equations (1), (2)  can be written in vector conservative form as,
\begin{equation}
{\bf Q}_t +{\bf F}({\bf Q})_x + {\bf G}({\bf Q})_y = {\bf S}({\bf Q}),
\end{equation}
where ${\bf Q} =(\rho, \rho u, \rho v )^{\rm T}$ and, the fluxes and sources are given as
\begin{eqnarray}
{\bf F}({\bf Q}) &=&  \left(\begin{array}{c}
\rho u \\
\rho u^2 + \rho C_0^2 \\
\rho uv
\end{array}\right), 
{\bf G}({\bf Q}) =  \left(\begin{array}{c}
\rho v \\
\rho uv \\
\rho v^2 + \rho C_0^2 \\
\end{array}\right),   \nonumber
\\
{\bf S} &=& \left(\begin{array}{c}
0 \\
\displaystyle{\frac{1}{\tau}\rho\left( -V(\rho)\frac{ \Phi_x}{\| \nabla \Phi \|}-u \right)} \\
\displaystyle{\frac{1}{\tau}\rho\left( -V(\rho)\frac{ \Phi_y}{\| \nabla \Phi \|}-v \right)}
\end{array}\right). 
\end{eqnarray}

The eigenvalues of the Jacobian matrices ${\bf A}({\bf Q}) = \displaystyle{\frac{\partial {\bf F}}{\partial {\bf Q}}}$ and ${\bf B}({\bf Q}) = \displaystyle{\frac{\partial {\bf G}}{\partial {\bf Q}}}$ are, respectively,
\begin{equation} 
\lambda_1^{A} = u - C_0, \; \lambda^A_2 = u, \;\lambda^A_3 = u+C_0 ,
\end{equation}
and
\begin{equation} \lambda_1^{B} = v - C_0, \;\lambda^B_2 = v, \;\lambda^B_3 = v+C_0, 
\end{equation}
with their corresponding
eigenvectors being linearly independent, i.e., the model
is strictly hyperbolic.  System (9) preserves its isotropic
nature,
meaning that information from all directions affect pedestrians'
motion and this is the most distinct characteristic
of this model. These are given as
\begin{equation}
    {\bf e}^A_1 = \left( \begin{array}{c} 1 \\ u-C_0 \\ v \end{array} \right), \; {\bf e}^A_2 = \left( \begin{array}{c} 0 \\ 0 \\ 1 \end{array} \right),\; {\bf e}^A_3 = \left( \begin{array}{c} 1 \\ u+C_0 \\ v \end{array} \right),
\end{equation}
and
\begin{equation}
    {\bf e}^B_1 = \left( \begin{array}{c} 1 \\ u \\ v-C_0 \end{array} \right), \; {\bf e}^B_2 = \left( \begin{array}{c} 0 \\ 1 \\ 0 \end{array} \right),\; {\bf e}^B_3 = \left( \begin{array}{c} 1 \\ u\\ v+C_0 \end{array} \right).
\end{equation}

\subsection{Epidemic Spreading Model}

The model utilized here is modified from \cite{salam1} and is based on a PDE macroscopic version of a SEISV model where each type of pedestrian (SEIV) moves with the crowd speed ${\bf v}$, as this results from the crowd flow model. The density of each type of  pedestrian satisfies  then the following 
\begin{eqnarray}
\rho^S_t + \nabla \cdot (\rho^S {\bf v})  &=& \kappa \rho^I -\beta_I \rho^S -\xi \rho^S,\\
\rho^E_t +\nabla \cdot (\rho^E {\bf v}) &=& \beta_I \rho^S -\theta \rho^E,\\
\rho^I_t + \nabla \cdot (\rho^I {\bf v})&=& \theta \rho^E  -\kappa \rho^I, \\
\rho^V_t + \nabla \cdot (\rho^V {\bf v})&=& \xi \rho^S,
\end{eqnarray}
 where $\rho^S, \rho^E$, $\rho^I$, and $\rho^V$ are densities of the susceptible, exposed, infected, and vaccinated/masked pedestrians, respectively, satisfying $\rho = \rho^S+\rho^E+\rho^I +\rho^V$. The model equations are supplied with relevant initial conditions $\rho_0^S, \rho_0^E$, $\rho_0^I$ and $\rho_0^V$, similar to equations (1)--(4), satisfying $\rho_0 = \rho_0^S + \rho_0^E +\rho_0^S + \rho_0^V$. Further, $\beta_I$ is the infection rate, $\kappa$ is the recovery rate, $\theta$ is the rate with which
exposed persons are becoming infected, and $\xi$ denotes vaccination rate. Pedestrians are potentially becoming
exposed, when they are in contact with infected pedestrians.
However,   on the time scales under consideration, $\kappa$, $\xi$ and $\theta$ are very small, and thus, they are set to
zero in our numerical simulations. Hence, the number of infected and vaccinated pedestrians
remains constant, considering that there is no inflow of infected and vaccinated pedestrians at the boundary entries.

Here, we compute $\beta_I = i_0\beta({\bf x},t)$, where  $\beta({\bf x},t)$ is the solution of the following drift-reaction-diffusion PDE in 2D following \cite{salam1} as
\begin{equation}
\beta_t + \nabla \cdot(\beta {\bf U}_G) = \nabla \cdot (\sigma\nabla \beta) - \nu\beta +\frac{\rho^I}{\rho},
\end{equation}
with $\beta({\bf x},0)=0$ as initial condition, ${\bf U}_G = [u_G, v_G]^{\rm T}$   a given velocity field of the surrounding air-flow in the computational region, and $\sigma$ the effective turbulent viscosity for the aerosol. The term $- \nu\beta$
models the fact that aerosol particles are settling due to the gravitational force \cite{salam1, Zhao2}.
In this work we assume that $\sigma$ is a constant equal to $1.2\cdot 10^{-3}$\cite{Zhao2, salam1,salam2} and  we get  the velocity field of the surrounding air-flow using potential flow theory, which will be described next.
The parameter $i_0$ is determined by the infectivity and is of $O(10^{-2})$ \cite{salam1, salam2}.

In this work, and in order to produce  a steady velocity field ${\bf U}_G$ of the surrounding air-flow in the computational region $\Omega$ for model (19), it is assumed that the air flow is inviscid, incompressible, and  irrotational \cite{potential}.  Thus,  the produced steady flow is governed by Laplace's equation as
\begin{equation}
\Delta  \Psi =0,
\end{equation}
where $\Psi({\bf x}$ is the so-called velocity potential function. 
By implementing Neumann boundary conditions and imposing conditions on the derivative of the potential with respect to the normal direction (the derivative perpendicular to the boundary), we introduce  air-flows for inflow and outflow boundaries imitating in such a way the inflows from  ventilation ducts  and outflows from exhaust  ducts, respectively.
Having computed the potential field  we  compute the resulting velocity field ${\bf U}_G$ as
\begin{equation}
{\bf U}_G = \nabla \Psi.
\end{equation}.

\section{Numerical Methodologies}

The numerical approaches adopted here are based on the implementation of well-known conservative finite volume schemes to discretize  model equations (9) for the crowd flow, equations (15)--(18) for the contagion model,  and the advection part in equation (19) for  the infection rate. Here we only give a brief presentation of them and we refer, for example, to the works in \cite{Leveque, Leveque2, Toro} for more details in hyperbolic PDEs and their numerical solution, and in \cite{Kah} for crowd flow models in particular.  To this end, the
solution space $(x, y, t)$ is split up into a uniform computational grid,
where the grid spaces in the $x$ and $y$ directions are given by $\Delta x$ and
$\Delta y$, respectively and in time direction by $\Delta t$.
The positions of the $i$th and $j$th nodes in the $x$ and $y$ directions, and
$n$th node in time direction  $(x_i, y_j, t^n)$, are given by $(i\Delta x, j\Delta y, n\Delta t)$, for $i=0,1,\ldots, N_x$ and $j=0,1,\ldots, N_y$, with the solution vector, denoted as ${\bf  Q}_{i,j}^n$, being defined
at the center of the computational cell $[x_{i-1/2,j}, x_{i+1/2,j}] \times [y_{i,j-1/2}, y_{i,j+1/2}]$.

By doing so, the generic finite volume numerical scheme in 2D  with explicit time integration can be written as
\begin{multline} {\bf Q}_{i,j}^{n+1} =  {\bf Q}_{i,j}^{n} - \frac{\Delta t}{\Delta x} [{\bf F}^n_{i+1/2,j}  -{\bf F}^n_{i-1/2,j} ] 
-\frac{\Delta t}{\Delta y} [{\bf G}^n_{i,j+1/2}  -{\bf G}^n_{i,j-1/2} ] +\Delta t{\bf S}_{i,j}^n, \end{multline}
where ${\bf F}^n_{i\pm 1/2,j}$ and ${\bf G}^n_{i,j\pm 1/2}$ are the numerical fluxes at the $x$ and $y$ (cell-boundary) directions, respectively.
However, in our implementation, for model equations (9), we apply dimensional  splitting by splitting the 2D problem into a sequence of two 1D problems along with  a splitting technique between the advective part and the source terms of the model  equations. This is performed as follows, along the $x-$direction
\begin{equation}
\begin{cases}
{\bf Q}_{i,j}^{*} =  {\bf Q}_{i,j}^{n} - \displaystyle{\frac{\Delta t}{\Delta x}} \left[{\bf F}^n_{i+1/2,j}  -{\bf F}^n_{i-1/2,j} \right]; \\
{\bf Q}_{i,j}^{n+1/2} = {\bf Q}_{i,j}^{*}  +\Delta t {\bf S}( {\bf Q}_{i,j}^{*} ),
\end{cases}
\end{equation}
and along the $y-$direction
\begin{equation}
\begin{cases}
{\bf Q}_{i,j}^{**} =  {\bf Q}_{i,j}^{n+1/2} - \displaystyle{\frac{\Delta t}{\Delta y}} \left[{\bf G}^{n+1/2}_{i,j+1/2}  -{\bf G}^{n+1/2}_{i,j-1/2} \right] \\
{\bf Q}_{i,j}^{n+1} = {\bf Q}_{i,j}^{**}  +\Delta t {\bf S}( {\bf Q}_{i,j}^{**} ). 
\end{cases}
\end{equation}

For computing the numerical fluxes in (23) and (24) we implement the well-known approximate Riemann solver of Roe  and we briefly describe  it here.  
The idea behind Roe’s scheme is to take a non-linear PDE system in
quasi-linear form, e.g., in 1D
\begin{equation} 
{\bf Q}_t + {\bf A}({\bf Q}){\bf Q}_x = 0, 
\end{equation}
and linearize locally by approximating the Jacobian matrix(ces) ${\bf A}({\bf Q})$ at a computational cell's interface using Roe averages and in every time step repeat the process.
For the purpose of this section we define ${\bf Q}^R$ and ${\bf Q}^L$ that represent
the states  on the right ($R$) and left ($L$) in the $x-$direction from the cell boundary 
interface $(i\pm1/2,j)$. 
For a first-order accurate in space scheme, and  for the $(i,j)$th cell, $R\equiv ( i+1,j) $ and $L\equiv (i,j)$. 
Thus, the numerical flux is defined as
\begin{multline}
{\bf F}^{\rm Roe}_{i+1/2,j}({\bf Q}^L,{\bf Q}^R) =
\frac{1}{2}[{\bf F}({\bf Q}^L) + {\bf F}({\bf Q}^R)] 
-\frac{1}{2}\tilde{\bf R}\tilde{\bf |\Lambda|}\tilde{\bf R}^{-1}[{\bf Q}^R - {\bf Q}^L].
\end{multline}
To obtain our approximated solution we write
$|{\bf A} | (\tilde{\bf Q}) = \tilde{\bf R}\tilde{\bf |\Lambda|}\tilde{\bf R}^{-1}$, where $\tilde{\bf R}$ is
the (average) eigenvectors' matrix and $\tilde{\bf \Lambda}$ is the diagonal matrix of the approximated (average) eigenvalues. 
Further, the Harten and Hyman \cite{Harten, Leveque} entropy fix has been implemented, which modifies the
approximated eigenvalues used in the Roe linearization by never allowing any eigenvalue to be too close to zero.
In the same way the numerical fluxes ${\bf G}_{i,j\pm 1/2}$ in the $y-$direction are also computed.
For the   crowd model (9) the Roe averages are 
\begin{eqnarray}
\tilde{\rho} = \sqrt{\rho^R\cdot\rho^L},    
\tilde{u} = \frac{u^L\sqrt{\rho^L} + u^R\sqrt{\rho^R}}{\sqrt{\rho^R} +\sqrt{\rho^L}},
\tilde{v} = \frac{v^L\sqrt{\rho^L} + v_R\sqrt{\rho^R}}{\sqrt{\rho^R} +\sqrt{\rho^L}},
\end{eqnarray} 
and are used to compute the approximated eigenvalues (and eigenvectors) in $\tilde{\bf R}$ and $\tilde{\bf \Lambda}$, as given in (11)--(14).
A  second-order spatial discretization can be obtained by properly defining the left ($L$) and right  ($R$) states at the computational cell  interfaces. To this end, we utilize the classical Monotone Upstream-Centered Scheme for Conservation Laws  (MUSCL) reconstruction technique; we refer, for example, in \cite{Toro, Leveque}. Details about the MUSCL reconstruction are given in the Appendix.

For the boundary conditions here we implement the following at a layer of grid points (called ghost cells) around the computational domain. 
\begin{itemize}
\item Free-slip at walls: $\rho_g = \rho_i$, ${\bf v}_g = {\bf v}_i -2(\vec{\bf v}_i\cdot \vec{n})\vec{n}$, with $i$ being the neighbor point inside the domain and $\vec{n}$ the unit normal vector at a boundary cell.
\item Outflow (at exits): $\rho_g = \rho_i$, $\vec{\bf v}_g = u_{\max}\vec{n}$ (assuming on exits pedestrians can leave with maximum speed).
\item Inflow boundary conditions can also be implemented e.g. $\rho_g = \rho_{in}({\bf x},t)$.
\end{itemize}
Finally, for the crowd flow model, and to numerically obtain the direction vector $\vec{\mu}$ in the relaxation source term, we numerically solve the Eikonal equation (7), (8) at each time step implementing the Fast-Sweeping Method (FSM) of \cite{Zhao}.

For each of the model  equations (15)--(18) for the contagion model and the advection part in equation (19)
we implement the well-known  Rusanov  numerical flux in scheme (22), again using the MUSCL reconstruction scheme,  which reads, for example, in the $x-$direction as
\begin{equation}
{\bf F}^{\rm Rus}_{i+1/2,j} = \frac{1}{2}\left[\bar{{\bf F}}(\bar{{\bf Q}}^R) + \bar{{\bf F}}(\bar{{\bf Q}}^L)\right] - \frac{C_{i+1/2,j}}{2}\left[\bar{\bf Q}^R-\bar{\bf Q}^L\right],
\end{equation}
where $\bar{{\bf F}}$ and $\bar{{\bf G}}$ are the advection fluxes in (15)--(18) and (19), $\bar{\bf Q}$ the relevant vector of unknowns, and  $C_{i+1/2,j} = \max\{|u^L|,|u^R|\}$, i.e., the maximum absolute advection velocity in the $x-$direction at the computational cell interface. Similar is the formulation in the $y-$direction and the numerical fluxes ${\bf G}^{\rm Rus}_{i,j\pm 1/2}$. Further, for the diffusion term $\nabla \cdot (\sigma\nabla \beta)$  in equation (19), as well as for solving the Laplace equation   (20), classical second-order central finite difference schemes are implemented \cite{Leveque2}. 

Since the  numerical solution of all models  is obtained at each time node using the same time step $\Delta t$, their computation is stable   under the global Courant-Friedrichs-Lewy (CFL) stability
condition \cite{Leveque, Leveque2} given as
\begin{equation}
\Delta t \le \min \left( \frac{  {\rm CFL}\cdot \min\{\Delta x, \Delta y\} }{ \displaystyle{\max_{i,j} }\left\{ \| {\bf v} \|_{i,j}+C_0  \right\} }, \frac{  {\rm CFL} \cdot\min\{\Delta x, \Delta y\} }{ \displaystyle{\max_{i,j} }\left\{ \| {\bf U}_G \|_{i,j}  \right\} }, \frac{\Delta x^2\Delta y^2}{2\sigma(\Delta x^2 +\Delta y^2)} \right),
\end{equation}
where the first therm in equation (29) stems from the Roe scheme, the second from the advective part of equation (19) using the Rusanov scheme, and the third from the dicretization of the diffusion term of (19). The value of ${\rm CFL } =0.8$ was implemented in all simulations in Section 4 next.

{\bf Remark.} There also exists another  potential source of numerical  instability for  second-order crowd flow models with relaxation terms present (and in general in PDEs having relaxation terms) when explicit integration methods are used.  Explicit methods may produce  numerical instabilities when approximating the relaxation if the  time step $\Delta t$ is greater than the smallest intrinsic time scale of the system, see, e.g., \cite{Treiber}. In our case this results in satisfying that $\Delta t < \tau$, which is always true for all simulations presented next.


\section{NUMERICAL TESTS AND RESULTS}

In this section we present numerical simulations for evaluating the behavior of the model 
and, in particular, for evaluating the disease spreading under different conditions. Emphasis is given on the effect of certain parameters in the model, in particular we study the effect of the ventilation rates and direction of the resulting air-flow field,  as well as we investigate the effect of variable maximum desired velocity $u_{\max}$, anticipation factor $C_0$, and of the presence of  vaccinated/masked population.
If not otherwise stated, we fix some of the parameters used in all the simulation scenarios presented below; these are  $\tau= 0.6$ $s$, $\alpha =7.5$, and $\rho_{\max} = 6$ ped/$m^2$, in the density-speed relation $V(\rho)$, 
$i_0 =0.04$ (as in \cite{salam1, salam2}) for the infectivity rate, and $\nu = 0.5$ in model (19) (i.e., we assume that $50\%$ of the aerosol particles are settling due to gravitational forces). We compute the total number of pedestrians inside a facility, at each time instant,  via relation
$R(t) = \displaystyle{\int_{\Omega} \rho( {\bf x},t) d{\bf x}}$.

\subsection{Crowd Flow Towards an Exit}
In this test we consider a closed room of size $\Omega = [0, 10\;m] \times [0,10\;m]$ with an exit $2$ $m$ wide at the right boundary centered at $y=5$ $m$. The initial density is $\rho_0({\bf x}) =2.5$ ped/$m^2$  in the region $[1, 5]\times [2.5,7.5]$, leading to 50 people in this region, with  ${\bf v}_0({\bf x})=0$ in the crowd flow model.  Further, we set $\rho_0^I=0.25\rho_0$, i.e., 25$\%$ of the pedestrians are considered infected, while initially $\rho_0^E=0$. This test scenario is similar to one  of the scenarios presented in \cite{Agnelli2015, Kim, Agnelli2023}. The values of $\Delta x$ and $\Delta y$ were set equal to $0.05$ $m$ and the value of $\Delta t = 2\cdot 10^{-3}$ $s$.

\begin{figure}[h!]
 \centering
 \includegraphics[width=6.34cm,height=4.75cm]{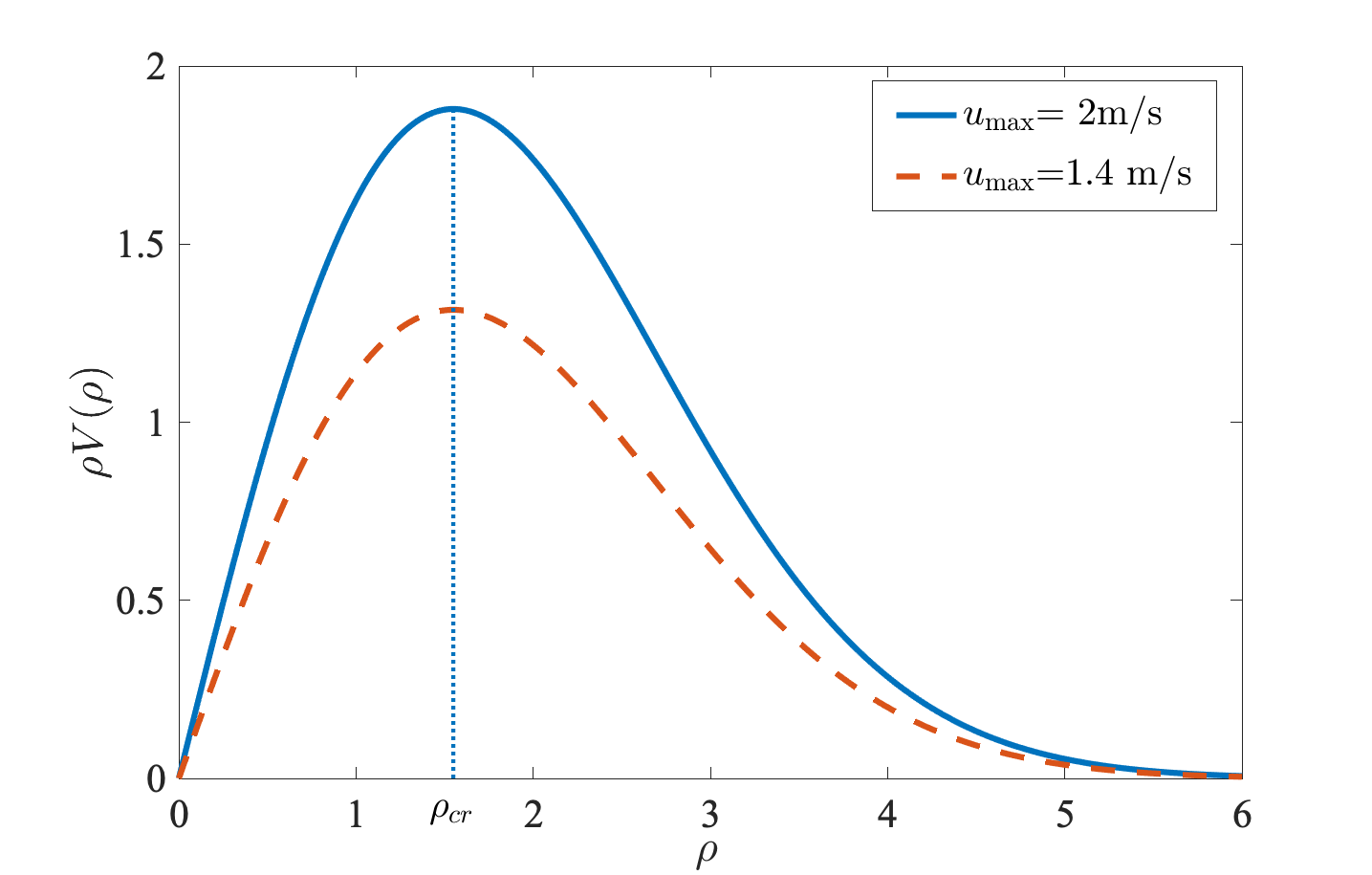} \hspace{-0.3in}
  \includegraphics[width=6.34cm,height=4.75cm]{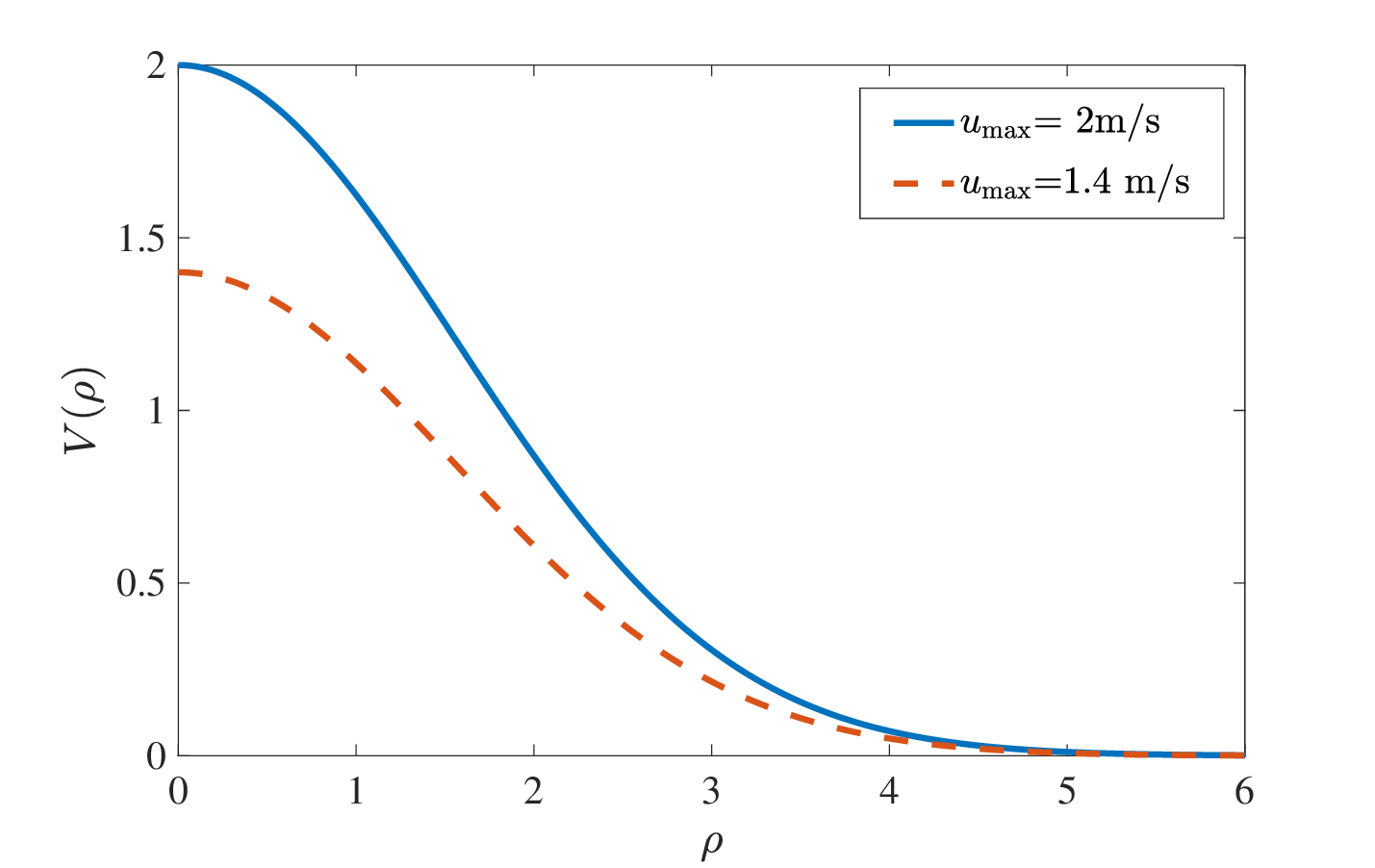}
\caption{Equilibrium speed-density relations for flow (left) and velocity (right) for two different values of $u_{\max}$. }
 \label{fig1}
 \end{figure}

In addition to studying the effect of ventilation air-flow field and of the presence of vaccinated/masked individuals to the percentage of total exposed individuals, we also study the effect of  the desired speed $u_{\max}$ and anticipation constant $C_0$ in the internal pressure-like function. In particular, we compute the respective evacuation times and the corresponding total number of exposed individuals within this time interval.  Empirical studies have shown that the average maximal reachable dimensionless velocity, i.e., the speed at which pedestrians walk when they are not hindered by others, is about $1.34$ $m/s$ with standard deviation of $0.37$ $m/s$ (see, for example, \cite{Buch2006, Goatin1}). However, due to impatience or in  emergency situations people tend to move faster. To this end we consider two cases here, one with $u_{\max}=1.4$ $m/s$ and one with $u_{\max}=2$ $m/s$. In Fig. \ref{fig1} we present the behavior of the speed density relation for the two different maximum velocities where for both cases the critical density value for maximum pedestrian flow is obtained at $\rho_{cr} = 1.549$ ped$/m^2$.
 
 As for the  anticipation constant $C_0$,  which is a macroscopic  indicator of the space concentration of  crowd density, see, e.g., \cite{Maity24},  smaller values of $C_0$ correspond to pedestrians coming closer to each other (as, for example, in emergency situations), while larger values correspond to larger average distances between them. Here we consider three cases, as $C_0=0.5, 0.8$, and $1.2$.
 
\subsubsection{Effect of  $u_{\max}$, $ C_0$, and Vaccinated/Masked Population}
\begin{figure}[htb]
\centering
 \includegraphics[width=6.32cm,height=5.4cm]{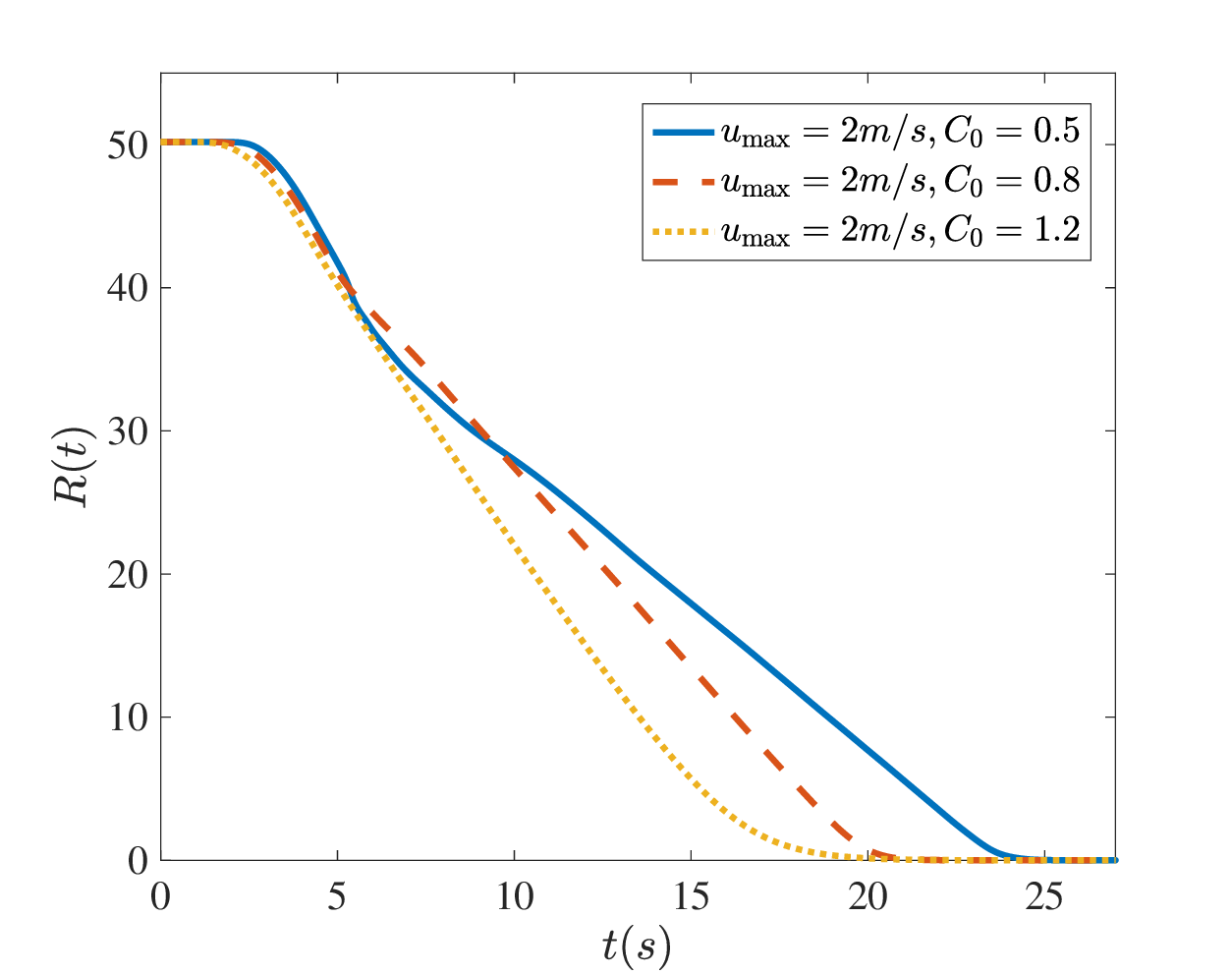} \hspace{-0.285in}
  \includegraphics[width=6.32cm,height=5.4cm]{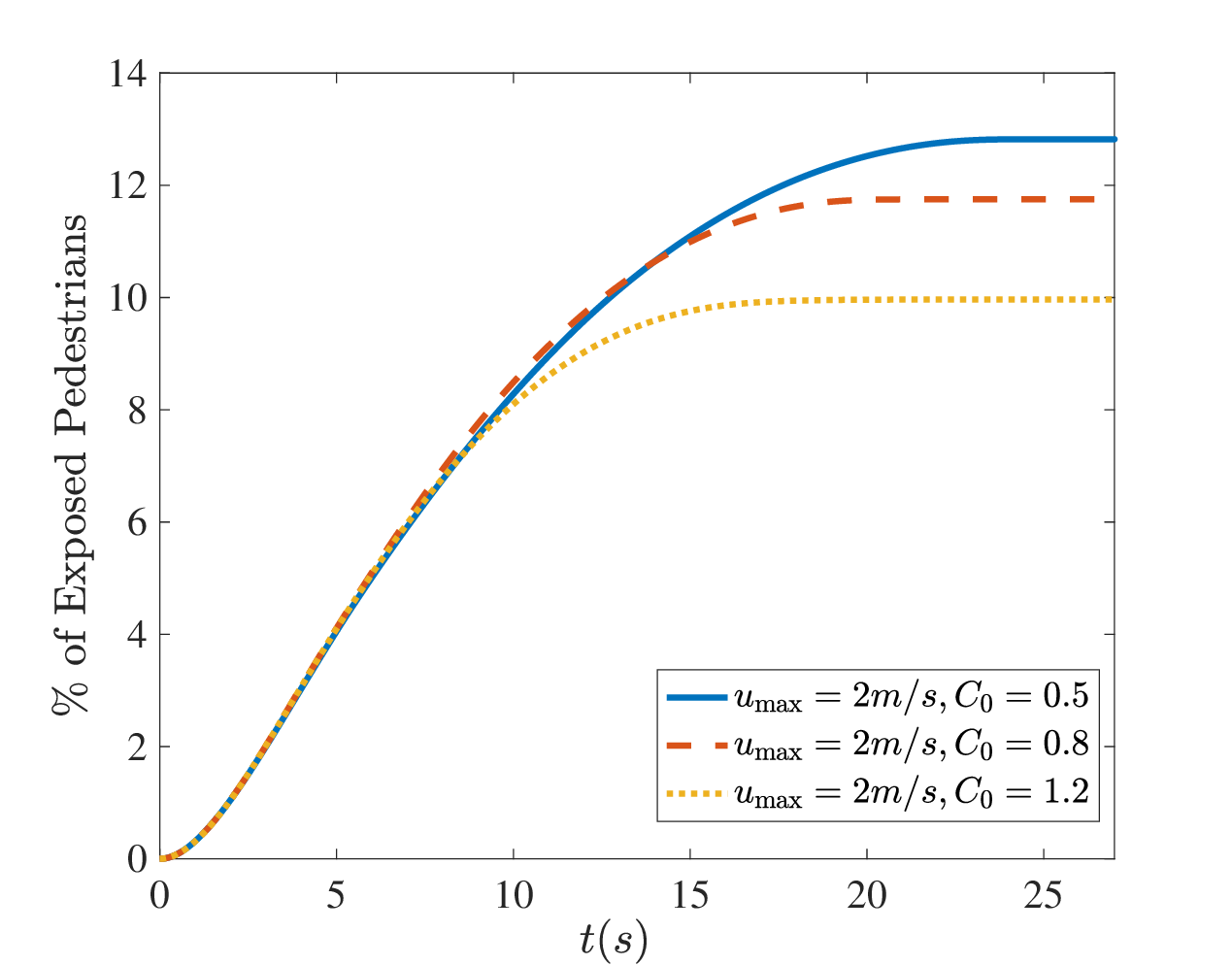}
   \includegraphics[width=6.35cm,height=5.4cm]{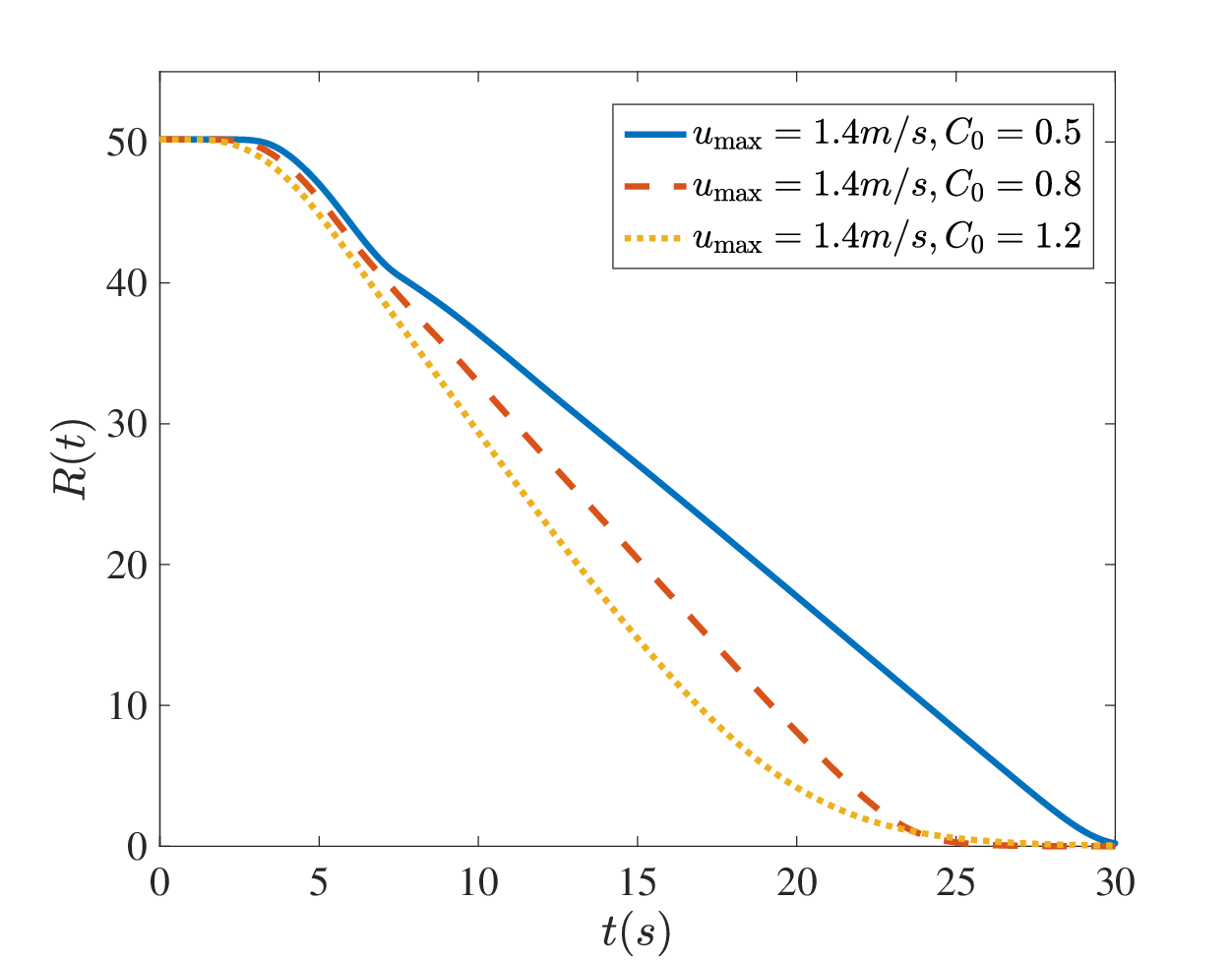}\hspace{-0.28in}
  \includegraphics[width=6.35cm,height=5.4cm]{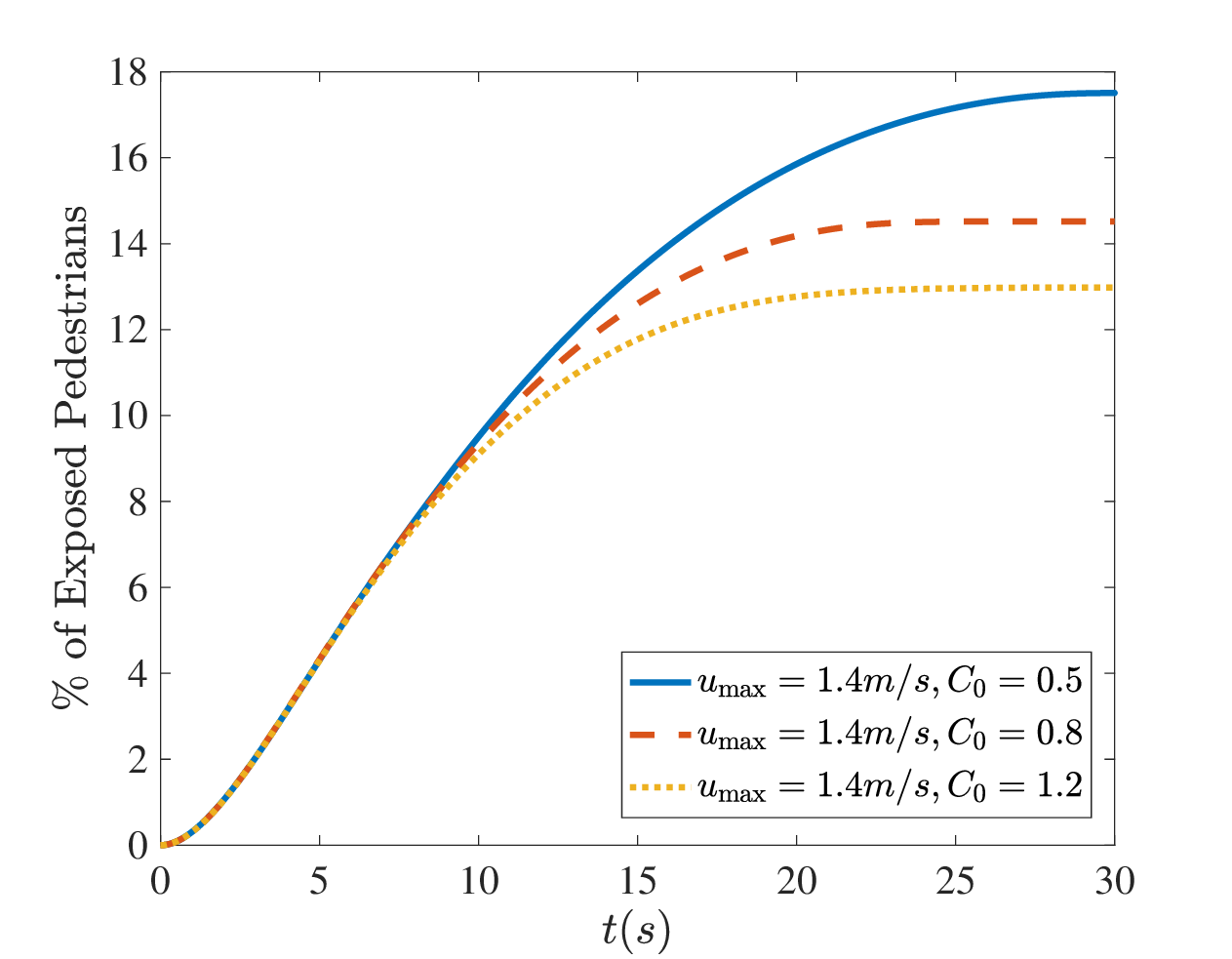}
\caption{Time evolution of the total number of pedestrians $R(t)$ (left) and percentage of exposed pedestrians (right) with $u_{\max}=2$ $m/s $ (top) and $u_{\max}=1.4$ $m/s $ (bottom),  for three different values of $C_0$ with $\rho_0^V = 0$. }
 \label{fig2}
 \end{figure}

Initially, we perform simulations in this test scenario, where  we assume that there is no ventilation imposed 
 in the facility. In Fig. \ref{fig2} we present the total evacuation times for	each	value	of	$C_0$ along with the predicted percentage of the exposed pedestrians when $\rho_0^V = 0$, i.e., with no masked/vaccinated pedestrians included.	As		it	can	be		observed  in Fig. \ref{fig2} , as the maximum desired	velocity increases the total percentage of exposed pedestrians decreases, around between 3$\%$ and 5$\%$, mainly as a result of the decreasing evacuation time, which is reasonably expected for increasing maximum speed. We note that, at an outflow boundary (exit)  we keep track of the amount of exposed pedestrians that has exited the domain.  
 As expected, after all pedestrians have exited the domain the total number of predicted exposed pedestrians remains constant	in	each	case.
 
\begin{figure}[h!]
\centering
 \includegraphics[width=4.cm,height=11cm]{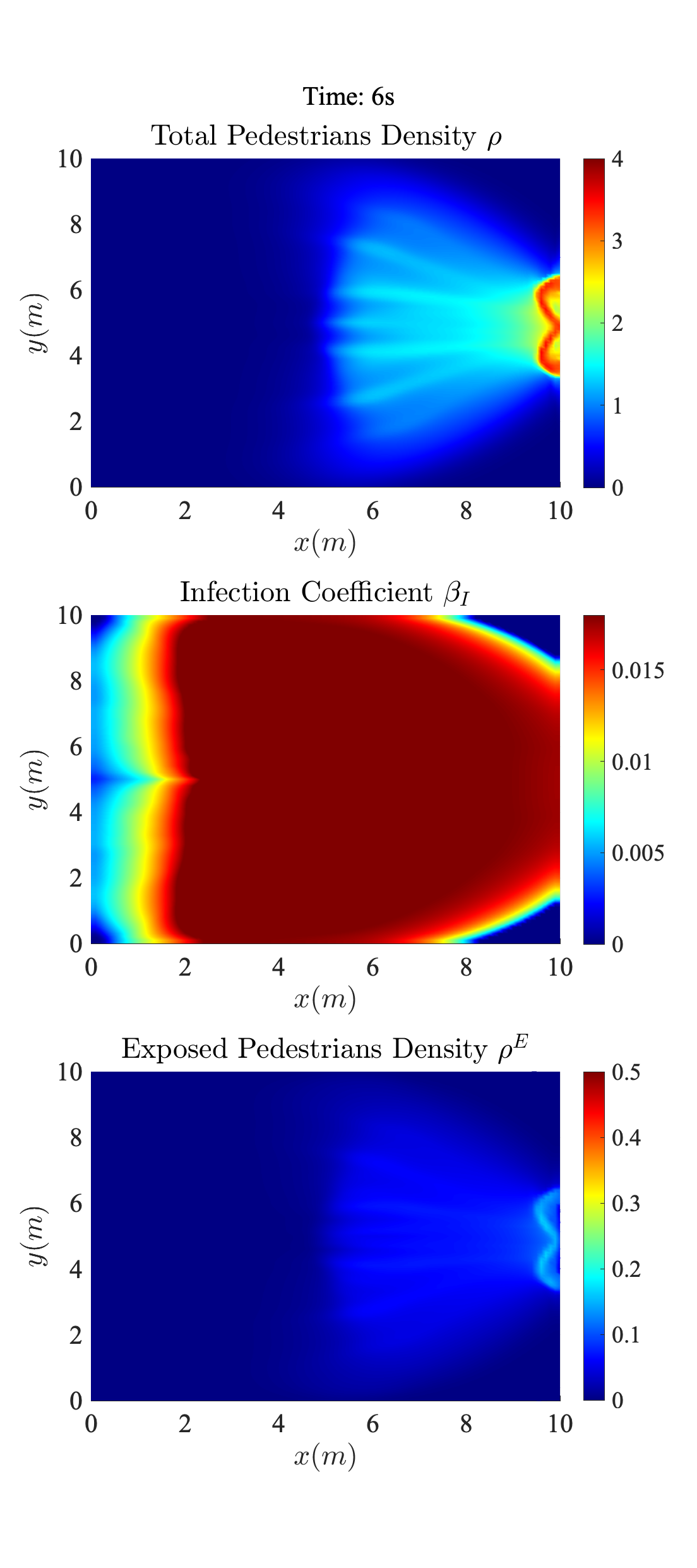} \hspace{-0.16in}
 \includegraphics[width=4.cm,height=11cm]{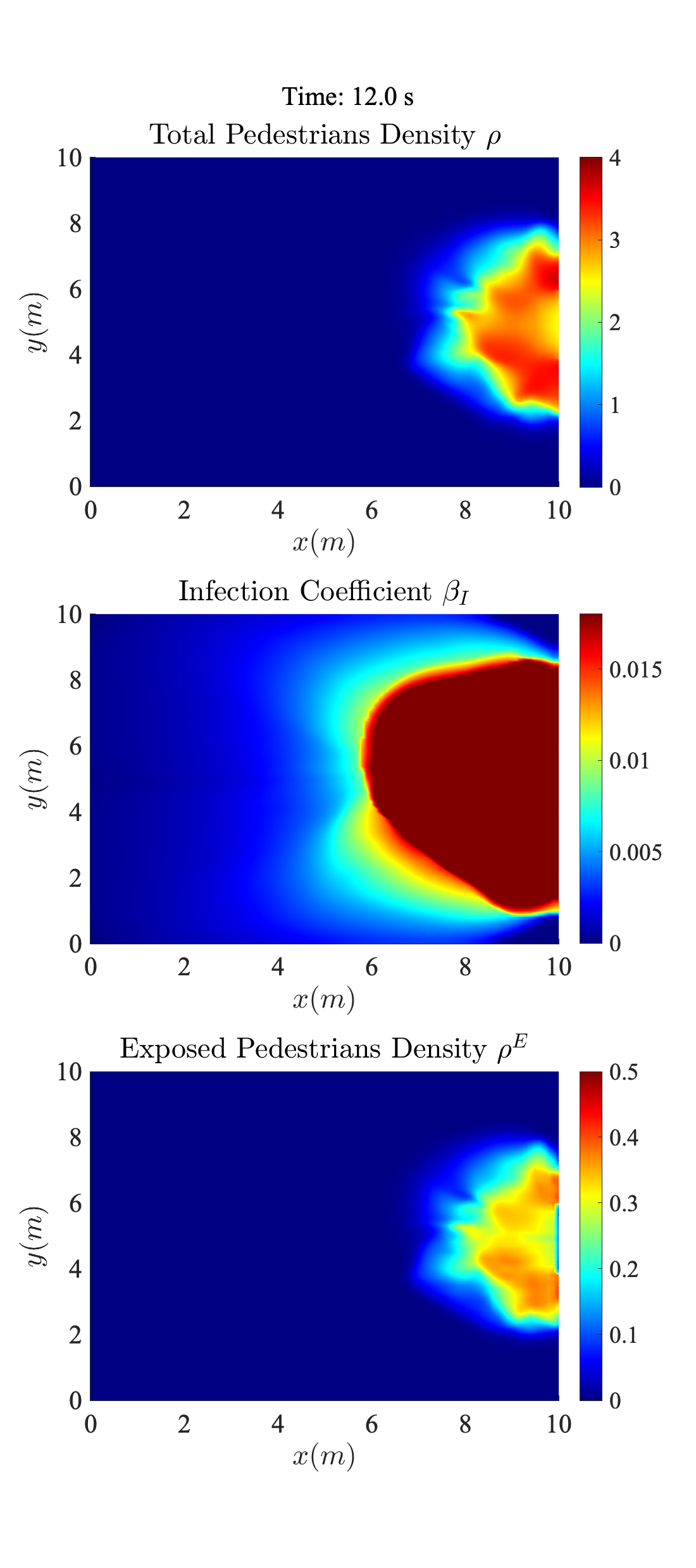}\hspace{-0.1in}
 \includegraphics[width=4.cm,height=11cm]{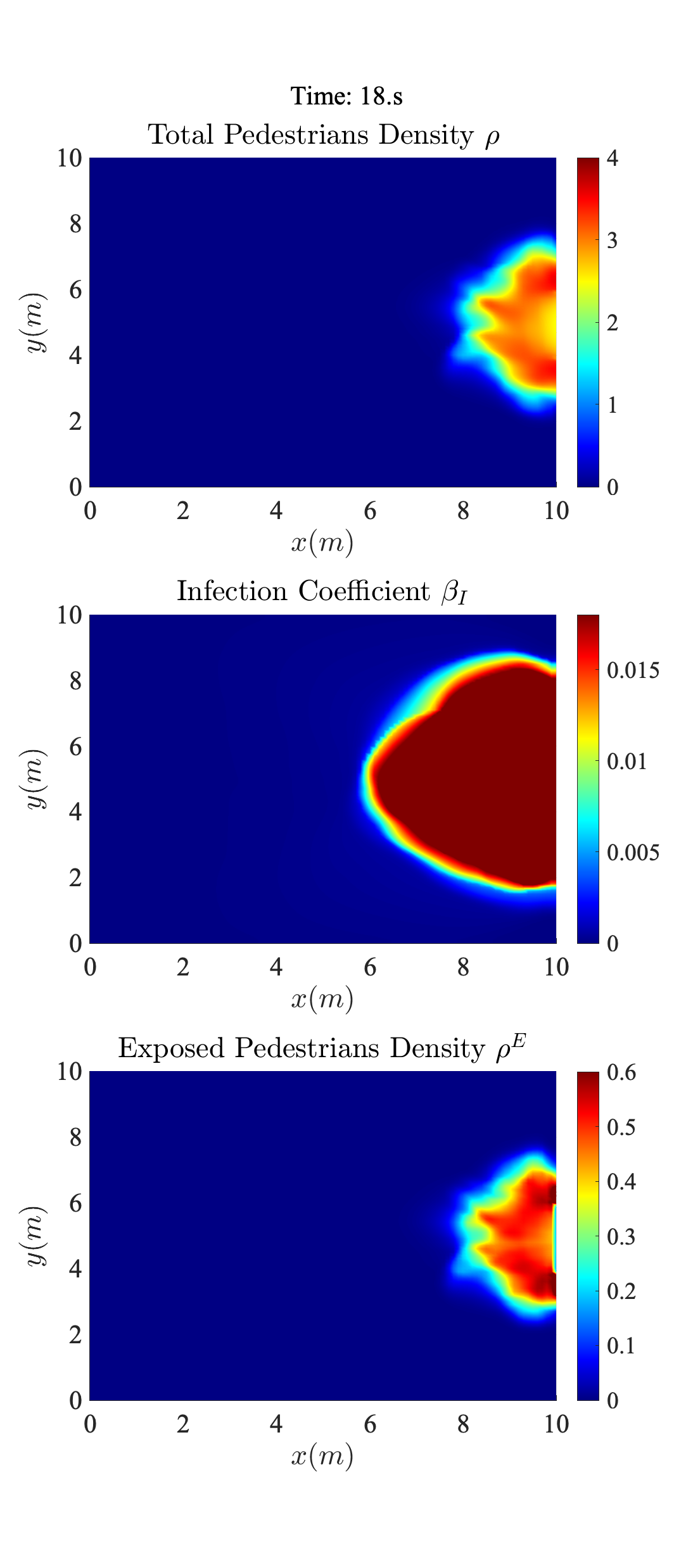}
\caption{Total density profiles (top), infection coefficient profile (middle) and exposed pedestrians' profile at different time instances with $u_{\max}=1.4$ $m/s $, $C_0=0.5$, with no ventilation and $\rho_0^V = 0$.}
 \label{fig2d1}
 \end{figure}

\begin{figure}[h!]
\centering
 \includegraphics[width=4.cm,height=10cm]{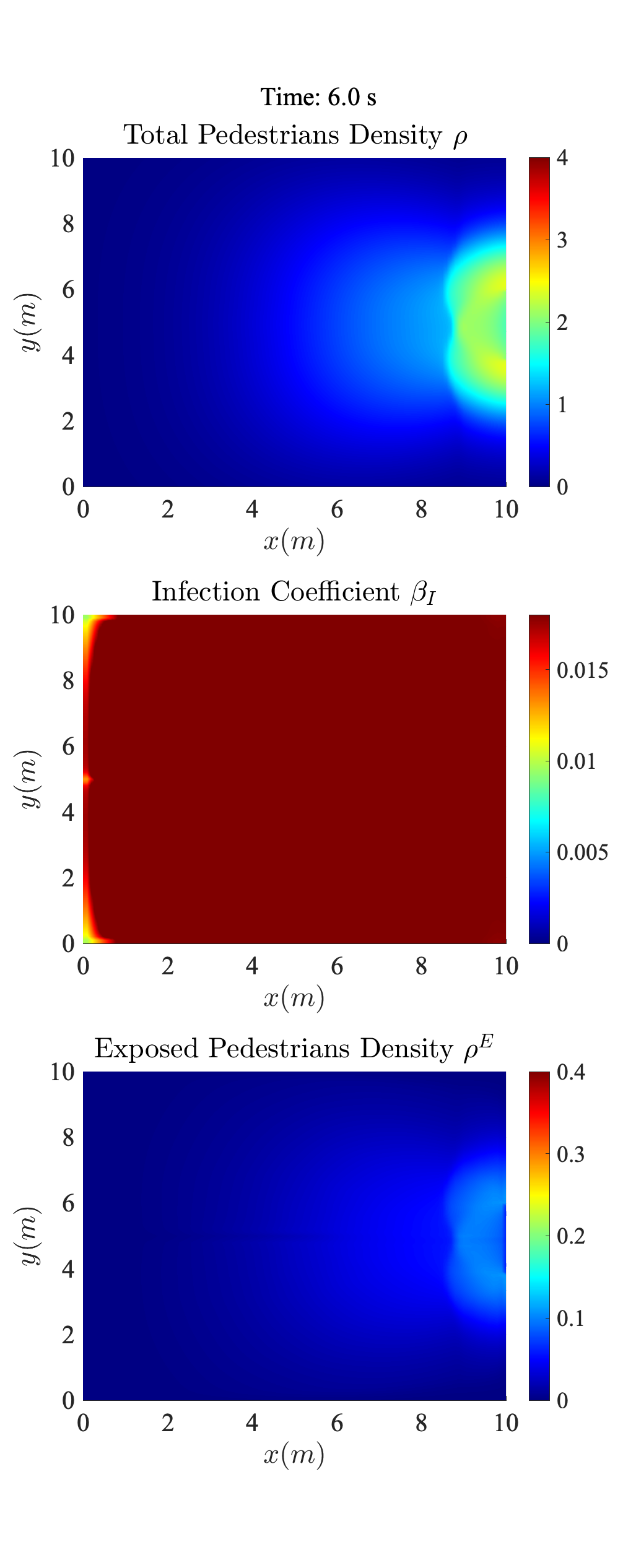} \hspace{-0.16in}
 \includegraphics[width=4.cm,height=10cm]{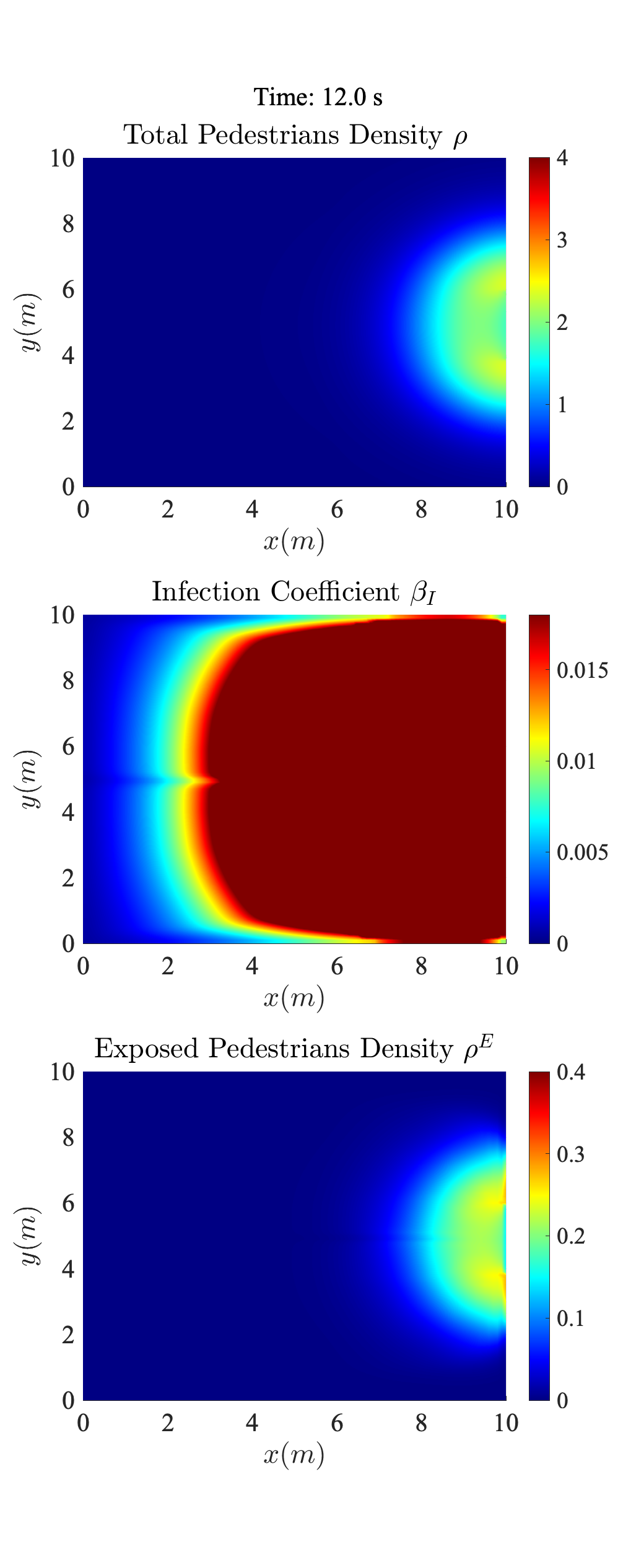}\hspace{-0.1in}
 \includegraphics[width=4.cm,height=10cm]{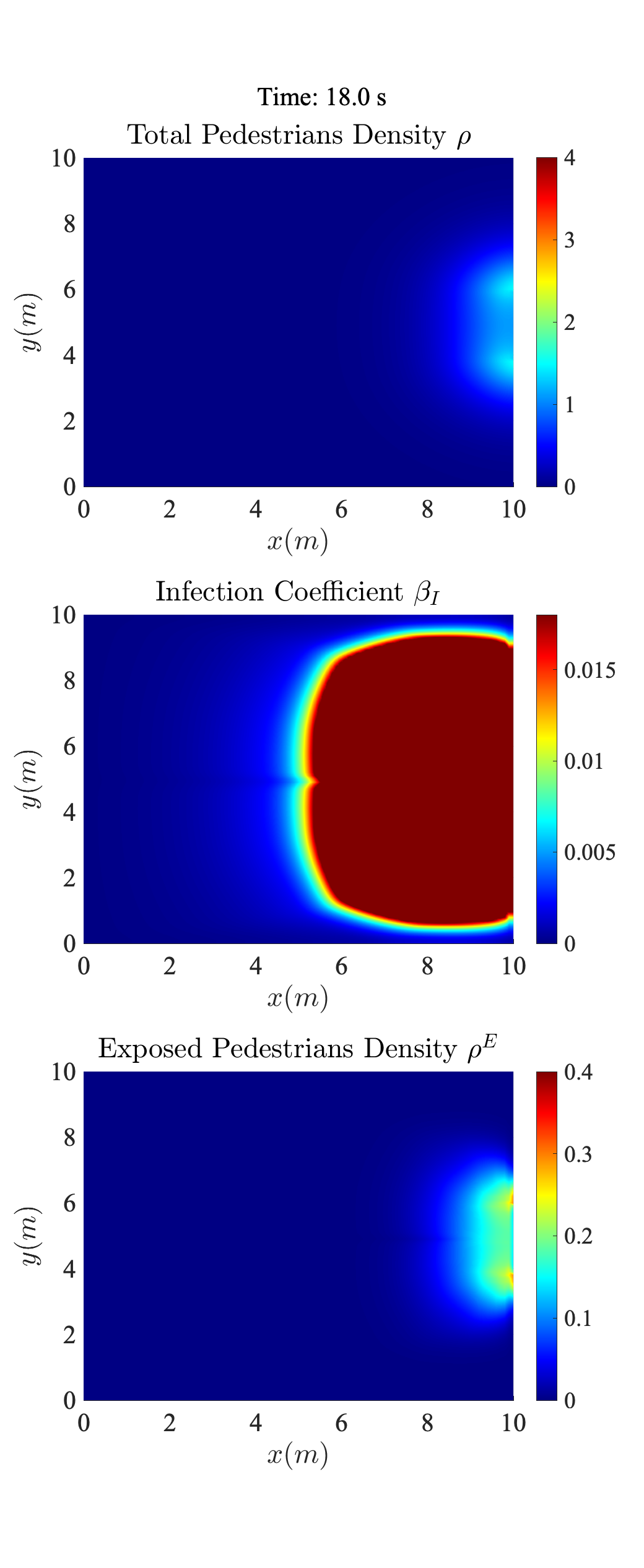}
\caption{Total density profiles (top), infection coefficient profile (middle) and exposed pedestrians' profile at different time instances with $u_{\max}=1.4$ $m/s $, $C_0=1.2$, with no ventilation and  $\rho_0^V = 0$. }
 \label{fig2d2}
 \end{figure}

  We further observe that as the anticipation factor increases the percentage of total exposed individuals decreases as a result of increasing average distances between individuals (dictated by the pressure function in the crowd flow model). Such an increase also results in a decrease in the evacuation time, which is explained as larger average distances between individuals may also lead to smoother overall flow, and thus, also, for example, to less evident clogging effects	at	the	exit.  This		is	evident also in Figs \ref{fig2d1} and \ref{fig2d2} where total density profile, infection coefficient profile, and exposed pedestrians' density, at different time instances with $u_{\max}=1.4$ $m/s $, are presented for two different values of $C_0$. Comparing the  results in Figs \ref{fig2d1} and \ref{fig2d2} we see that for smaller values of $C_0$, i.e., allowing smaller distances between  pedestrians, causes stronger interactions and formation of inhomogeneities in the density distribution and, as a result,  we observe much higher densities, especially near the exit, which persist in time.	 On the other hand, increasing the value of $C_0$, higher repealing forces between pedestrians prevent high densities and formation  of  inhomogeneities in the density distribution, allowing for a smoother and faster evacuation. Further, in Figs  \ref{fig2d1} and \ref{fig2d2}, we can observe the evolution of the infection coefficient $\beta^I$ that follows the pedestrians' movement and, at the same time, spreads in the domain, due to diffusion effects as given in model (19).  

\begin{figure}[h!]
\centering
 \includegraphics[width=6.3cm,height=5.5cm]{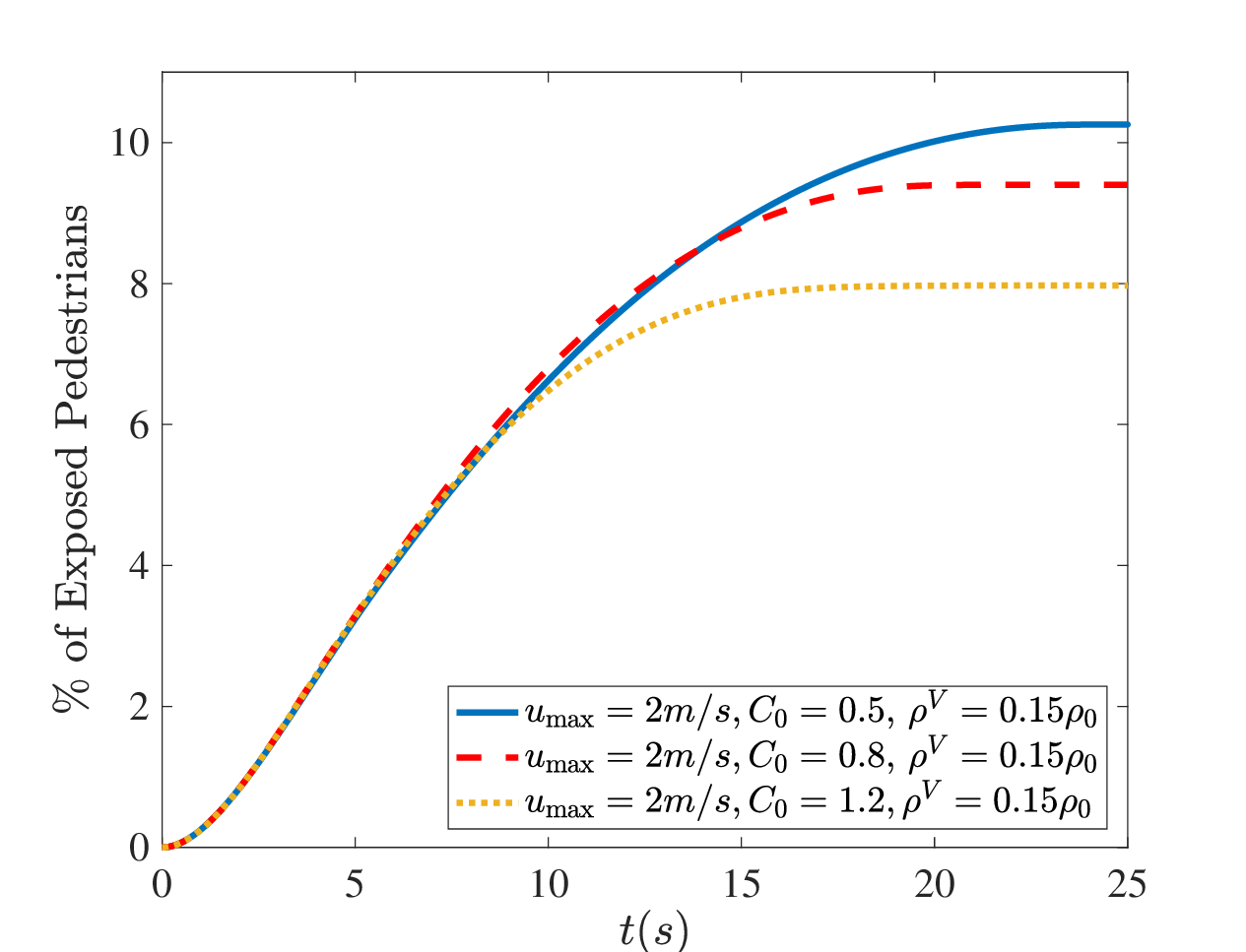} \hspace{-0.28in}
  \includegraphics[width=6.3cm,height=5.5cm]{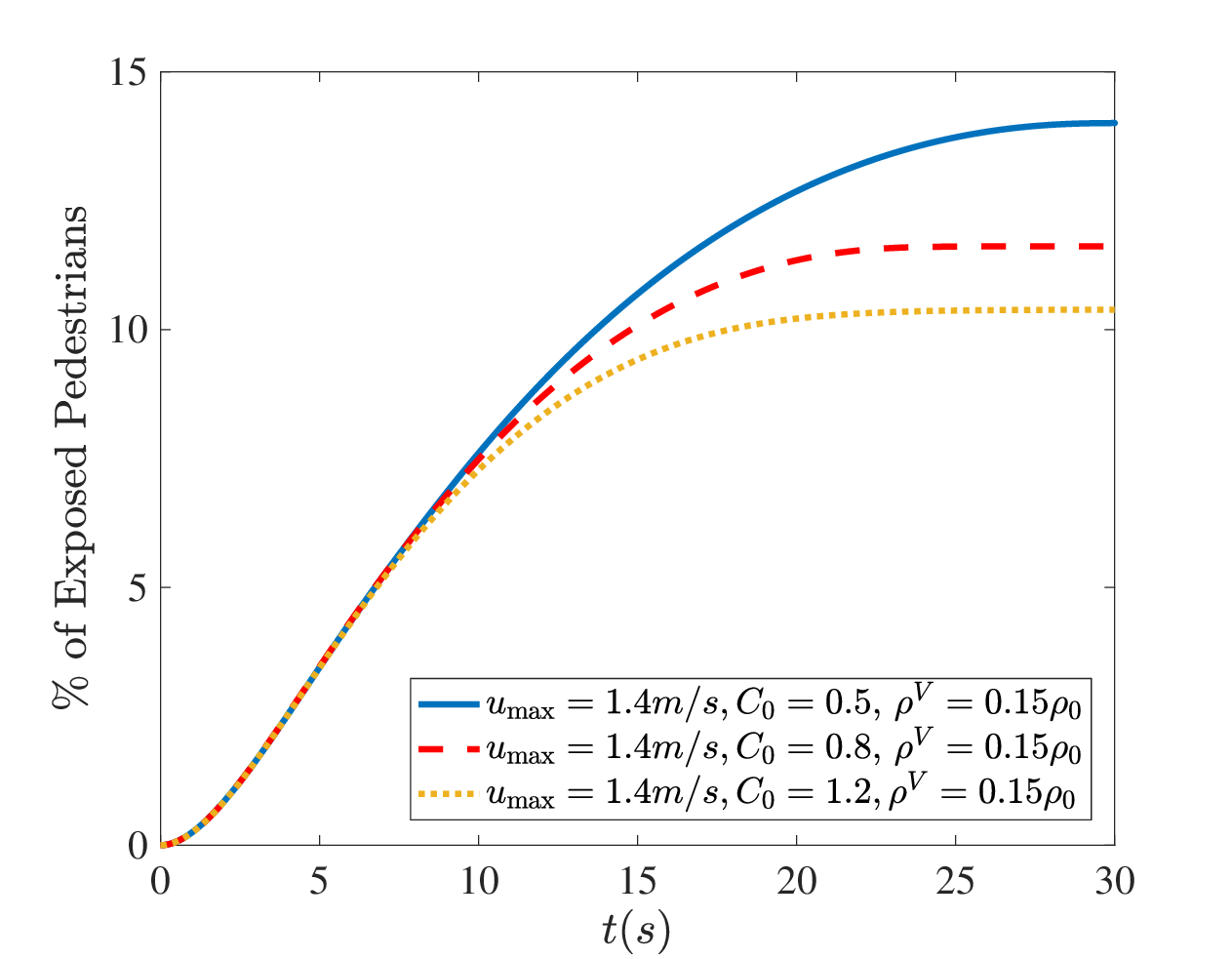}
  \caption{Time evolution of  the percentage  of exposed pedestrians  with $u_{\max}=2m/s $ (left) and $u_{\max}=1.4m/s $ (right), for three different values of $C_0$ for $\rho_0^V=0.15\rho_0$. }
 \label{fig3}
 \end{figure}

Next,  in Fig. \ref{fig3} we present the same simulations  but now assuming that  $15\%$ of the pedestrians are masked or vaccinated, i.e., $\rho_0^V = 0.15\rho_0$. The effect of incorporating masked/vaccinated individuals is a decrease in the total number of exposed individuals, which is reasonably expected. Comparing with the percentage of exposed pedestrians in Fig. \ref{fig2}, a decrease of about 2-5$\%$ is observed.

 \subsubsection{ Effect of Different Ventilation Air-Flow Rates and Directions}
\begin{figure}[h!]
\centering
 \includegraphics[width=6.3cm,height=5.5cm]{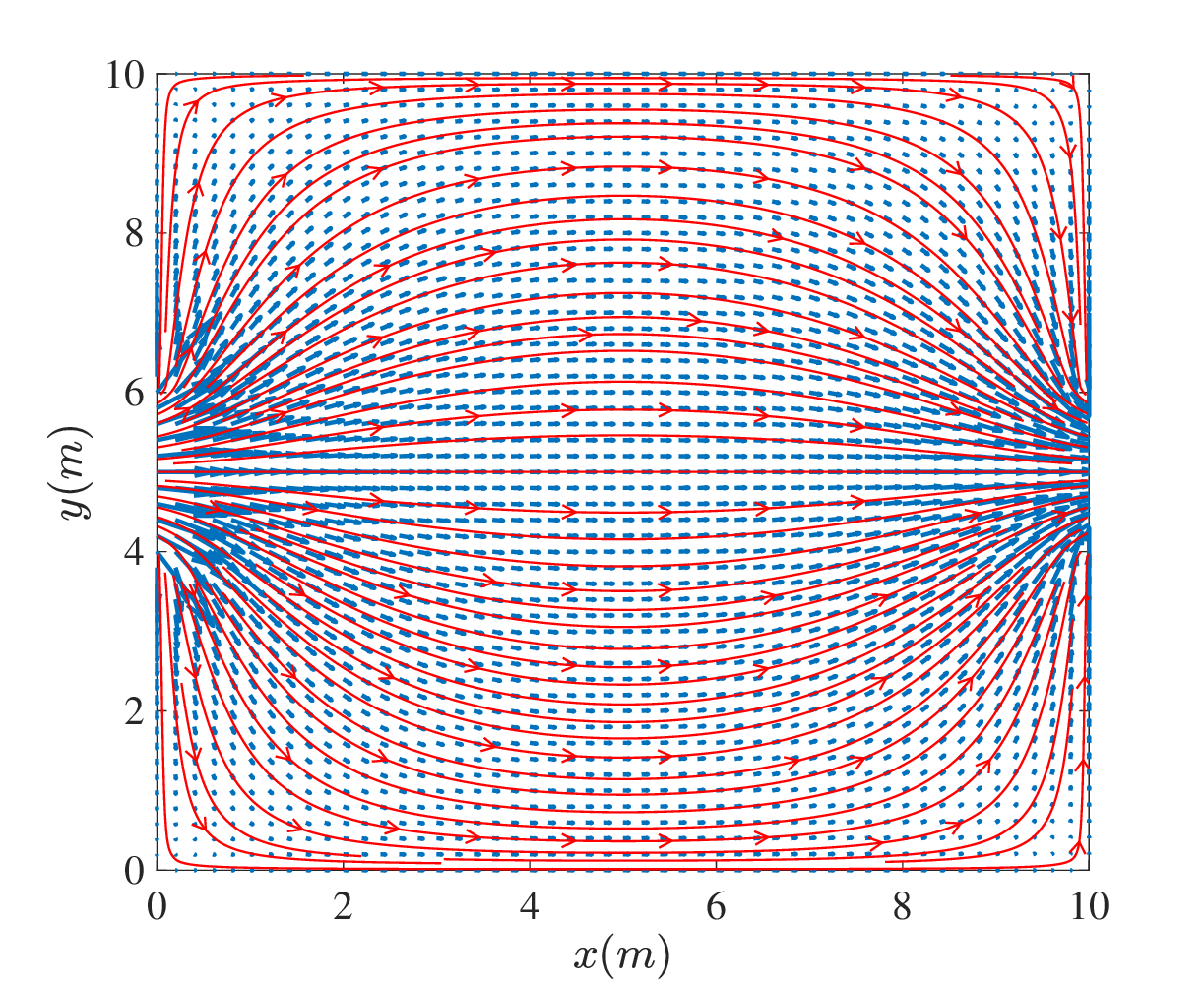} \hspace{-0.29in}
  \includegraphics[width=6.3cm,height=5.5cm]{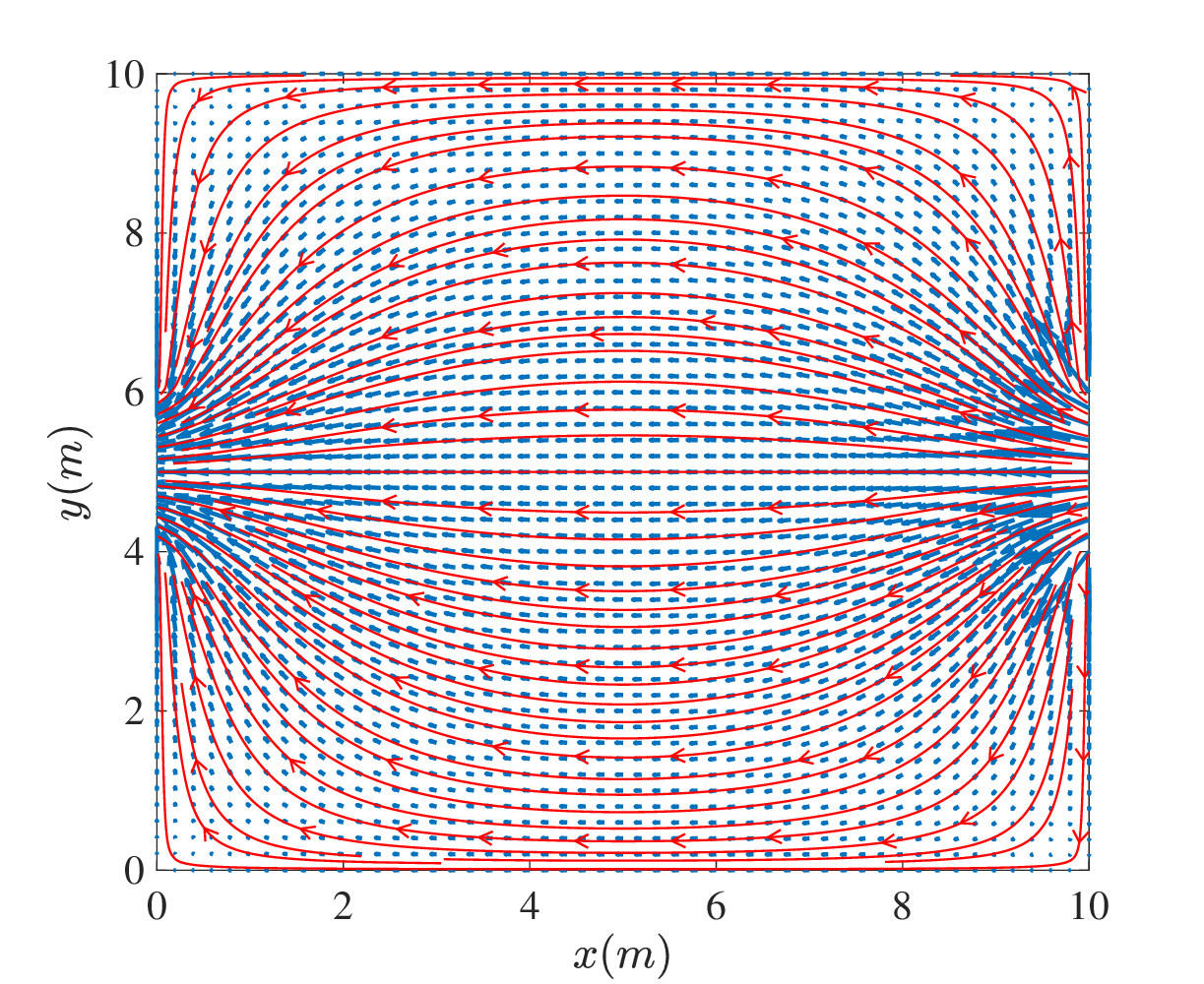}
  \caption{Velocity field ${\bf U}_G$ vectors (in blue) and the corresponding streamlines  for a ventilated room along the pedestrians' moving direction (left) and against (right)  for $u_{in}=10$ $ m/s$.}
 \label{fig4}
 \end{figure}

To study the effect of ventilation  imposed in the room, we assume  one case with an inflow ventilation duct at the middle of the left boundary and one exhaust  duct at the middle of the right boundary,  both being $2$ $m$ wide, producing an air-flow field along the pedestrians motion towards the exit. The second case reverses the position of inflow and outflow ducts producing an air-flow field against the pedestrians' motion.  To produce the steady velocity field ${\bf U}_G$  from (21), in the first case,  the derivative of the potential $\Psi$ normal to the ventilation boundaries was set equal to ${\bf u}_{in}=[u_{in}, 0]^{\rm T}$ for the left boundary and ${\bf u}_{out}=[-u_{in}, 0]^{\rm T}$ for the right boundary, as  to introduce the air-flow in the computational domain.  The opposite  is done for the second case. We consider two settings for the  ventilation speed, one  with $u_{in}=5$ $m/s$, and one with $u_{in}=10$ $m/s$ (that are practically realistic values, see, e.g., \cite{Ashare}). The resulting air-flow fields in both cases, along the pedestrian's motion and against it, are  shown in Fig. \ref{fig4} for $u_{in}=10$ $m/s$.
 
\begin{figure}[h!]
\centering
 \includegraphics[width=4.cm,height=10cm]{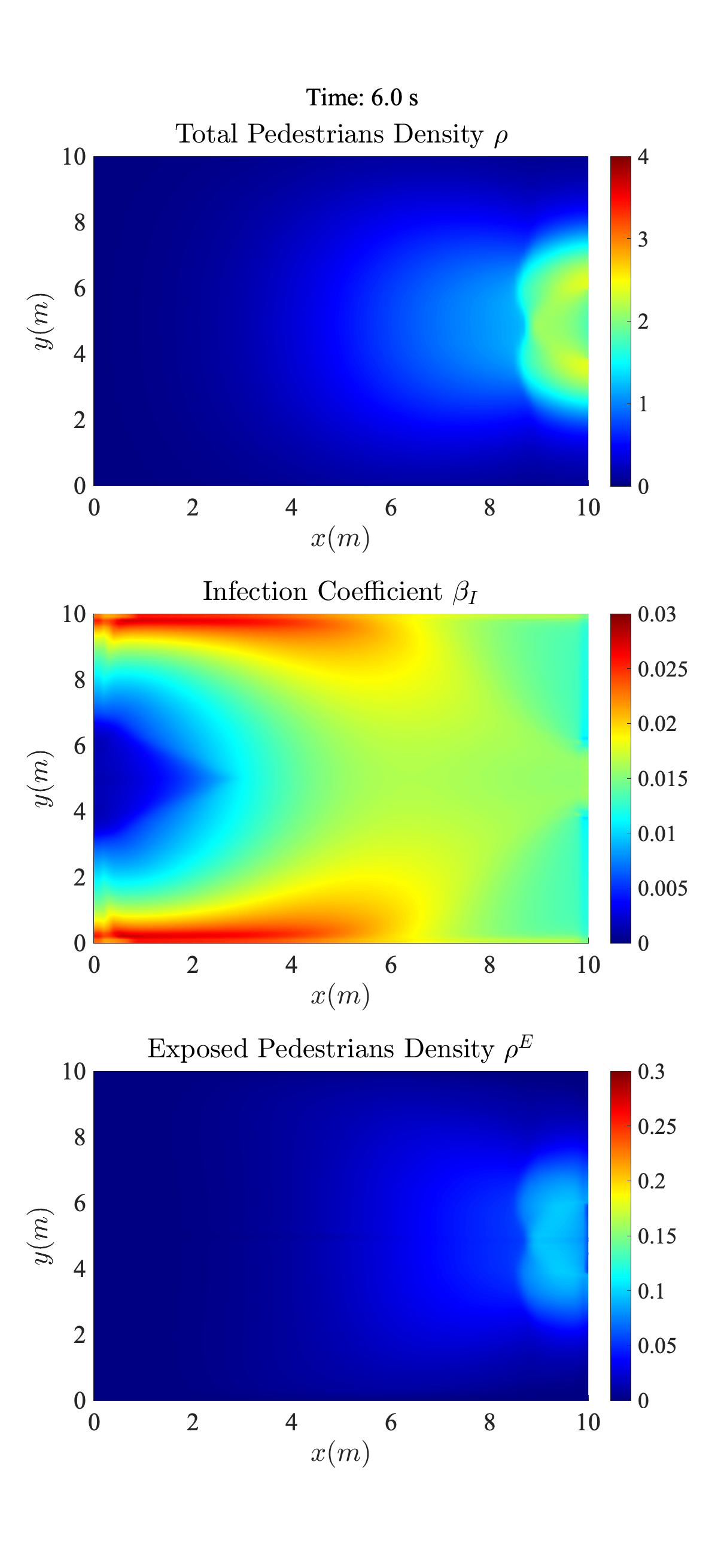} \hspace{-0.16in}
 \includegraphics[width=4.cm,height=10cm]{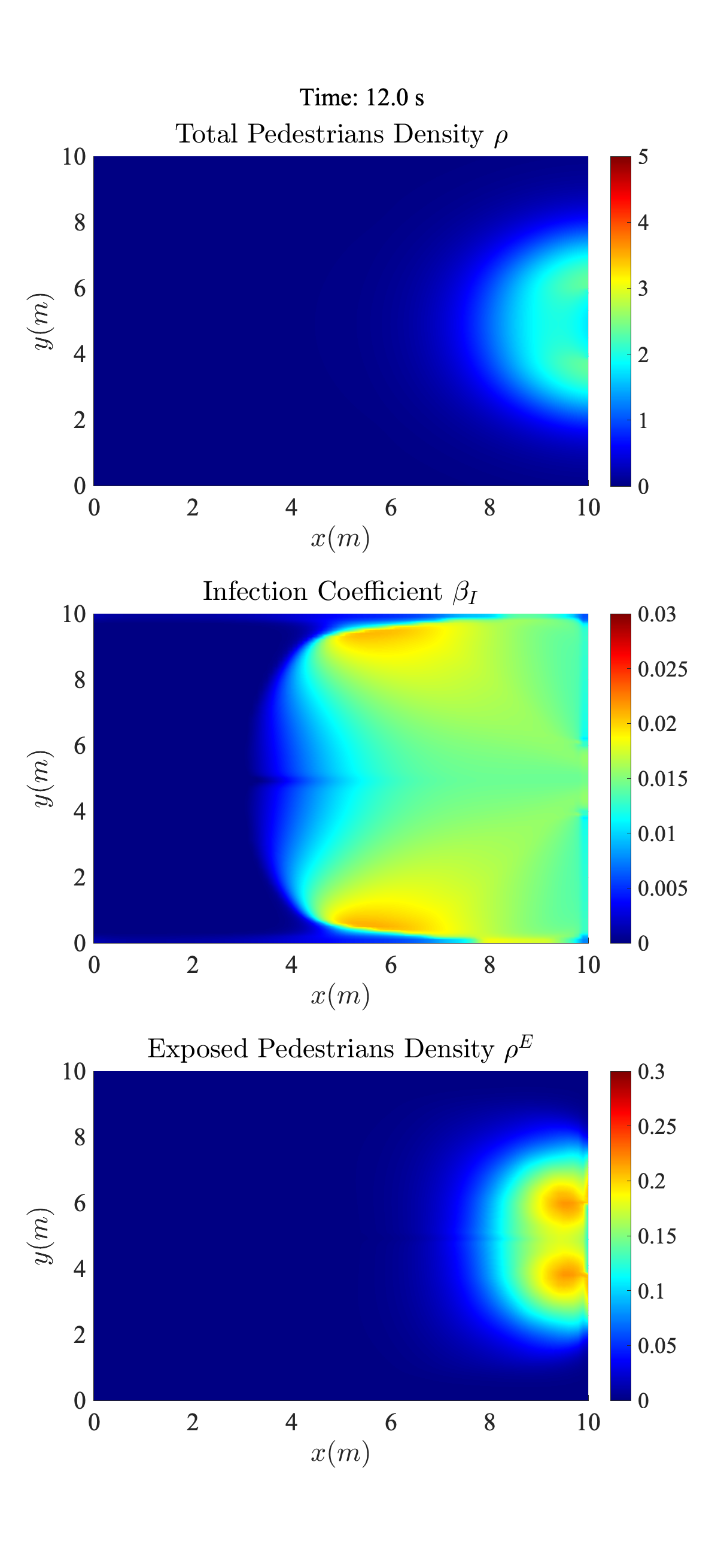}\hspace{-0.1in}
 \includegraphics[width=4.cm,height=10cm]{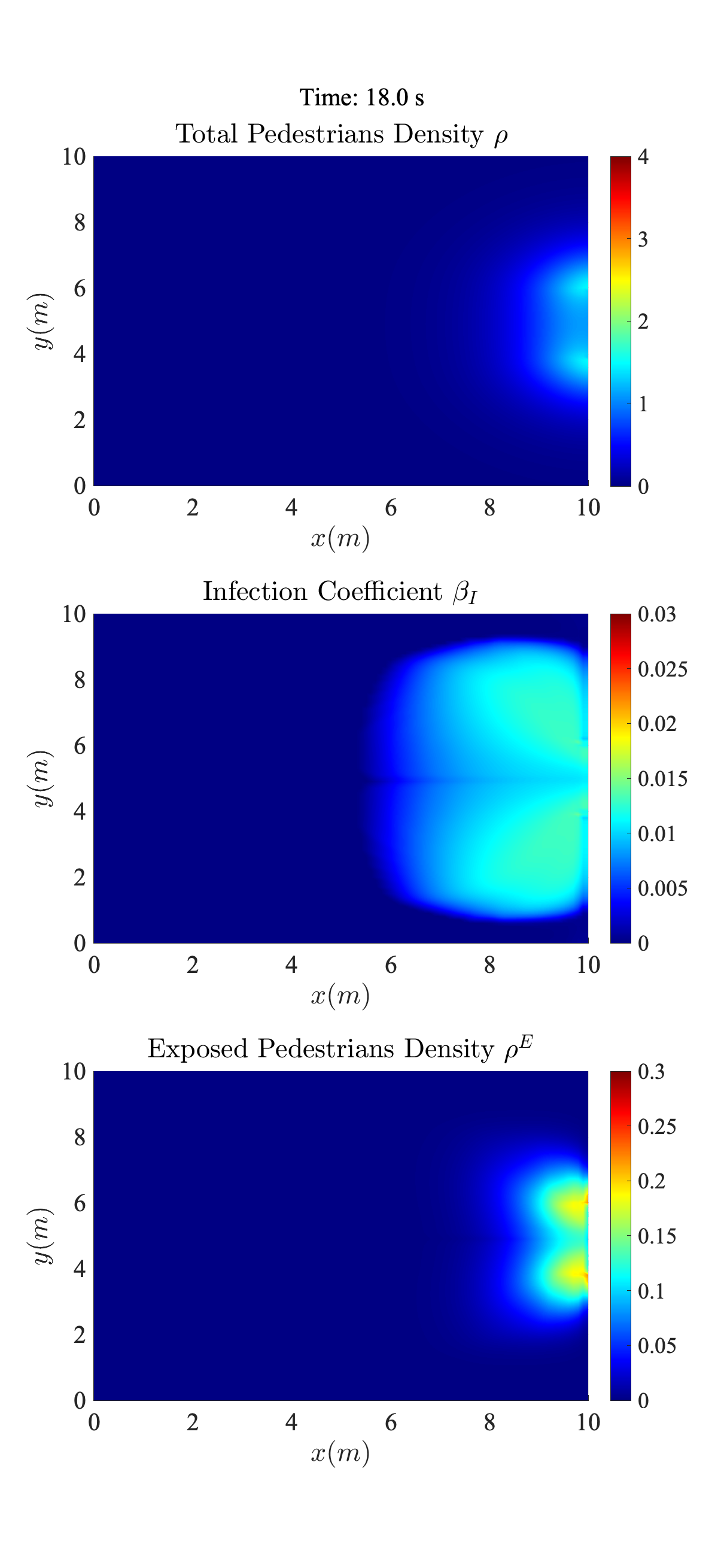}
\caption{Total density profiles (top), infection coefficient profile (middle), and exposed pedestrians' profile at different time instances with $u_{\max}=1.4$ $m/s $, $C_0=1.2$, and ventilation along the pedestrians'  flow direction  with $u_{in}=10$ $ m/s$.}
 \label{fig2d3}
 \end{figure}

\begin{figure}[h!]
\centering
 \includegraphics[width=4.cm,height=10cm]{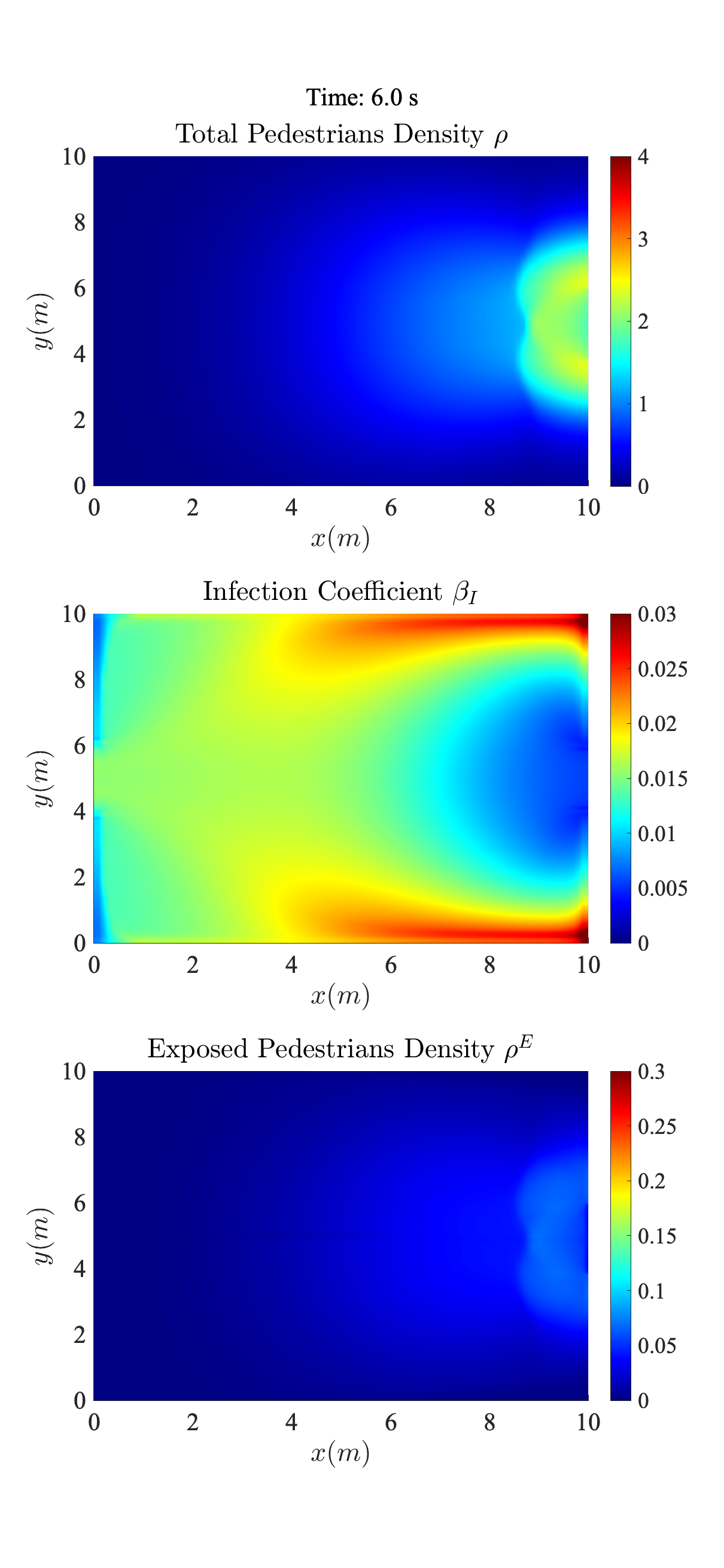} \hspace{-0.16in}
 \includegraphics[width=4.cm,height=10cm]{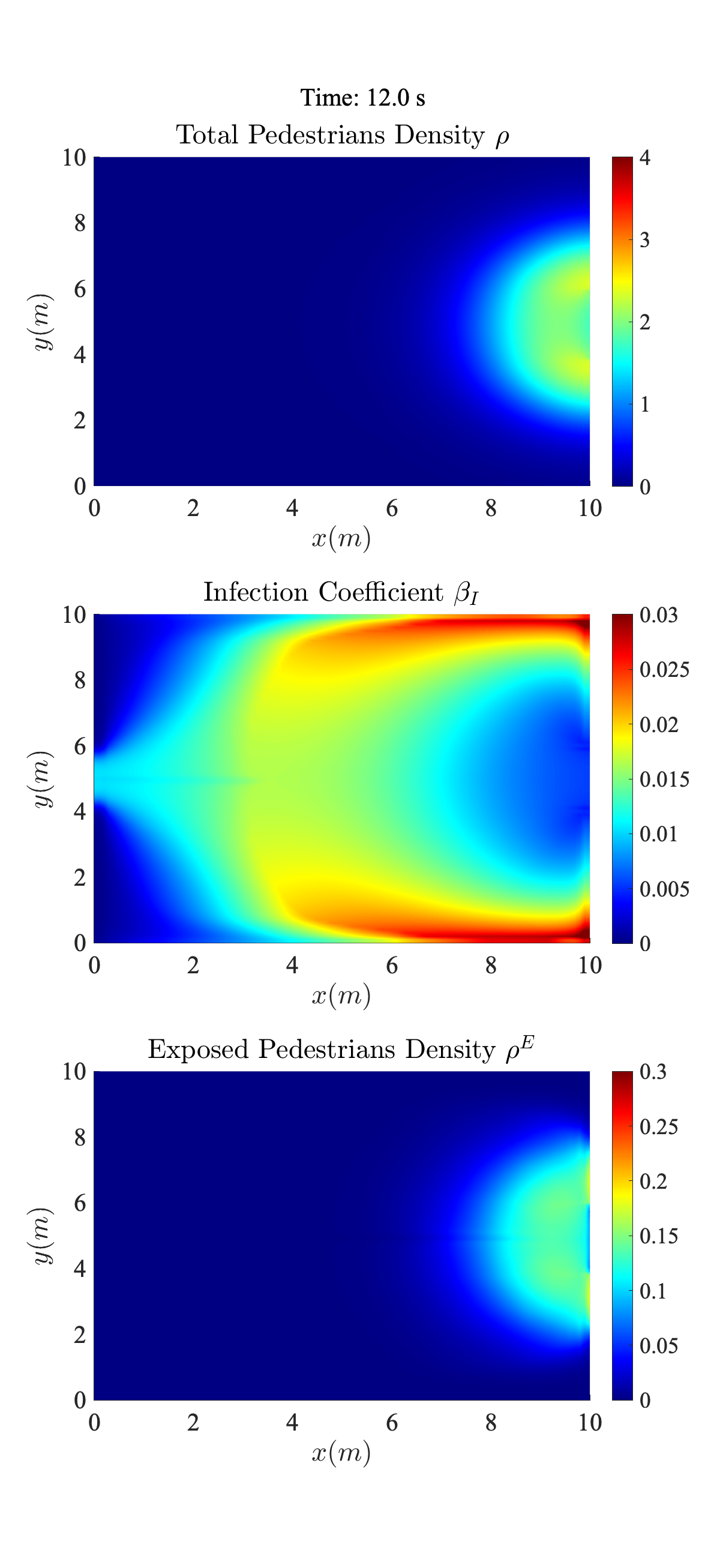}\hspace{-0.1in}
 \includegraphics[width=4.cm,height=10cm]{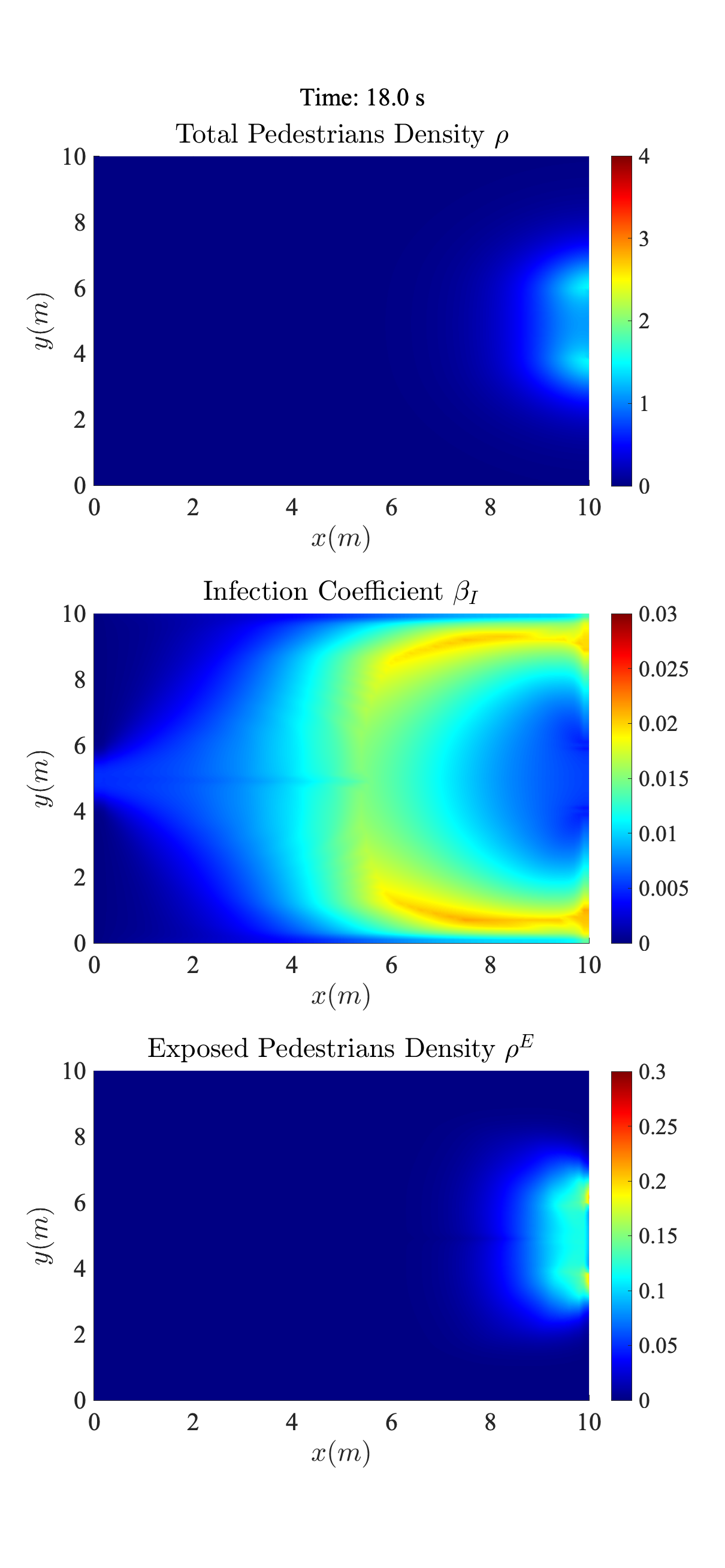}
\caption{Total density profiles (top), infection coefficient profile (middle), and exposed pedestrians' profile at different time instances with $u_{\max}=1.4$ $m/s $, $C_0=1.2$, and ventilation against the pedestrians' flow direction  with $u_{in}=10$ $m/s$.}
 \label{fig2d4}
 \end{figure}

 In Figs \ref{fig2d3} and \ref{fig2d4} we present, respectively,  with ventilation along  and against the pedestrians' flow direction,  the total pedestrians' density profiles, infection coefficient profile,  and exposed pedestrians' profile at different time instances for $u_{\max}=1.4$ $m/s $, $C_0=1.2$, and  $u_{in}=10$ $m/s$.  Although the evolution of the total pedestrians' movement is the same in each case, significant differences can be observed in the evolution of the infection coefficient in the room, affected by the  air-flow fields,  and, as a result, in that of the predicted exposed pedestrians.  In particular, we observe from Fig. \ref{fig2d4}  that people in this case, with air-flow against pedestrians' direction, move at locations where the values of $\beta^I$ is smaller, which is a result of air being cleaner at locations close to the inlet ventilation duct; while the opposite is true in the case where the air-flow is along their direction, as in Fig. \ref{fig2d3}. Further, since there exists a significant congestion (clogging) near the exit, especially for smaller values of $C_0$, this ``pocket" of clean air near the exit is expected to have a profound effect in the number of  exposed pedestrians. This observation is also consistent with, for example, \cite{Qian}.

\begin{figure}[h!]
\centering
 \includegraphics[width=6.32cm,height=5.35cm]{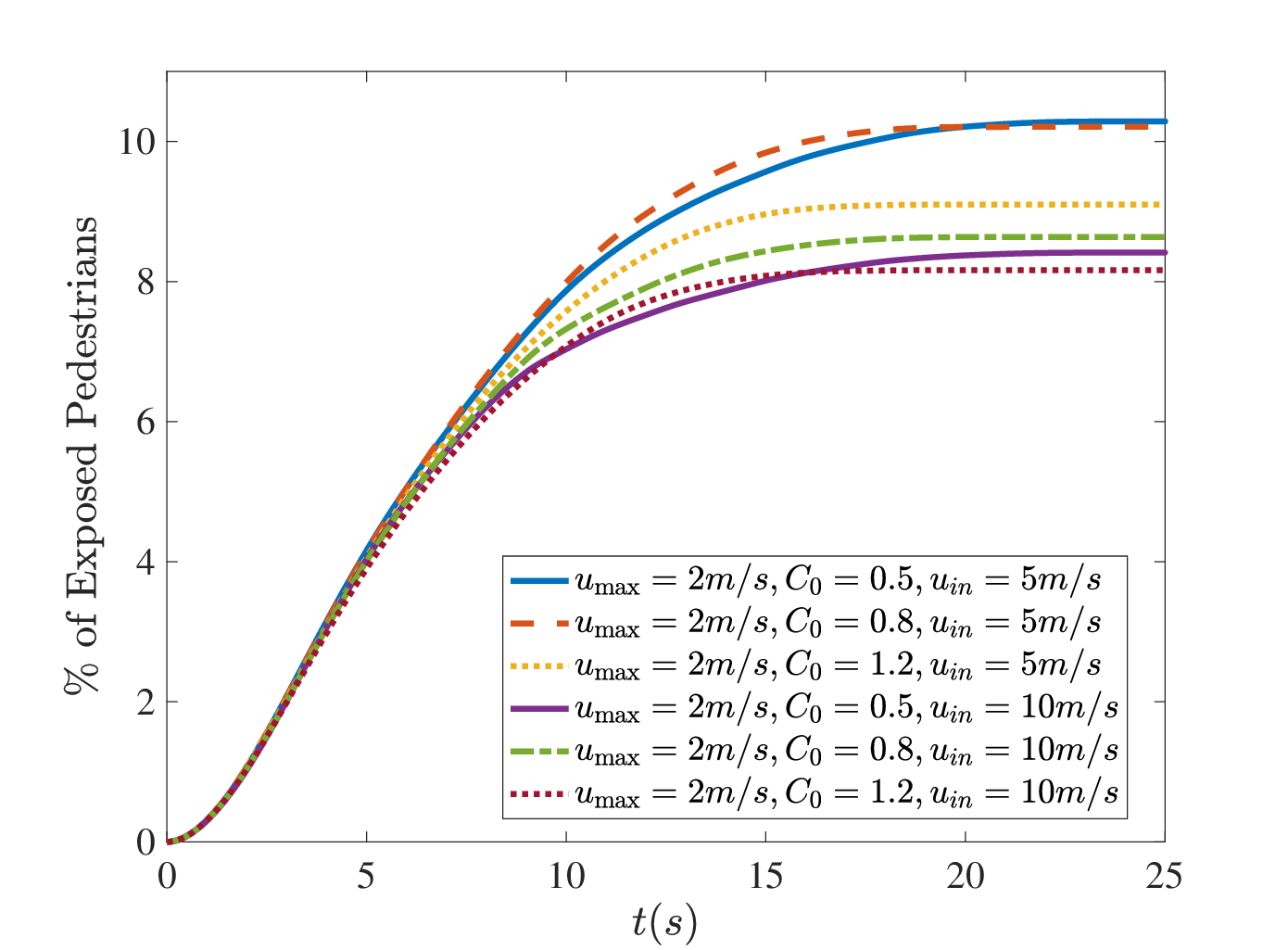} \hspace{-0.29in}
  \includegraphics[width=6.32cm,height=5.35cm]{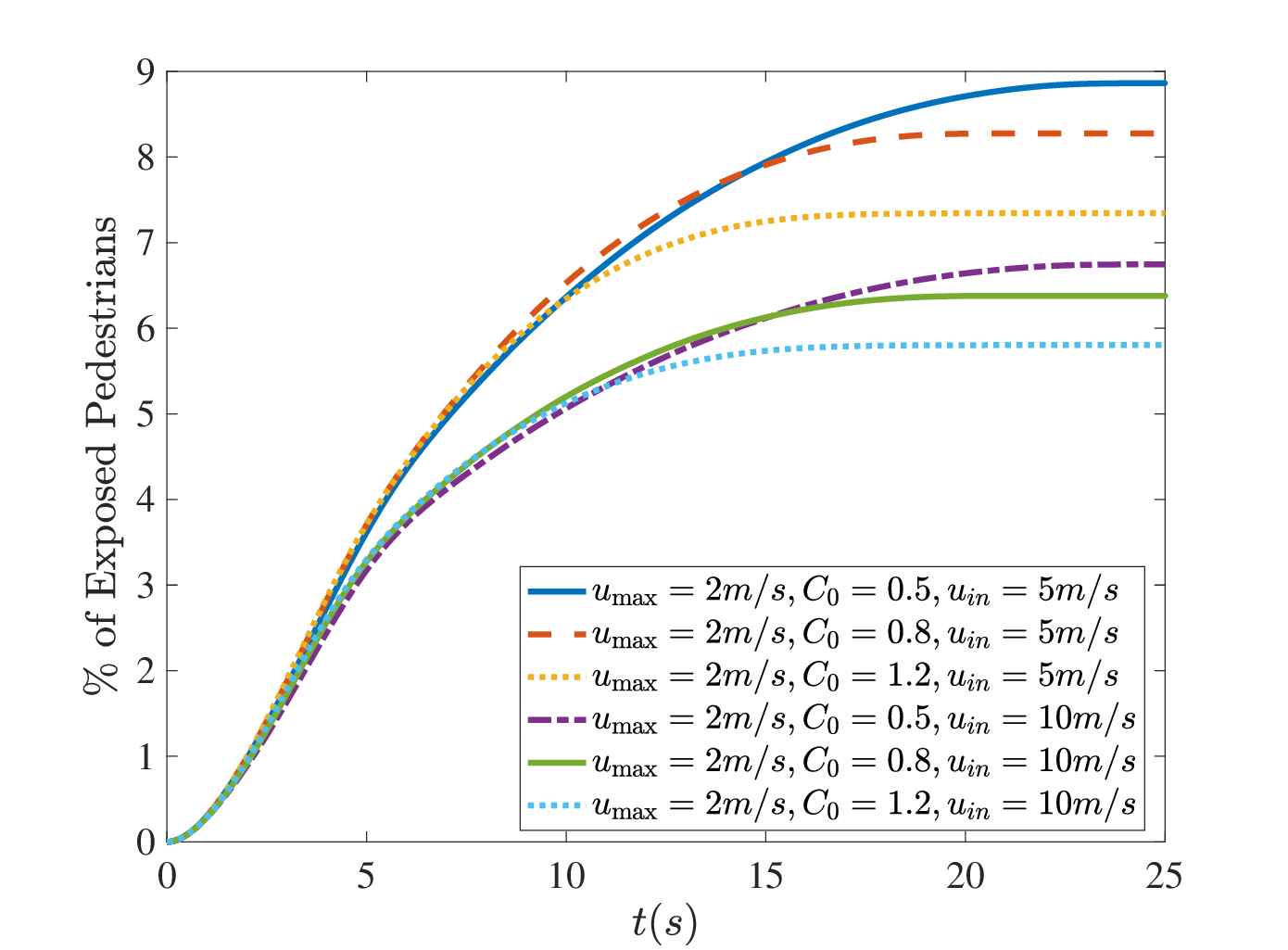}
  \caption{Percentage of exposed pedestrians in a ventilated room for two different ventilation speeds, along the pedestrians' direction (left) and against (right,) with $\rho_0^V=0$ and  $u_{\max}=2$ $m/s$. }
 \label{fig5}
 \end{figure}

\begin{figure}[h!]
\centering
 \includegraphics[width=6.32cm,height=5.35cm]{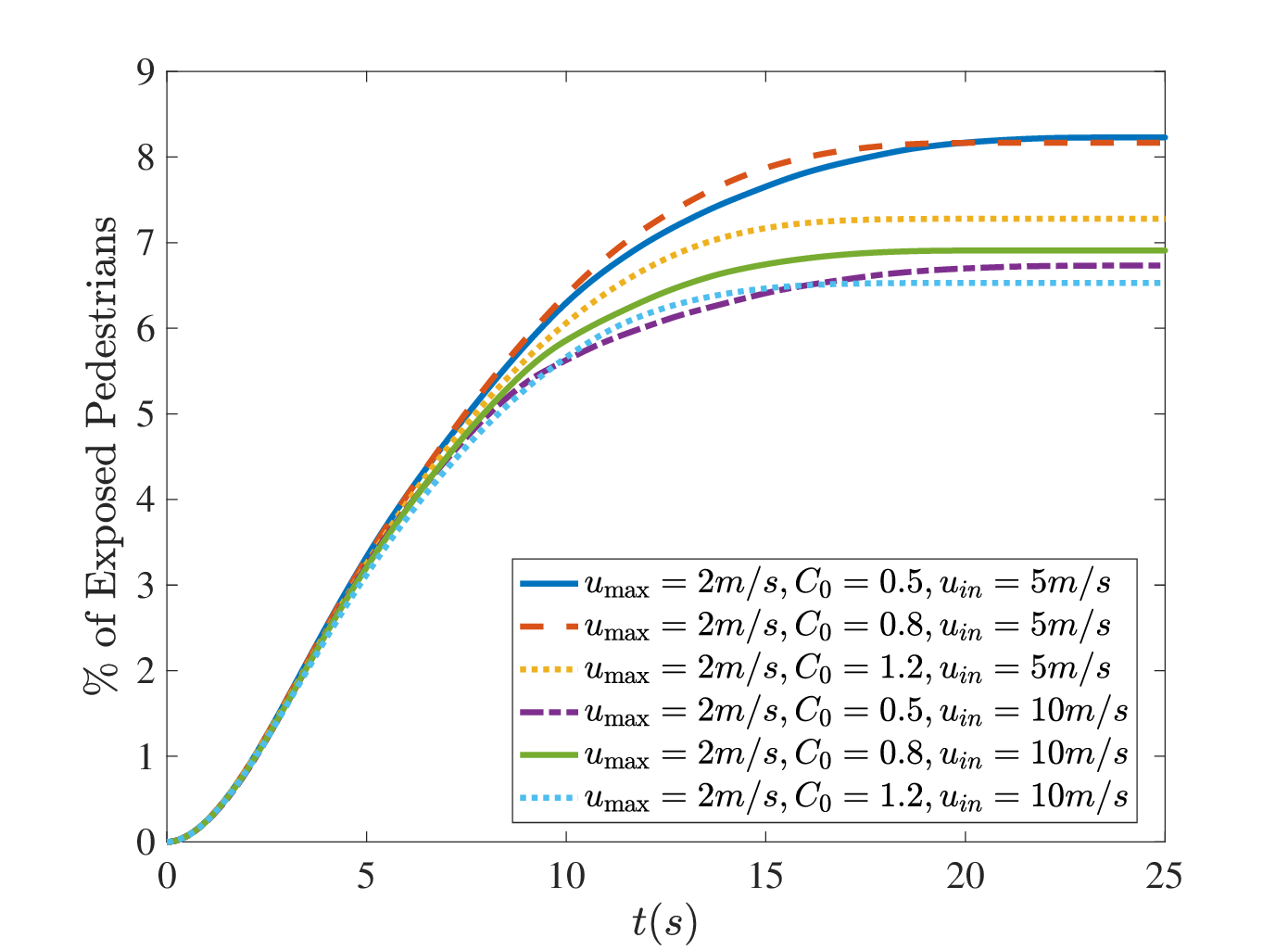} \hspace{-0.29in}
 \includegraphics[width=6.32cm,height=5.35cm]{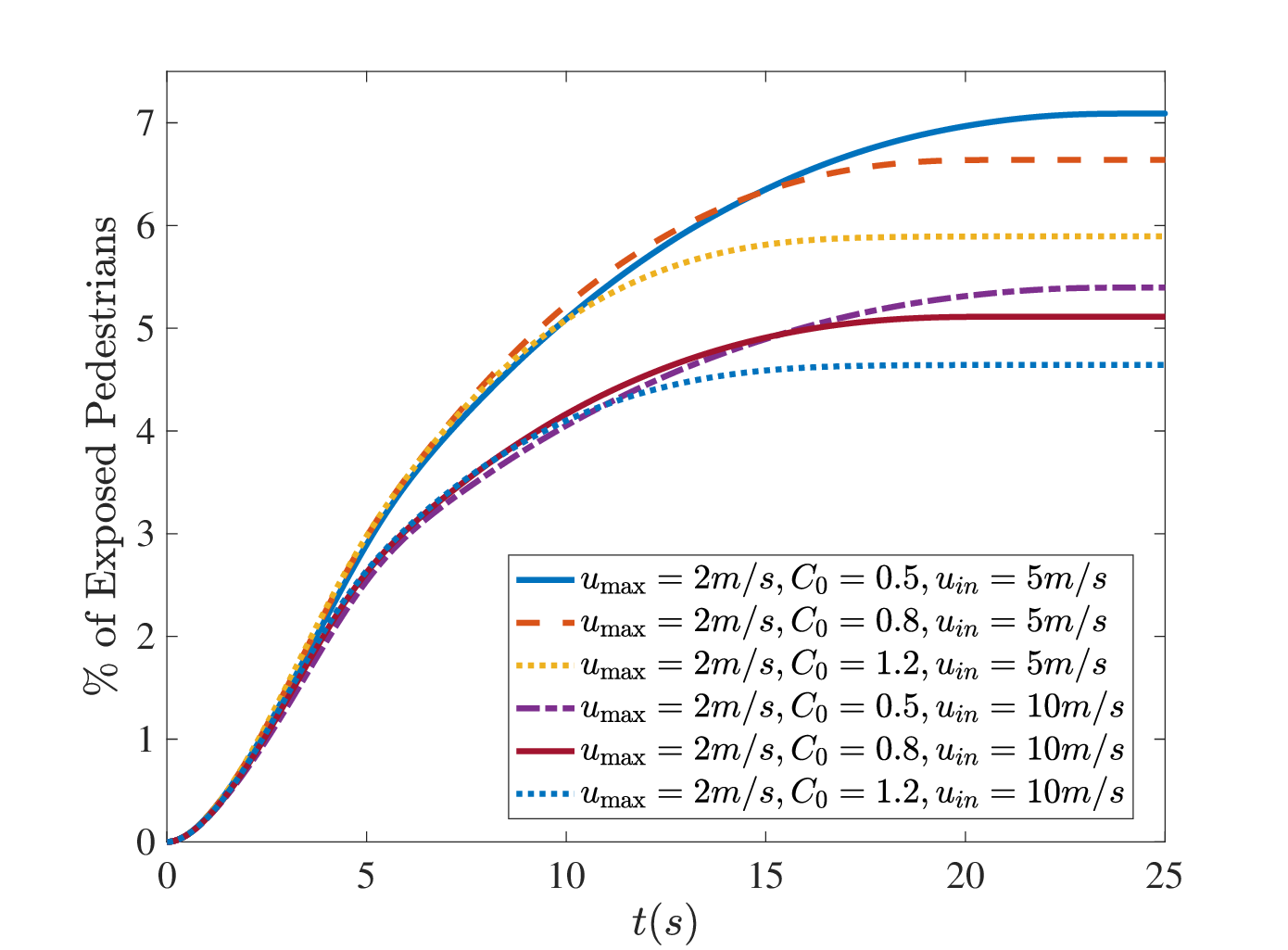}
  \caption{Percentage of exposed pedestrians in a ventilated room for two different ventilation speeds, along the pedestrians' direction (left) and against (right), with $\rho_0^V=0.15\rho_0$  and $u_{\max}=2$ $m/s$. }
 \label{fig6}
 \end{figure}

\begin{figure}[h!]
\centering
 \includegraphics[width=6.32cm,height=5.35cm]{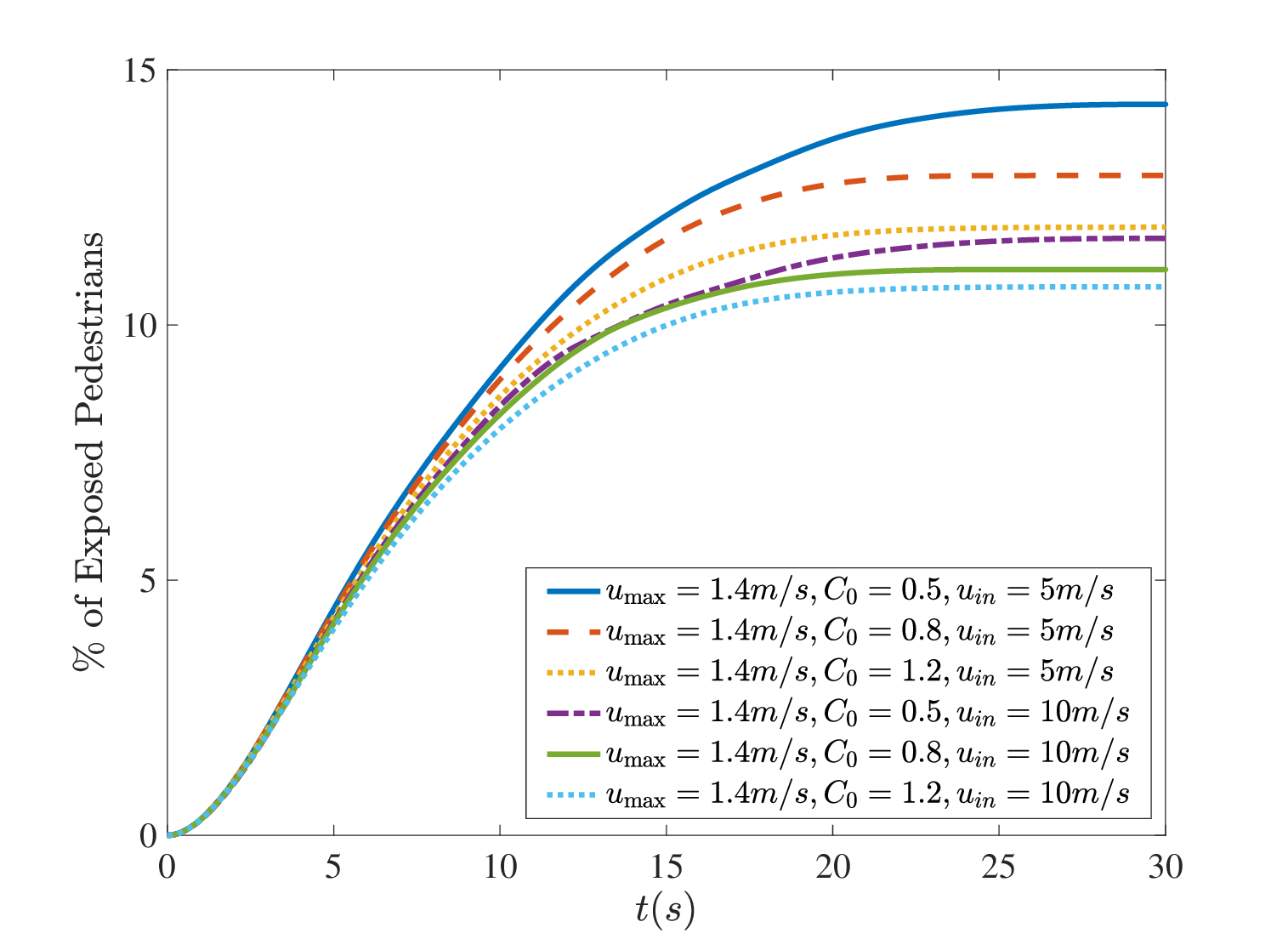} \hspace{-0.29in}
  \includegraphics[width=6.32cm,height=5.35cm]{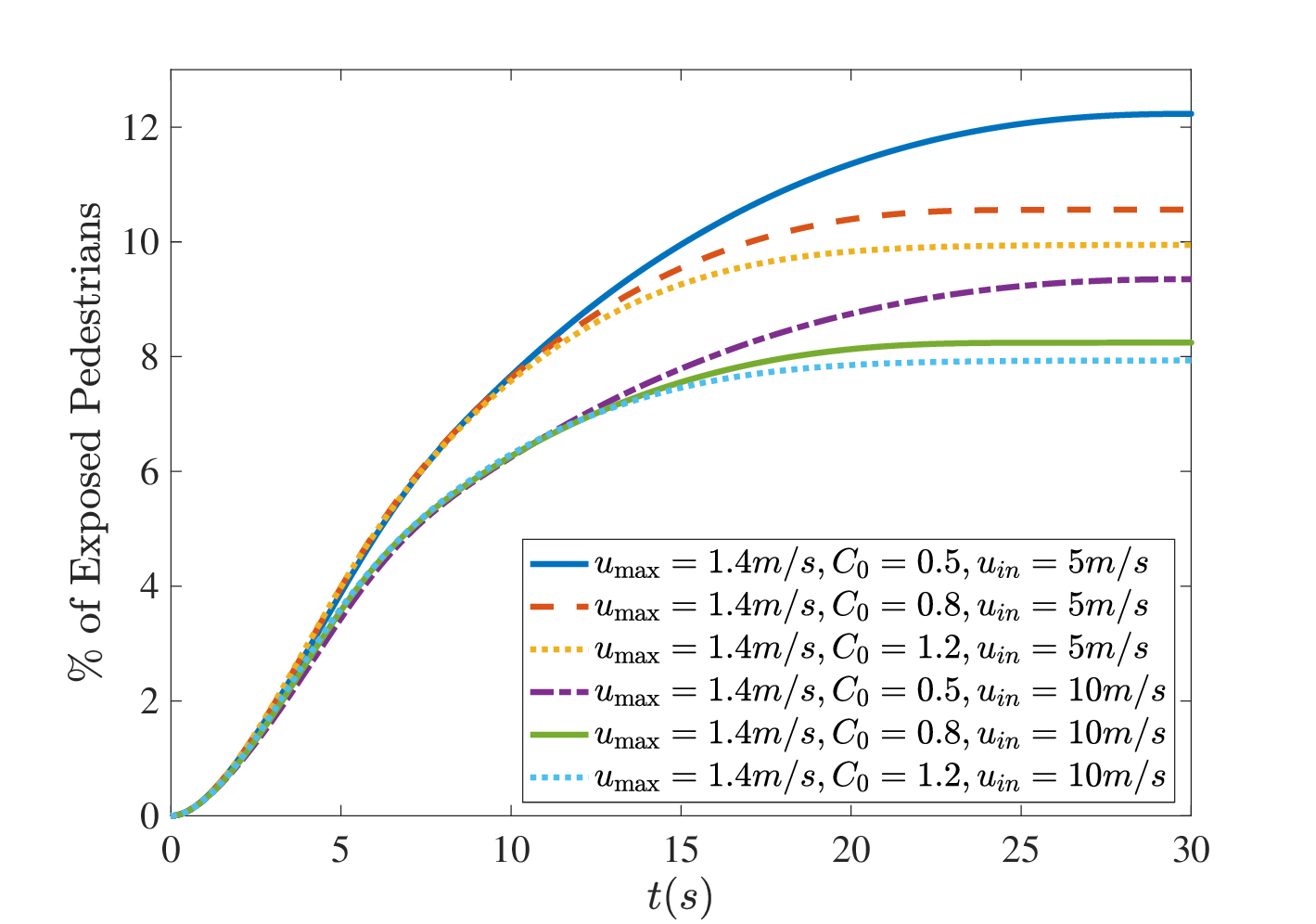}
  \caption{Percentage of exposed pedestrians in a ventilated room for two different ventilation speeds along the pedestrians' direction (left) and against (right), with $\rho^V=0$ and $u_{\max}=1.4$ $m/s$. }
 \label{fig7}
 \end{figure}

\begin{figure}[h!]
\centering
 \includegraphics[width=6.32cm,height=5.35cm]{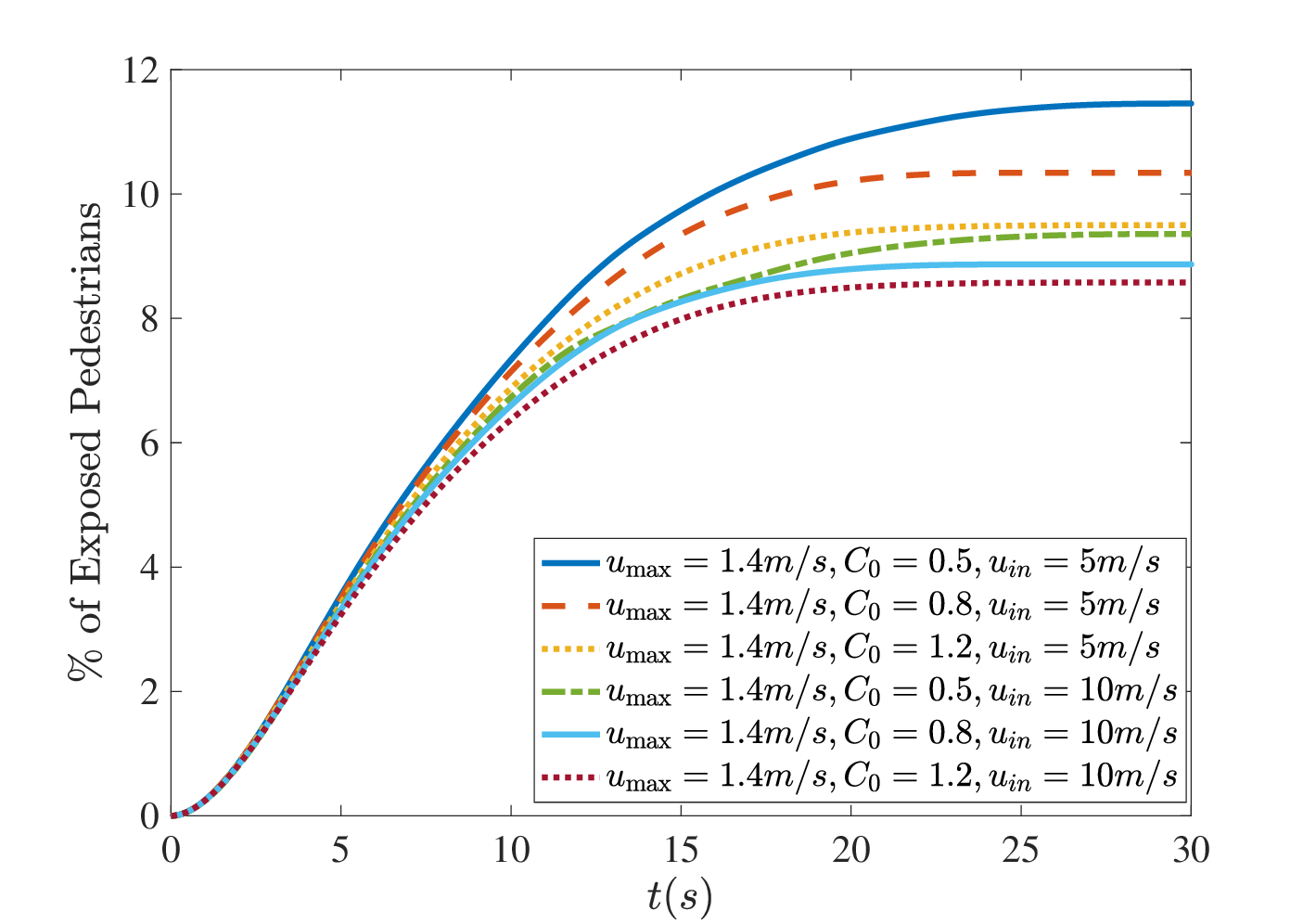} \hspace{-0.29in}
 \includegraphics[width=6.32cm,height=5.35cm]{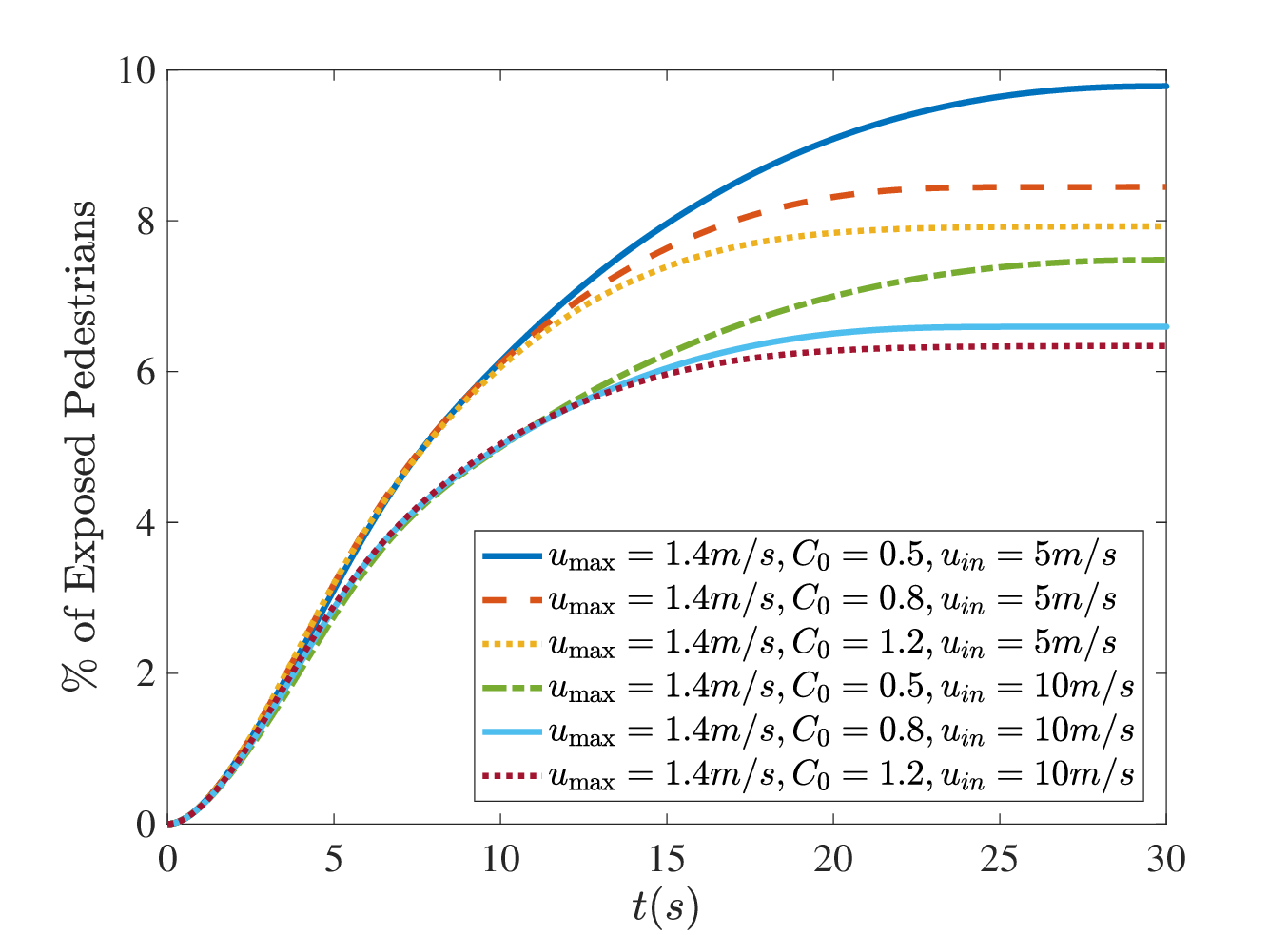}
  \caption{Percentage of exposed pedestrians in a ventilated room for two different ventilation speeds along the pedestrians' direction (left) and against (right), with $\rho^V=0.15\rho_0$ and $u_{\max}=1.4$ $m/s$.}
 \label{fig8}
 \end{figure}

 Indeed, the results as shown in Figs \ref{fig5}-\ref{fig8} for the two different values of $u_{\max}$ (that include also the effect of masked/vaccinated individuals) verify this.  More precisely, first we observe that the number of exposed individuals decreases with increasing ventilation rate, which is reasonably expected.  In particular, as model (19) predicts, increasing ventilation rate results in a larger area of smaller average values for the $\beta^I$ coefficient, which describes the degree of infection transmission.  Comparing the results presented in Figs. \ref{fig5} and \ref{fig7} with the results presented in Fig. \ref{fig2} there is a significant decrease, up to $6\%$ in the  expected number of exposed pedestrians. In all cases it can be   observed that the number of exposed individuals decreases when people move against the ventilation air-flow field in this test case. The same can be concluded if one compares Figs \ref{fig6} and \ref{fig8} with Fig. \ref{fig3} when a small percentage of vaccinated/masked pedestrians is included. In this case, the percentage of exposed pedestrians reaches its lowest value, of approximately 4$\%$ of the total number of pedestrians, when  $C_0=1.2$ and  $u_{\max}=2$ $m/s$, while it is approximately 7$\%$ when  $C_0=1.2$ and $u_{\max}=1.4$ $m/s$, both for an induced air-field against the pedestrians' movement with $u_{in}=10$ $m/s$.
  
\subsection{Crowd Flow Towards Two Exits}
Here we consider again a closed room of size $\Omega = [0, 10\; m] \times [0,10\; m]$ but now with two exits, each being $2$ $m$ wide at the right boundary centered at $y=3$ $m$ and $y=8$ $m$. The initial density is $\rho_0({\bf x}) =2.5$ ped/$m^2$  in the region $[1, 5]\times [2.5, 7.5]$, leading again to 50 people in this region, with  ${\bf v}_0({\bf x})=0$.  Further, $\rho_0^I=0.25\rho_0$, while initially $\rho_0^E=0$. In this test scenario we mainly focus on the effect of adding an extra exit (along with the exits' location), as well as on the effect of the direction of the imposed air-flow field. 

\begin{figure}[h]
\centering
 \includegraphics[width=6.32cm,height=5.35cm]{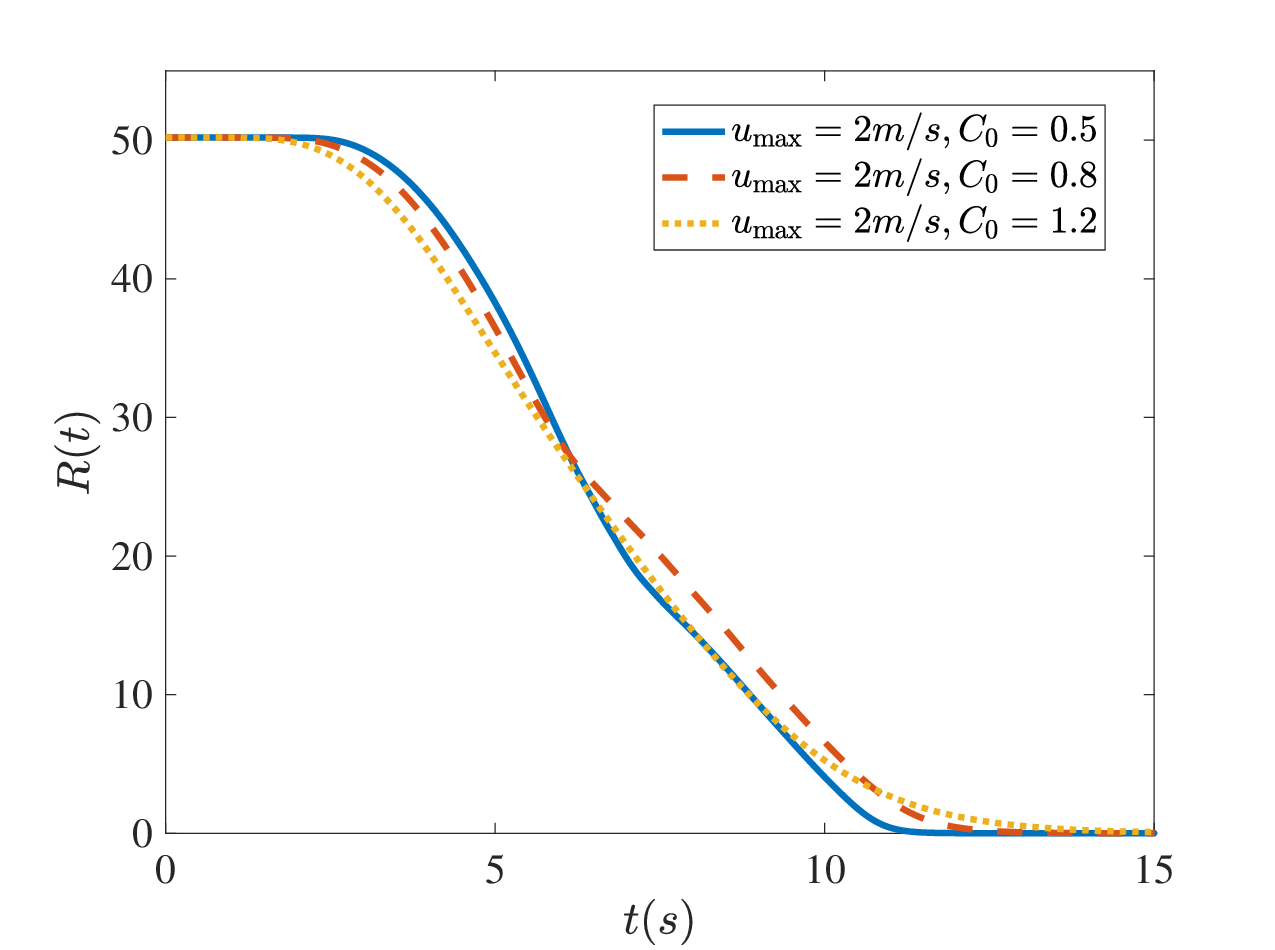} \hspace{-0.29in}
  \includegraphics[width=6.32cm,height=5.35cm]{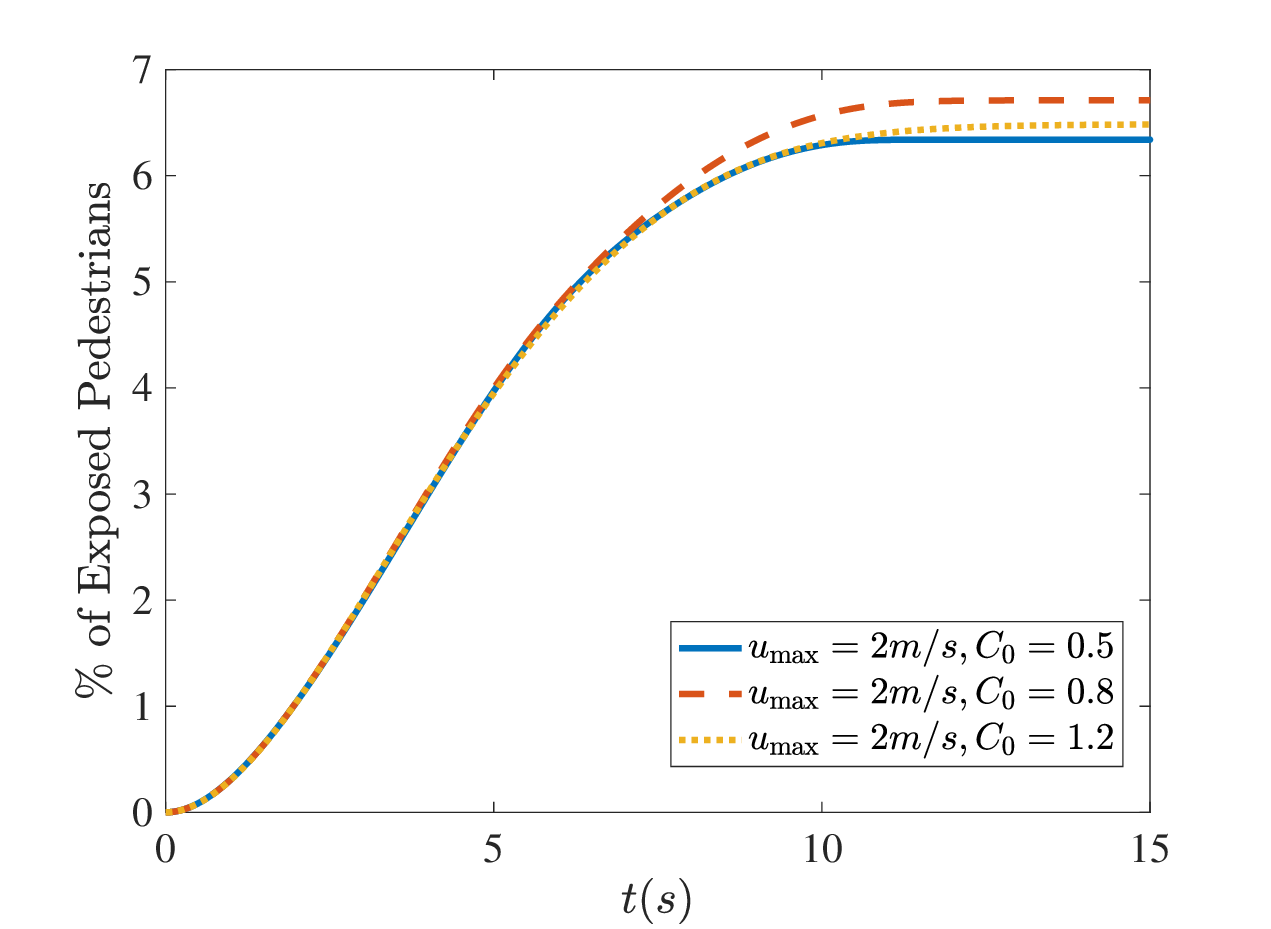}
\includegraphics[width=6.3cm,height=5.35cm]{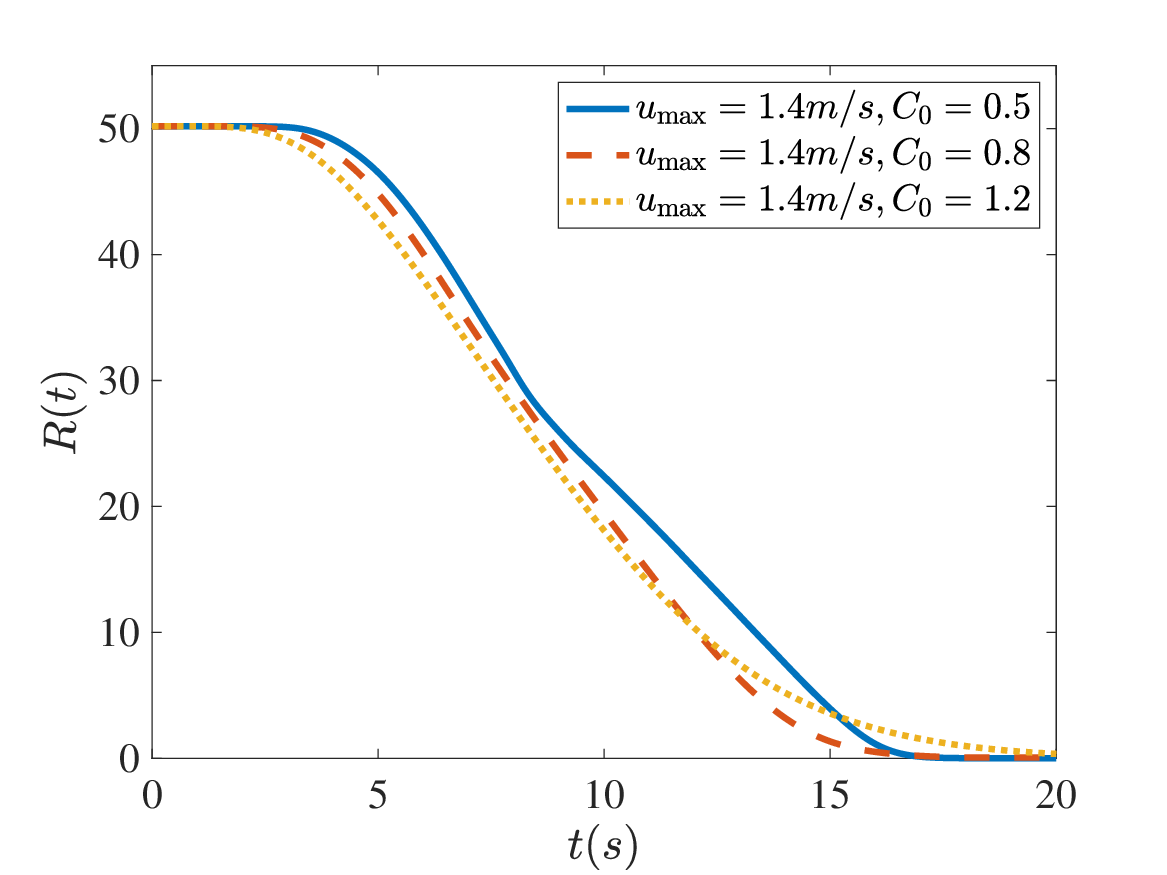}\hspace{-0.24in}
 \includegraphics[width=6.3cm,height=5.35cm]{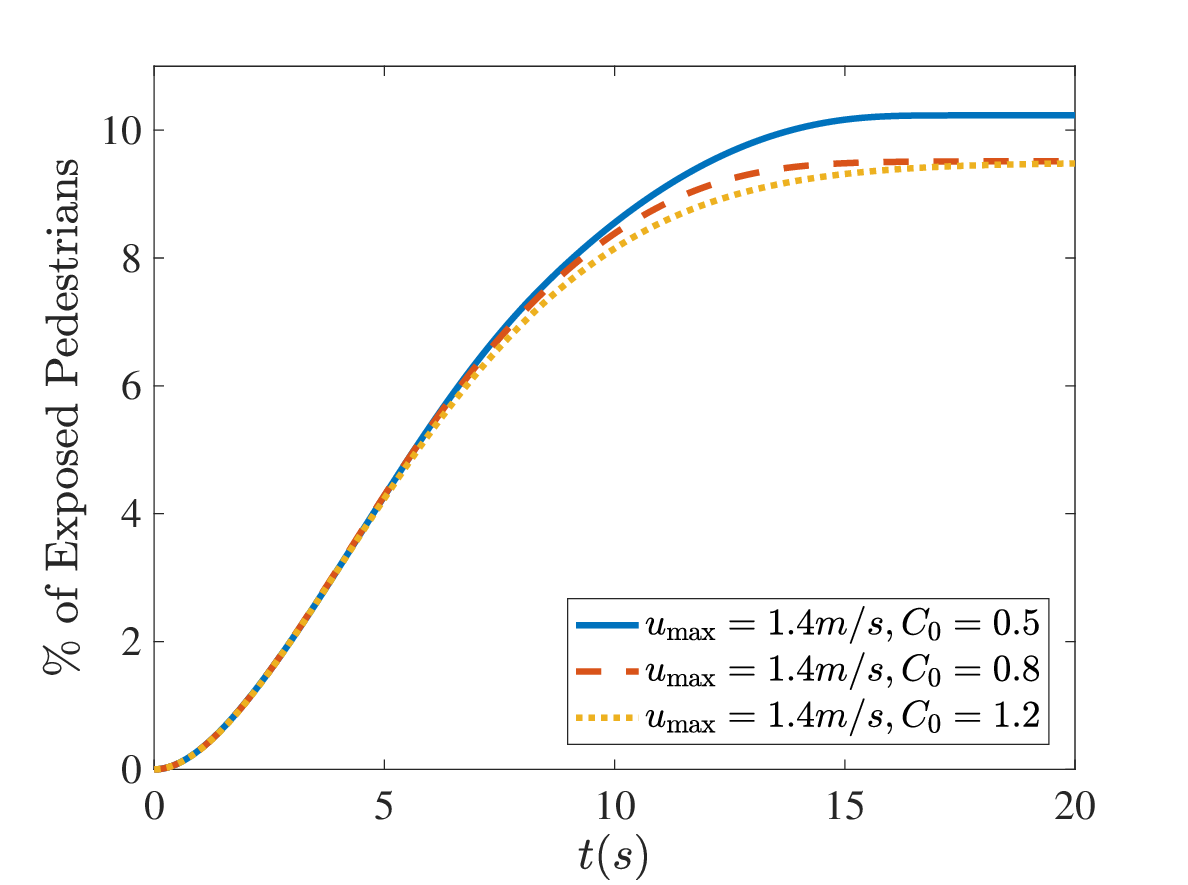}
\caption{Two exits: Time evolution of the total number of pedestrians $R(t)$ (left) and  percentage of exposed pedestrians (right) with $u_{\max}=2$ $m/s $ (top) and $u_{\max}=1.4$ $m/s $, for three different values of $C_0$. }
 \label{fig9}
 \end{figure}

\begin{figure}[h!]
\centering
 \includegraphics[width=4.cm,height=11cm]{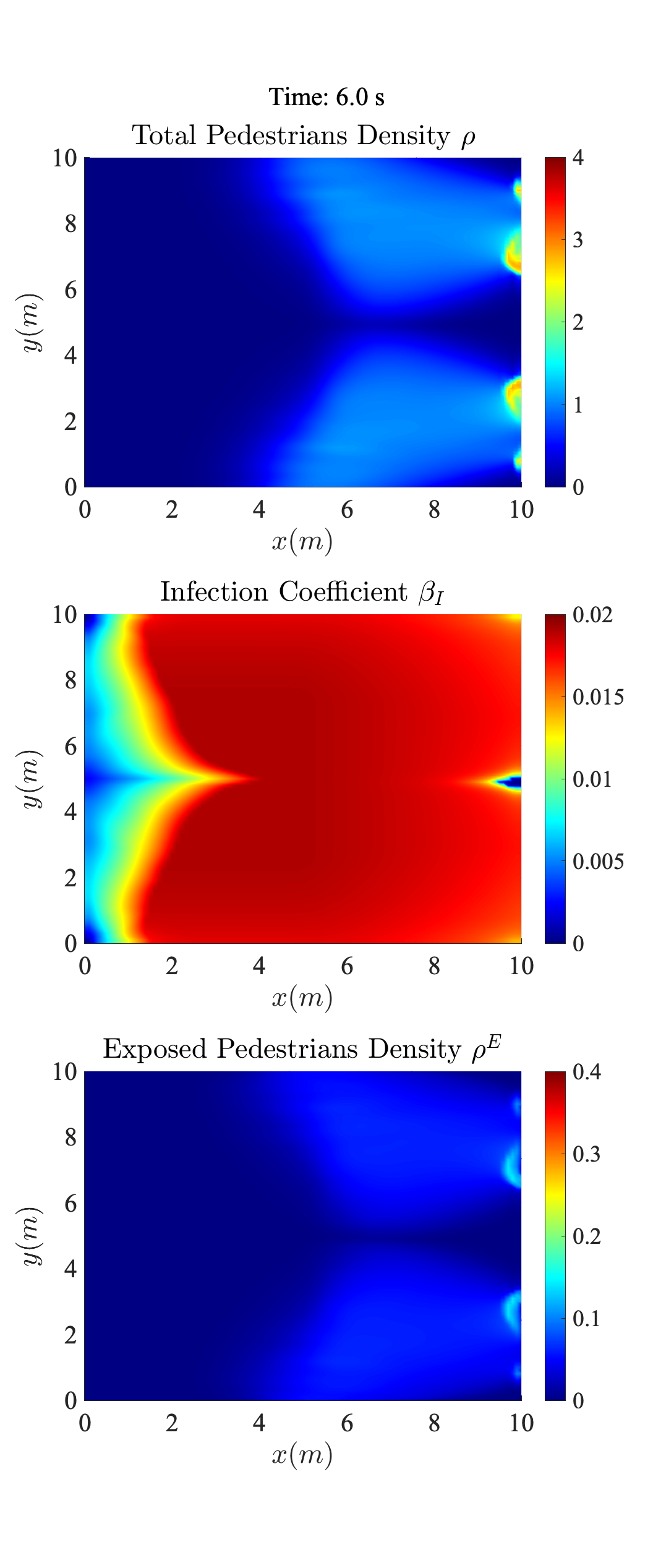} \hspace{-0.16in}
 \includegraphics[width=4.cm,height=11cm]{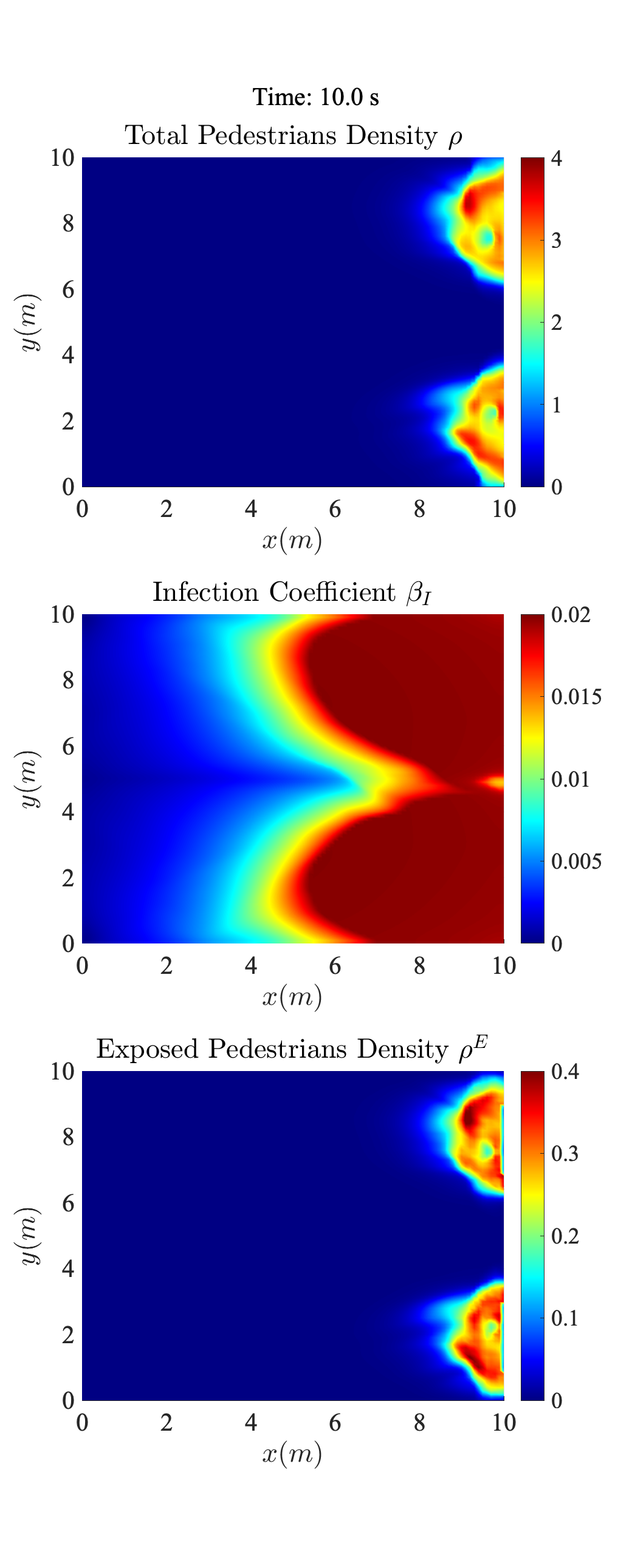}\hspace{-0.1in}
 \includegraphics[width=4.cm,height=11cm]{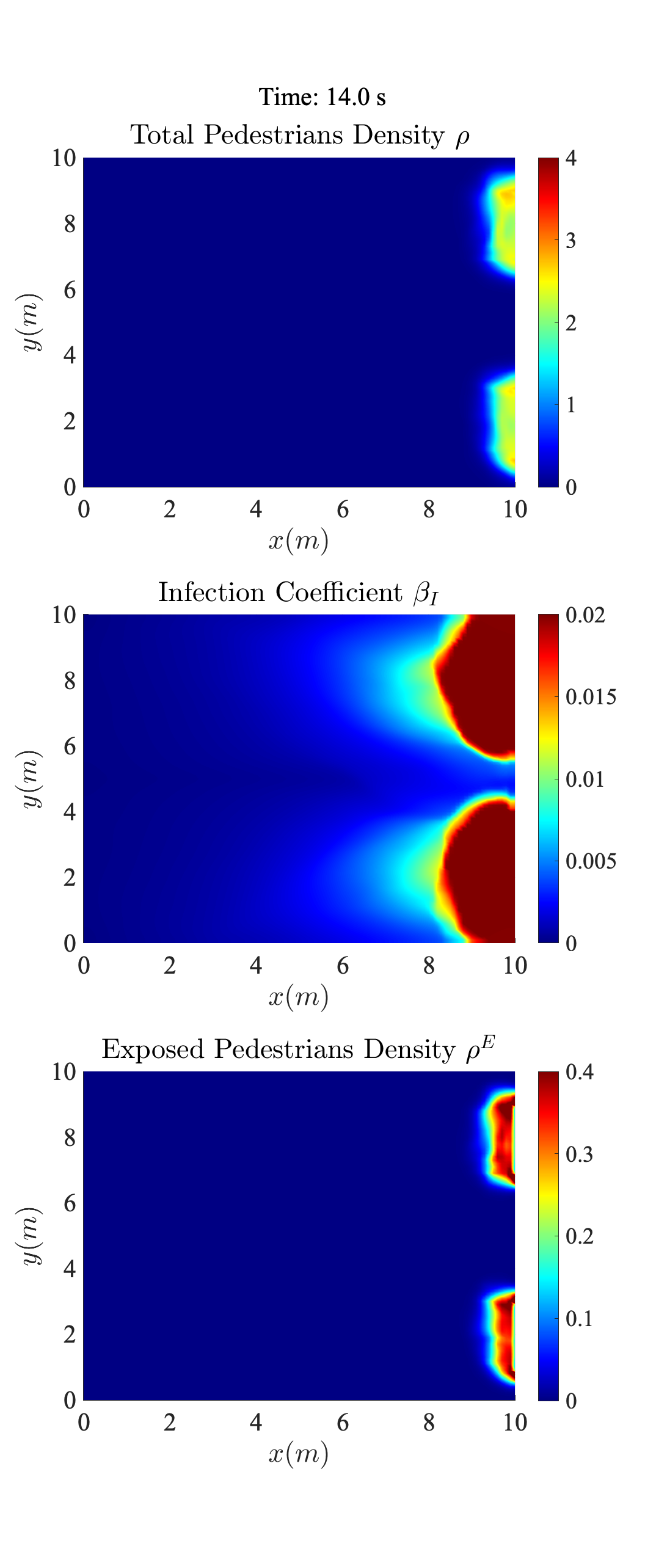}
\caption{Two exits: Total density profiles (top), infection coefficient profile (middle), and exposed pedestrians' profile at different time instances with $u_{\max}=1.4$ $m/s $, $C_0=0.5$, and no ventilation.}
 \label{fig2d5}
 \end{figure}

First, from Figure \ref{fig9}, where the time evolution  of pedestrians and  percentage of exposed pedestrians  with $u_{\max}=2$ $m/s $ and $u_{\max}=1.4$ $m/s $ for three different values of $C_0$ are presented, we observe that, in this  two-exit case, the evacuation time, and consequently, the total number of exposed individuals, are significantly reduced when compared with the results for the one exit scenario in Fig. \ref{fig2}. However, the effect of variable pressure coefficient $C_0$ is less evident, in comparison with the case of a single exit. This can be explained by the fact that two exits increase the capacity flow (i.e., the maximum discharge flow) at the exits of the space considered, and thus, larger outflows can be accommodated, which in turn implies that a clogging (i.e., a congestion) phenomenon may be avoided even for smaller average distances between pedestrians (that may imply a larger outflow at the exits). In addition, since pedestrians are roughly split into two groups (towards each exit as shown, for example, in Fig. \ref{fig2d5}) smaller densities are observed (i.e., smaller numbers of people within a certain area), and thus, increasing the pressure coefficient to impose stricter average distances between people does not have a significant effect on spreading. Further in this case, the effect of the increase in the maximum allowable speed $u_{\max}$ to $2$ $ m/s$ when compared to that of $1.4$ $ m/s$ can be clearly seen, in terms of a reduced evacuation time and a  lower percentage of exposed individuals.
  In Fig. \ref{fig2d5}, where  we show total density profile, infection coefficient profile, and exposed pedestrians' density, at different time instances with $u_{\max}=1.4$ $m/s $ for $C_0=0.5$, we can observe the evolution of the infection coefficient $\beta^I$ that follows the pedestrians' movement towards the two exits. For this value of $C_0$ some strong interactions and formation of inhomogeneities in the density distribution are still evident.

\begin{figure}[h!]
\centering
 \includegraphics[width=4.cm,height=11cm]{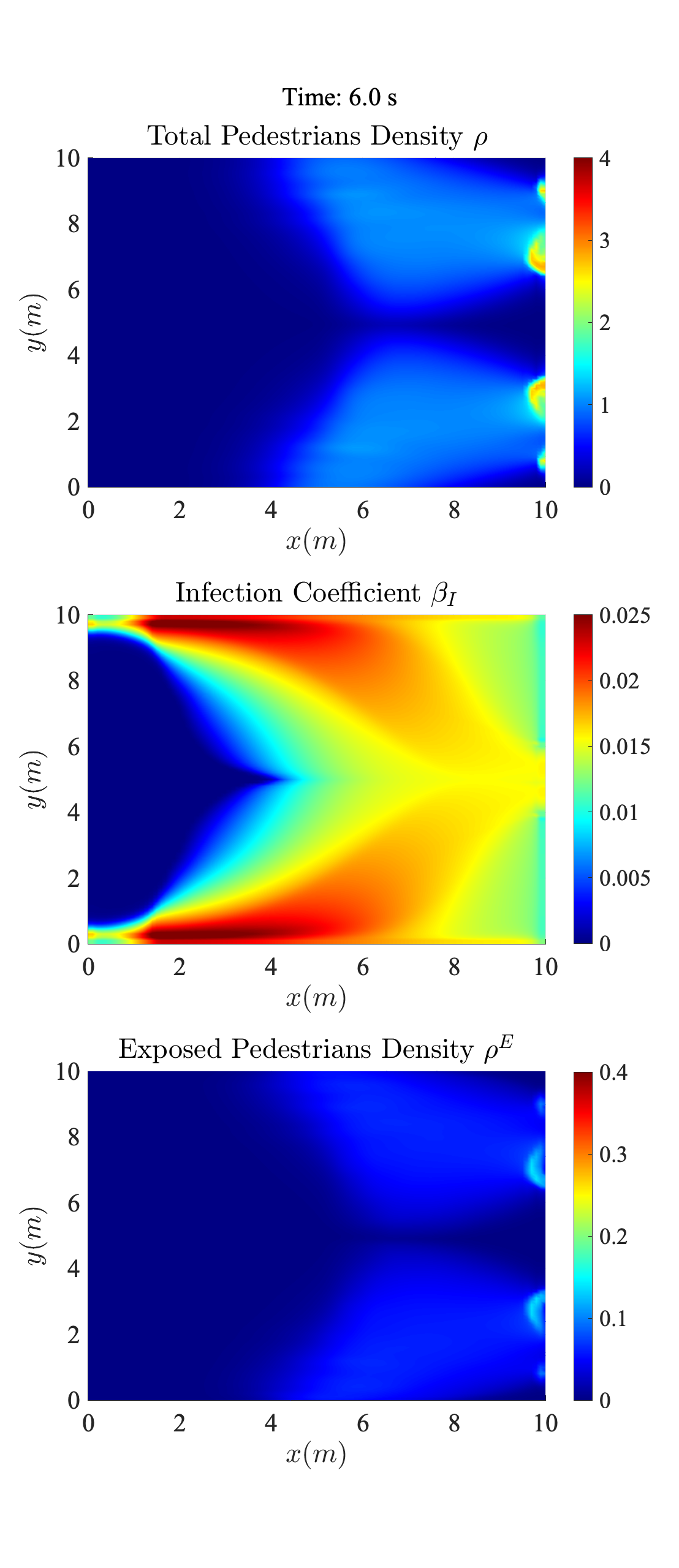} \hspace{-0.16in}
 \includegraphics[width=4.cm,height=11cm]{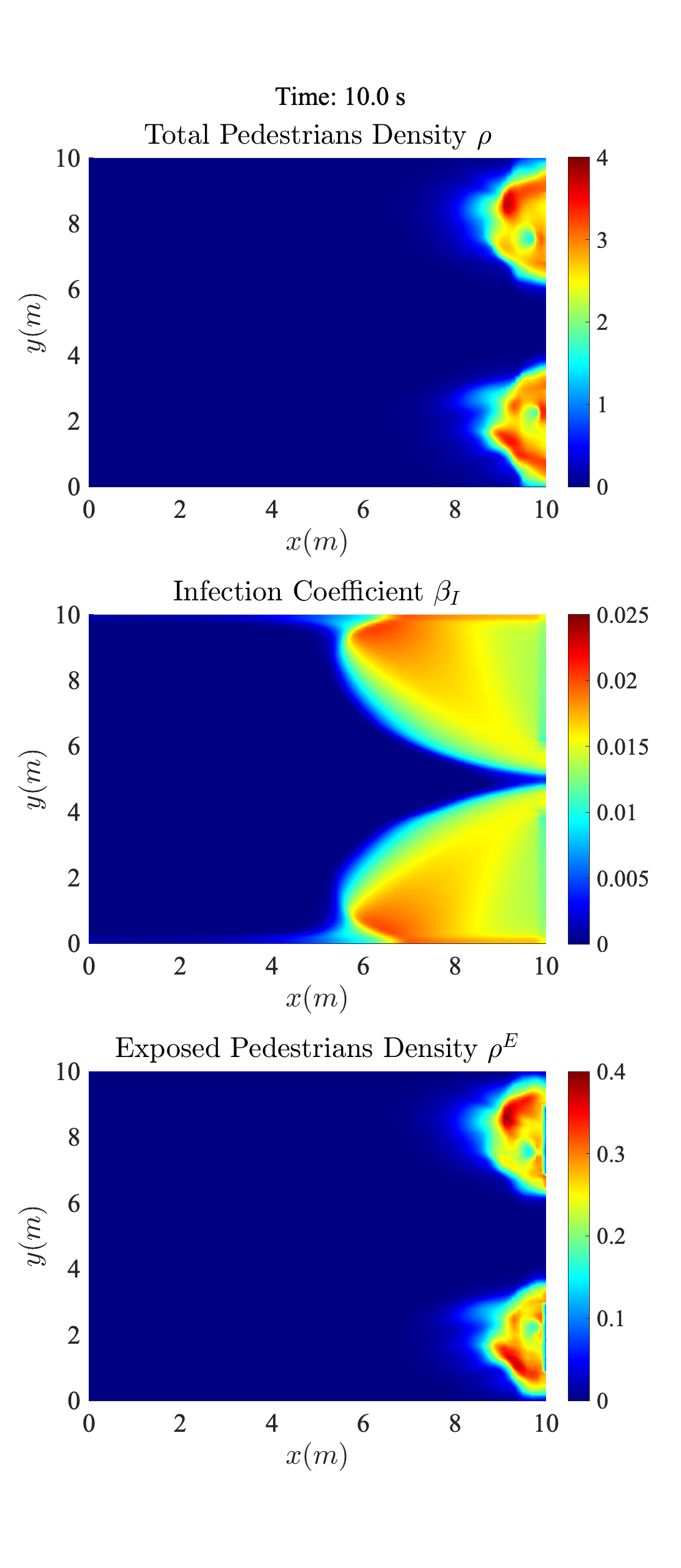}\hspace{-0.1in}
 \includegraphics[width=4.cm,height=11cm]{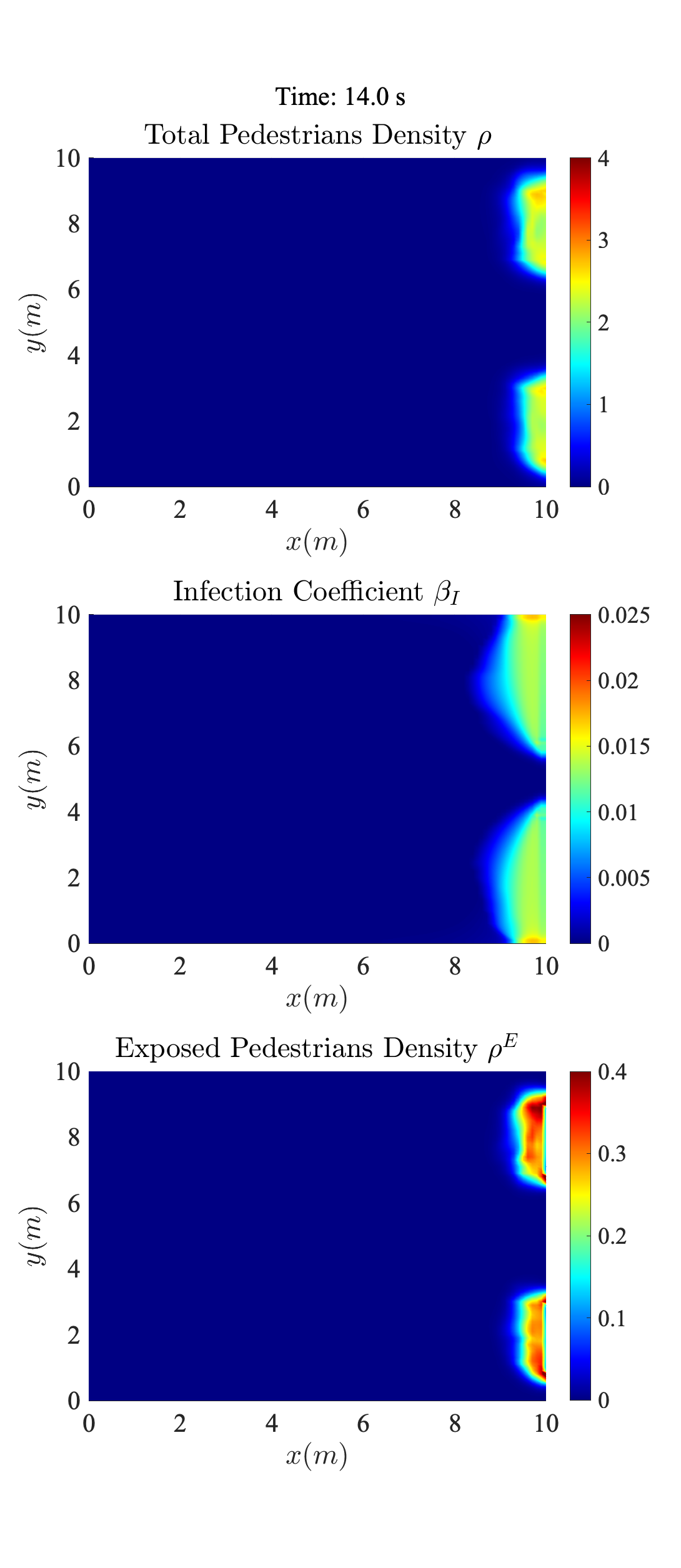}
\caption{Two exits: Total density profiles (top), infection coefficient profile (middle), and exposed pedestrians' profile at different time instances with $u_{\max}=1.4m/s $, $C_0=0.5$, and  with ventilation air-flow field along the pedestrians'  flow direction  with $u_{in}=10$ $m/s$.}
 \label{fig2d6}
 \end{figure}

\begin{figure}[h!]
\centering
 \includegraphics[width=4.cm,height=11cm]{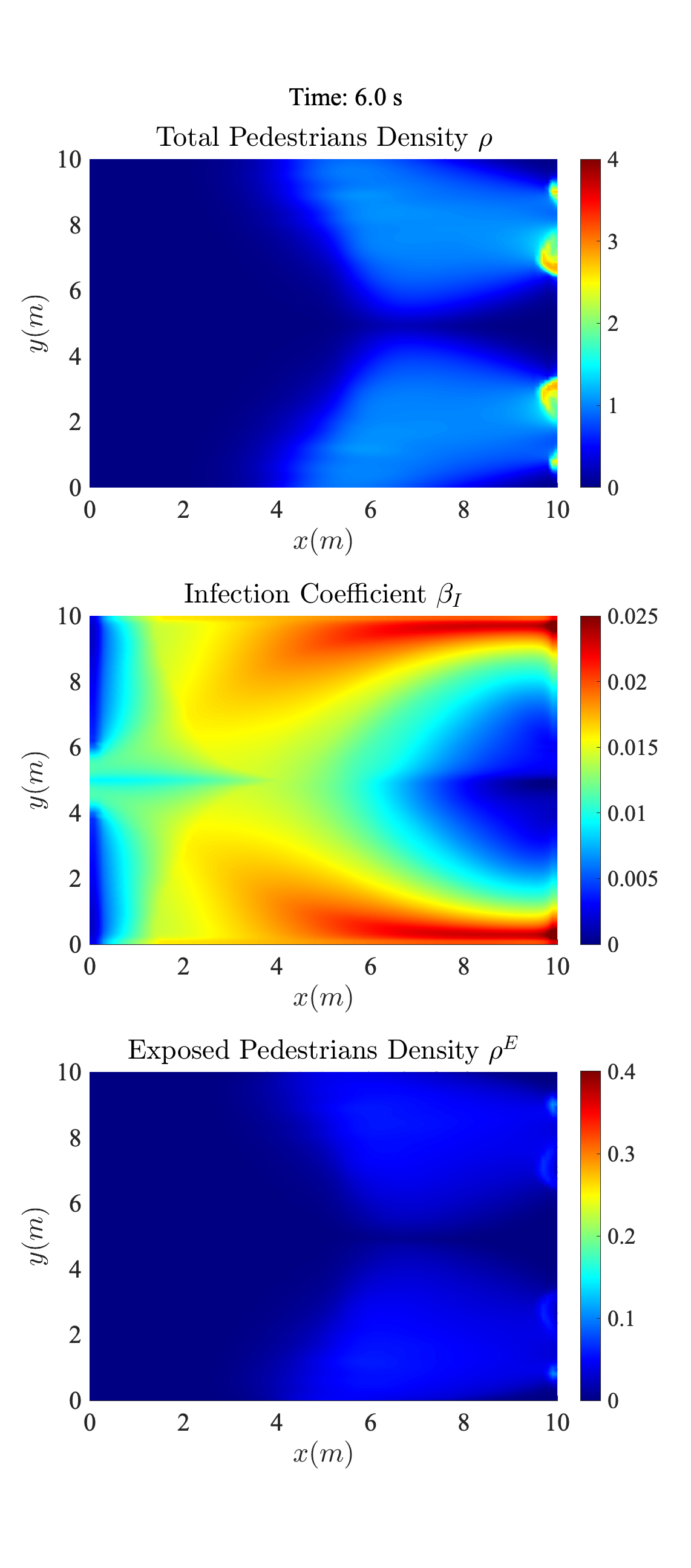} \hspace{-0.16in}
 \includegraphics[width=4.cm,height=11cm]{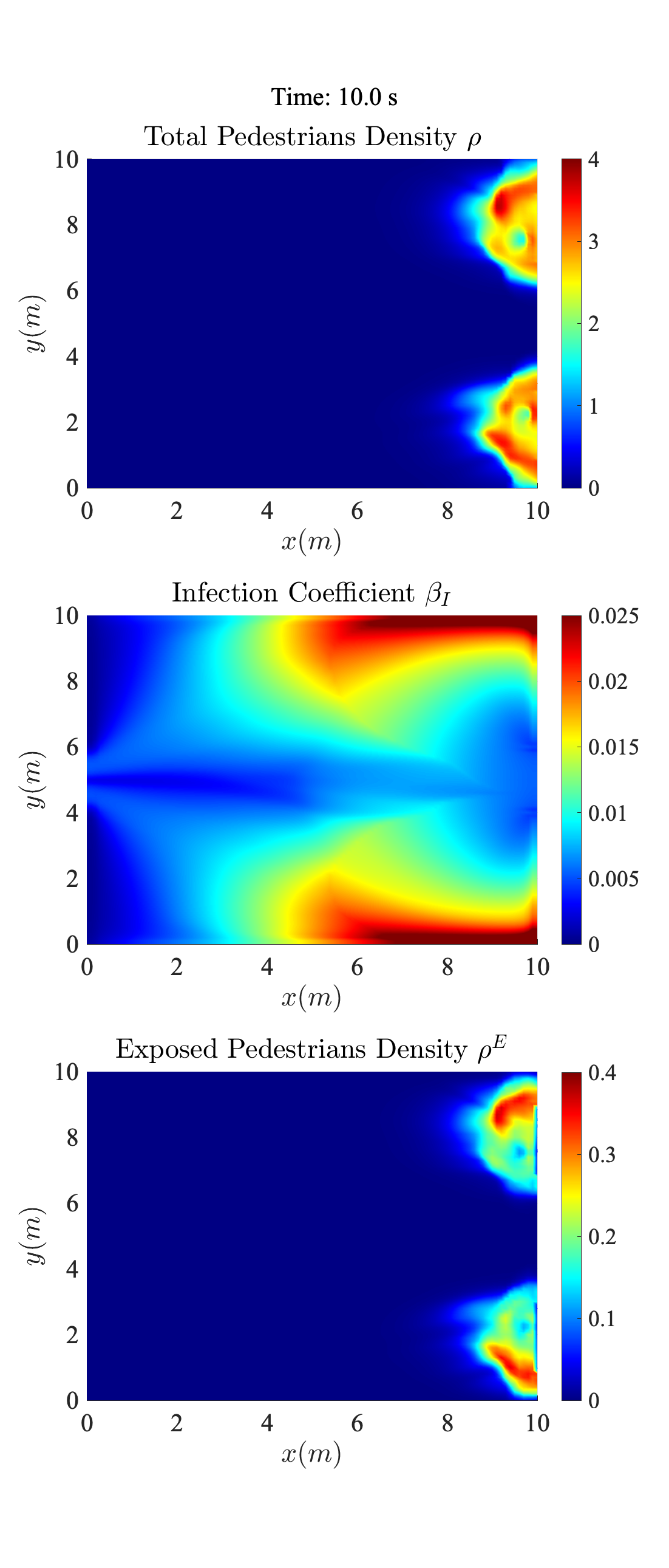}\hspace{-0.1in}
 \includegraphics[width=4.cm,height=11cm]{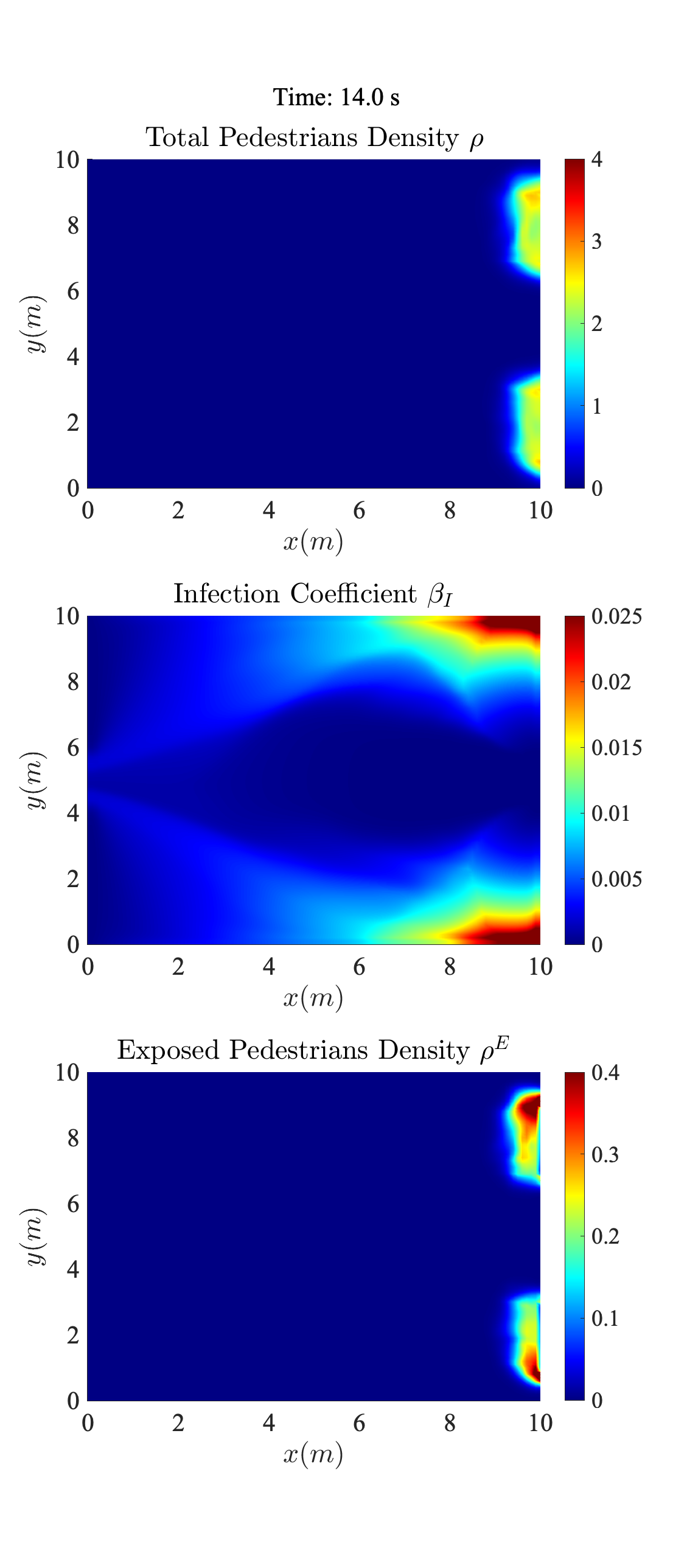}
\caption{Two exits: Total density profiles (top), infection coefficient profile (middle), and exposed pedestrians' profile at different time instances with $u_{\max}=1.4m/s $, $C_0=0.5$, and  with ventilation air-flow filed against the pedestrians'  flow direction  with $u_{in}=10$ $m/s$.}
 \label{fig2d7}
 \end{figure}

\begin{figure}[h!]
\centering
 \includegraphics[width=6.3cm,height=5.35cm]{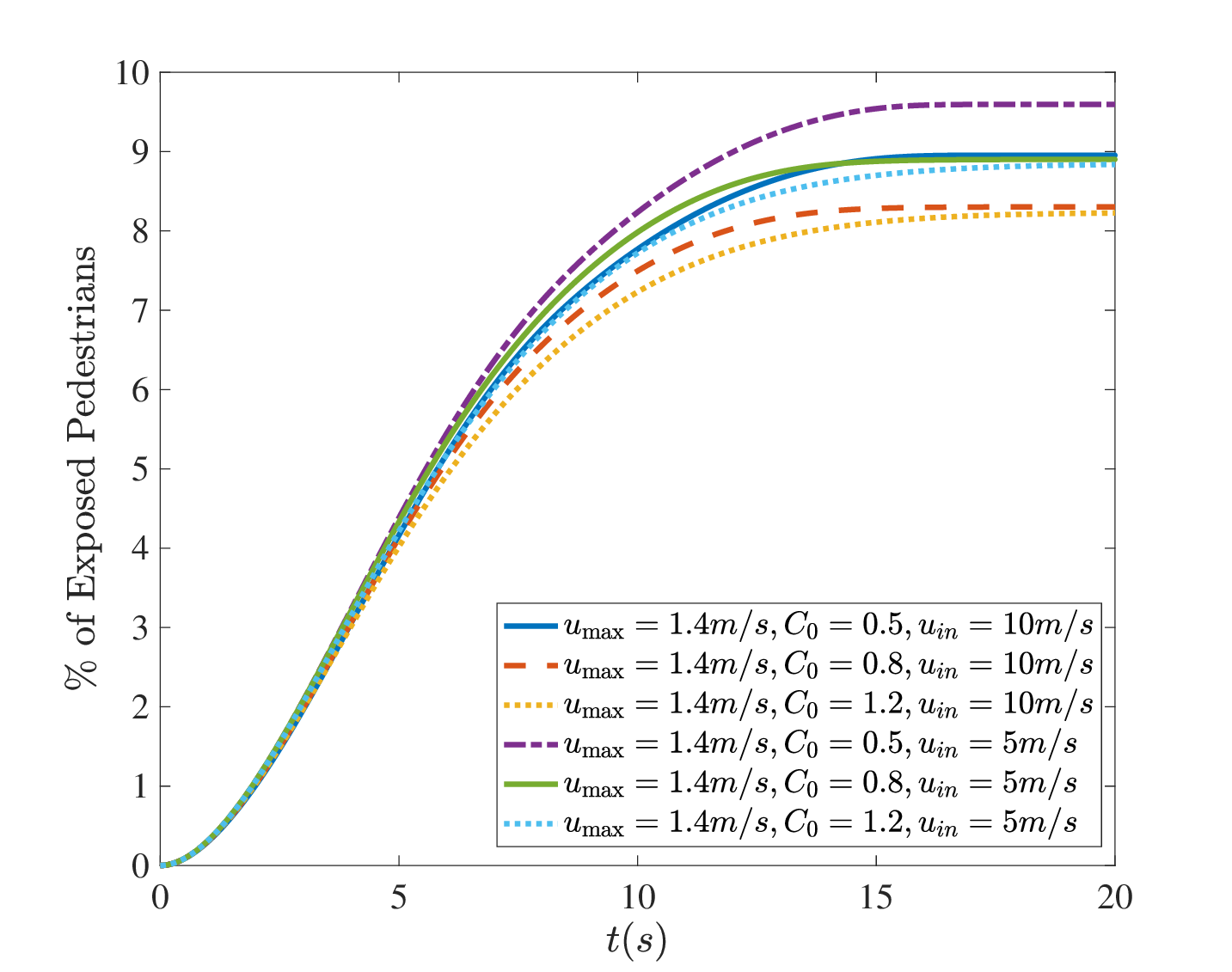} \hspace{-0.29in}
  \includegraphics[width=6.3cm,height=5.35cm]{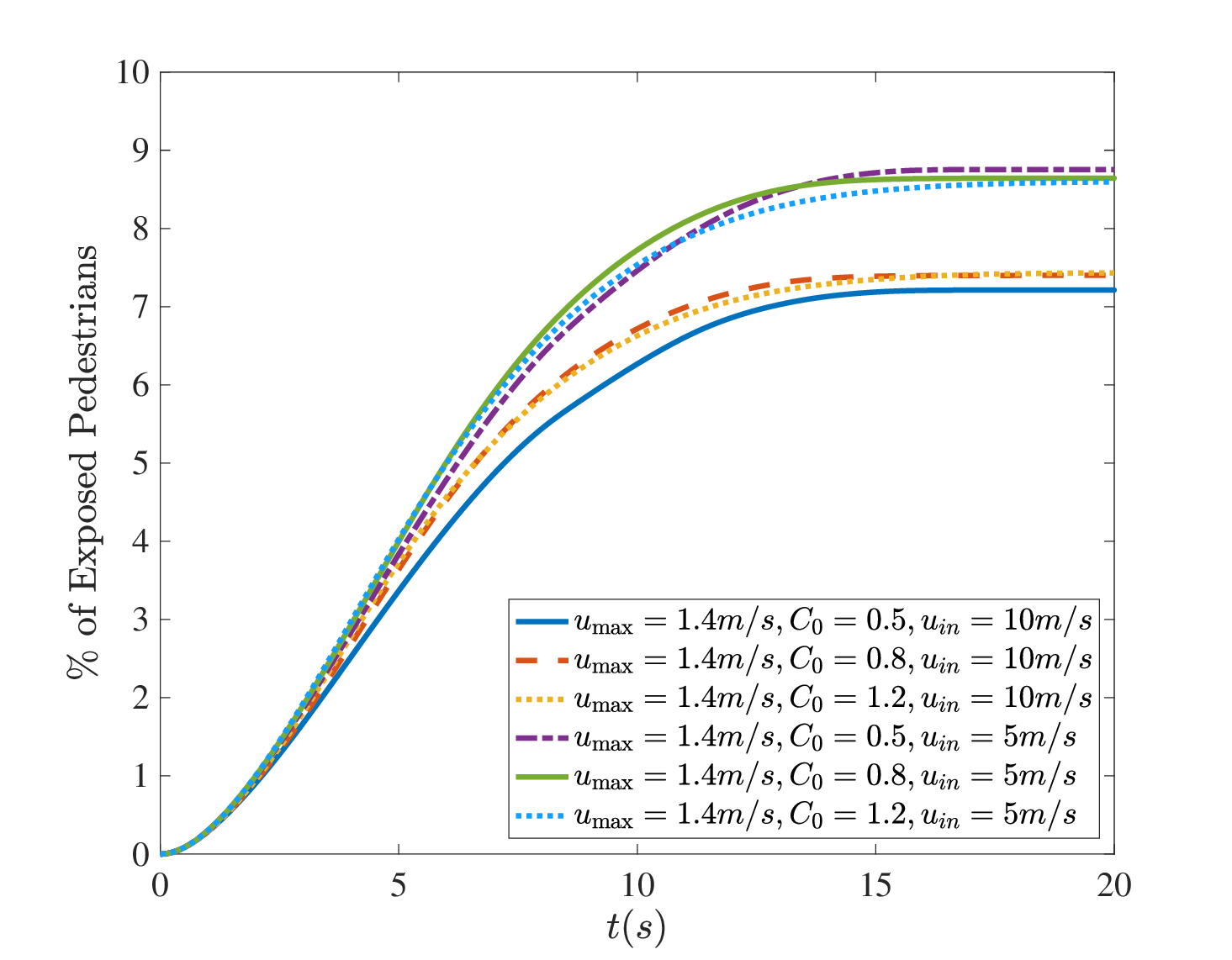}
  \caption{Two exits: Percentage of exposed pedestrians in a ventilated room for two different ventilation speeds along the pedestrians' direction (left) and against (right) with $\rho_0^V=0$ and $u_{\max}=1.4$ $m/s$. }
 \label{fig10}
 \end{figure}

  Next, in this two-exit case, we focus on the effect of imposing the two ventilation scenarios as in Section 4.1.2 for the case $u_{\max} = 1.4$ $m/s$, for which  the exposed pedestrians' number is in general larger, and $u_{in} = 10$ $m/s$ in the ventilation. Initially we present, as indicative results, in  Figs \ref{fig2d6} and \ref{fig2d7},  the total pedestrians' density profiles, infection coefficient profile,  and exposed pedestrians' profile at different time instances for air-flow  along and against the pedestrians' flow direction for $C_0=0.5$.
  From Fig. \ref{fig10} we observe that an increasing ventilation rate in general implies smaller percentages of exposed individuals, which is reasonably expected. However, the relevant improvement with respect to the no ventilation case is not as significant as in the case of one exit, at least for low ventilation rates. This can be explained by the fact that, in the case of two exits, the evacuation time is roughly half of the respective evacuation time in the one-exit case. This in turn implies that large ventilation rates may not result in significant improvements with respect to spreading as pedestrians exit the space fast enough. Still, when compared with the obtained results in Fig. \ref{fig9}, when ventilation is not imposed,  one can observe  a 3-4$\%$ decrease in the number of exposed individuals, in the case of an imposed air-flow  against the pedestrians direction for $u_{in}=10$ $m/s$. We also note here that, while the ventilation ducts (for both cases of the air-flow field) are placed at the middle of the right (exit) boundary, pedestrians are directed to two exits that are relatively far from these. In return it is expected that the effect of the imposed ventilation  will be less profound as compared with the respective case in Section 4.1.2.
 
 \subsection{Large Crowd Flow Towards an Exit}
With this test case we aim to investigate the performance of the model in the case of a larger number pedestrians moving  in larger spaces/distances. Our main target here is to study the effect of different ventilation patterns induced in the facility.
To this end,  we consider a walking facility of size $\Omega = [0, 50\;m] \times [0,20\;m]$ with an exit at the right boundary centered at $y=10\;m$. The initial density is $\rho_0({\bf x}) =1$ ped/$m^2$  in the region $[0, 20]\times [0,20]$, leading to around 400 people in this region, with  ${\bf v}_0({\bf x})=0$.  Further, we set $\rho_0^I=0.1\rho_0$, while initially $\rho_0^E=0$ and $\rho_0^V=0$. In this case we fix the value of $u_{\max}$ to $1.4 m/s.$ The values of $\Delta x$ and $\Delta y$  were set to $0.1$ and the value of $\Delta t$ to $5\cdot 10^{-3}$.

\begin{figure}[h]
\centering
  \includegraphics[width=6.25cm,height=5.25cm]{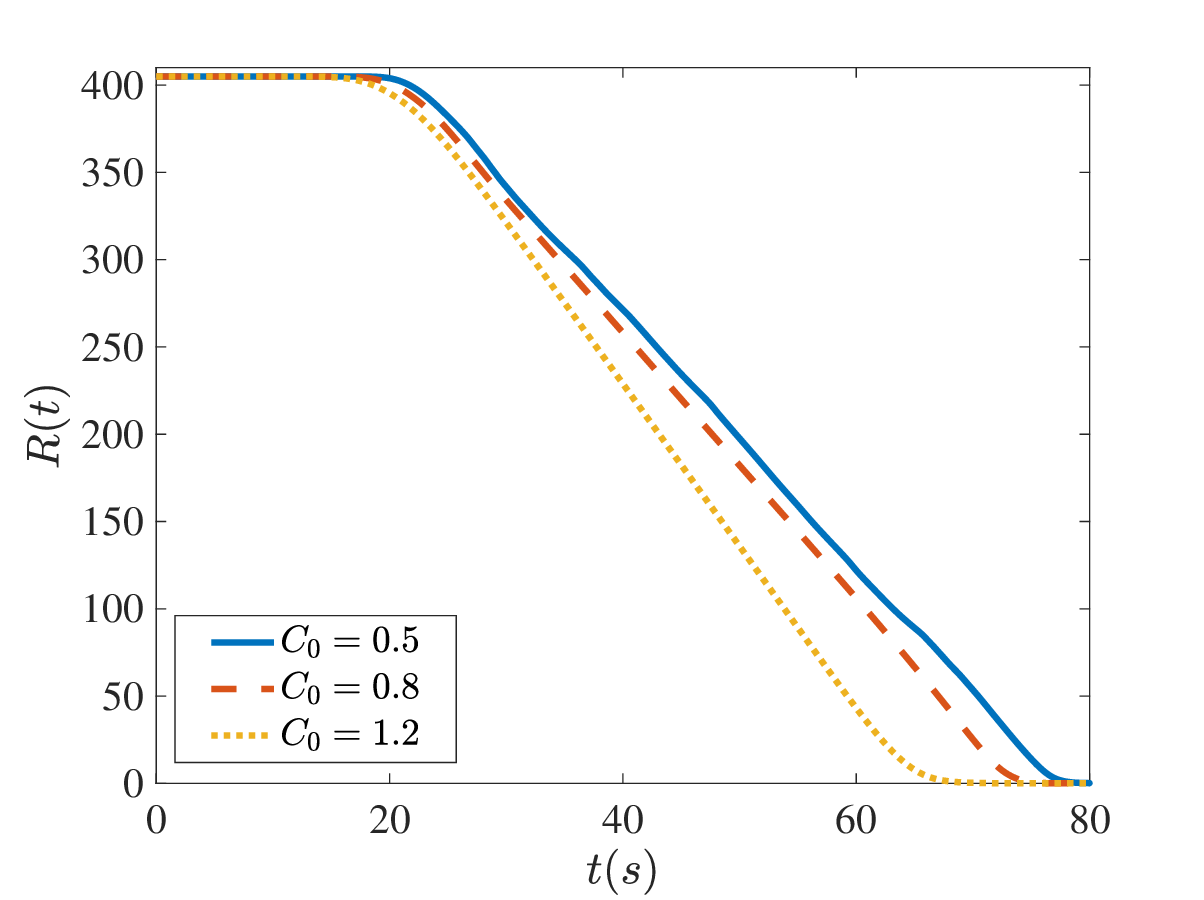} \hspace{-0.25in}
\includegraphics[width=6.25cm,height=5.25cm]{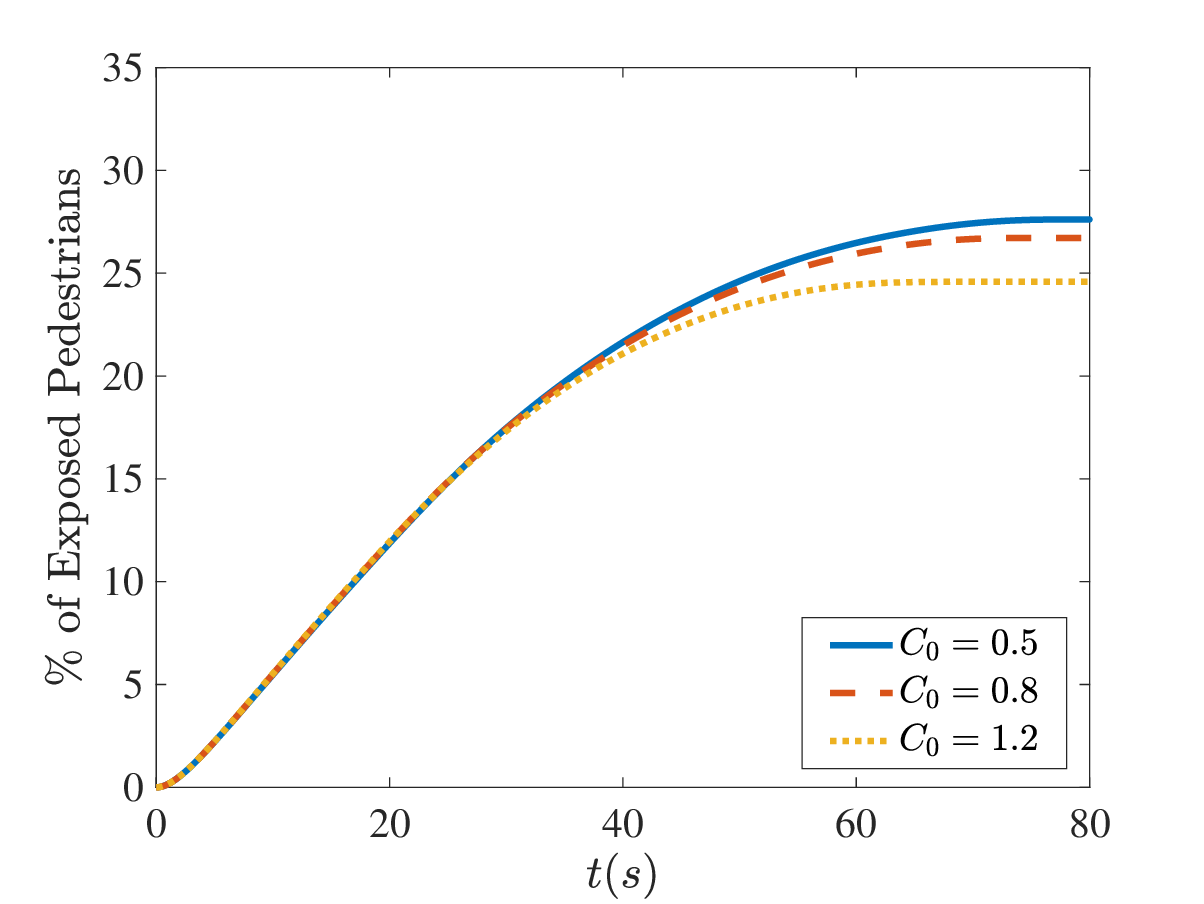} \\
\includegraphics[width=6.25cm,height=5.25cm]{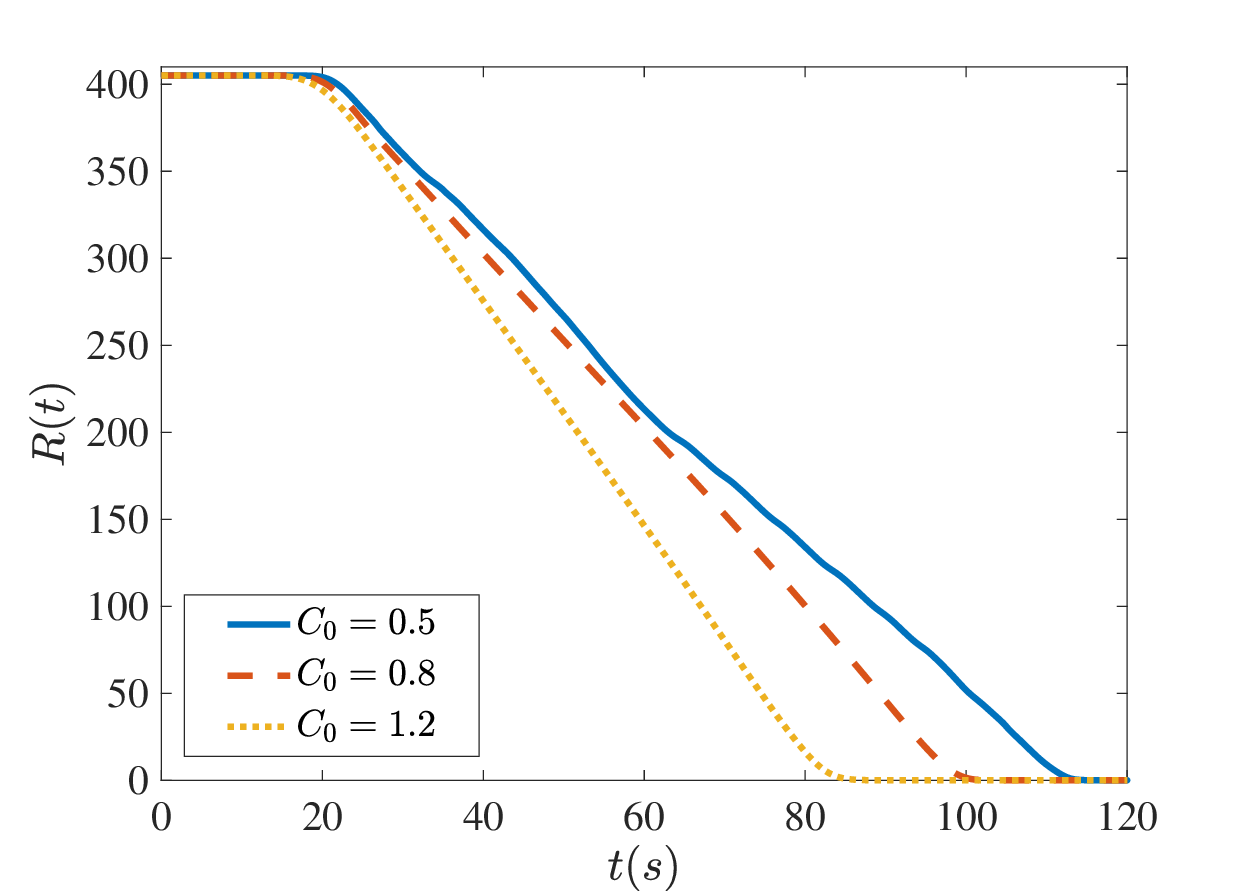} \hspace{-0.25in}
\includegraphics[width=6.25cm,height=5.25cm]{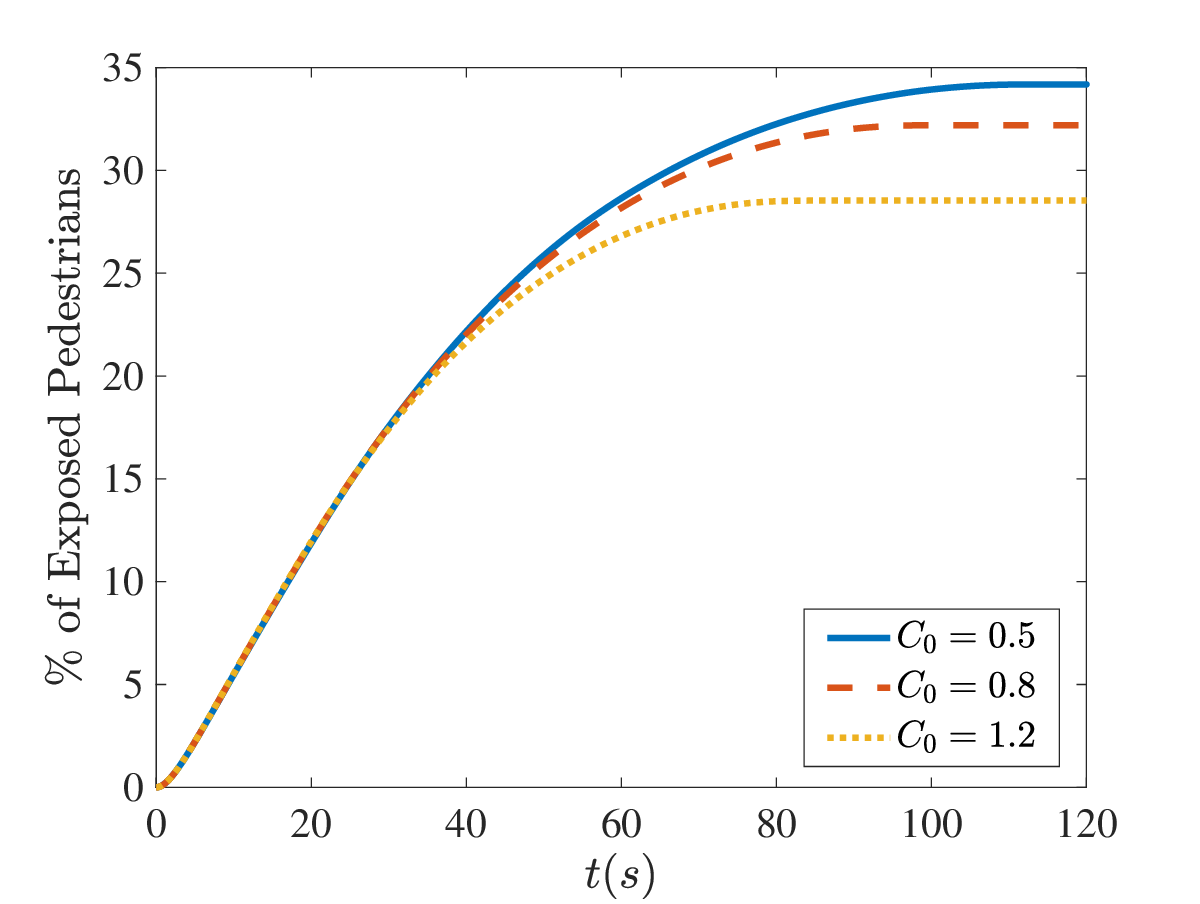} \\
  \caption{Corridor: Time evolution of the total mass $R(t)$ of pedestrians (left) and  percentage of exposed pedestrians (right) with  $u_{\max}=1.4$ $m/s $,  for three different values of $C_0$, with one exit $6\;m$ wide (top) and $4\;m$ wide (bottom), with no ventilation.}
 \label{fig11}
 \end{figure}

Initially, we compare the effect of the size of the exit for the three different values of $C_0$ as shown in Fig. \ref{fig11}, in terms of the exiting time, as well as the percentage of the exposed pedestrians predicted. We assume first an exit $6\;m$ wide and then one  $4\;m$ wide,  both for $u_{\max}=1.4$ $m/s$ and no ventilation present. As it was expected in this setup, the evacuation time is much larger for the smaller exit, for all values of $C_0$. This leads to an increase of the predicted percentages of exposed pedestrians, especially for the case of $C_0=0.5$ where an approximately 10$\%$ increase can be observed while for $C_0=1.2$ the  increase is approximately 5$\%$, as compared with the case of an exit $6\;m$ wide.  
 
\begin{figure}[h!]
\centering
 \includegraphics[width=4.2cm,height=9cm]{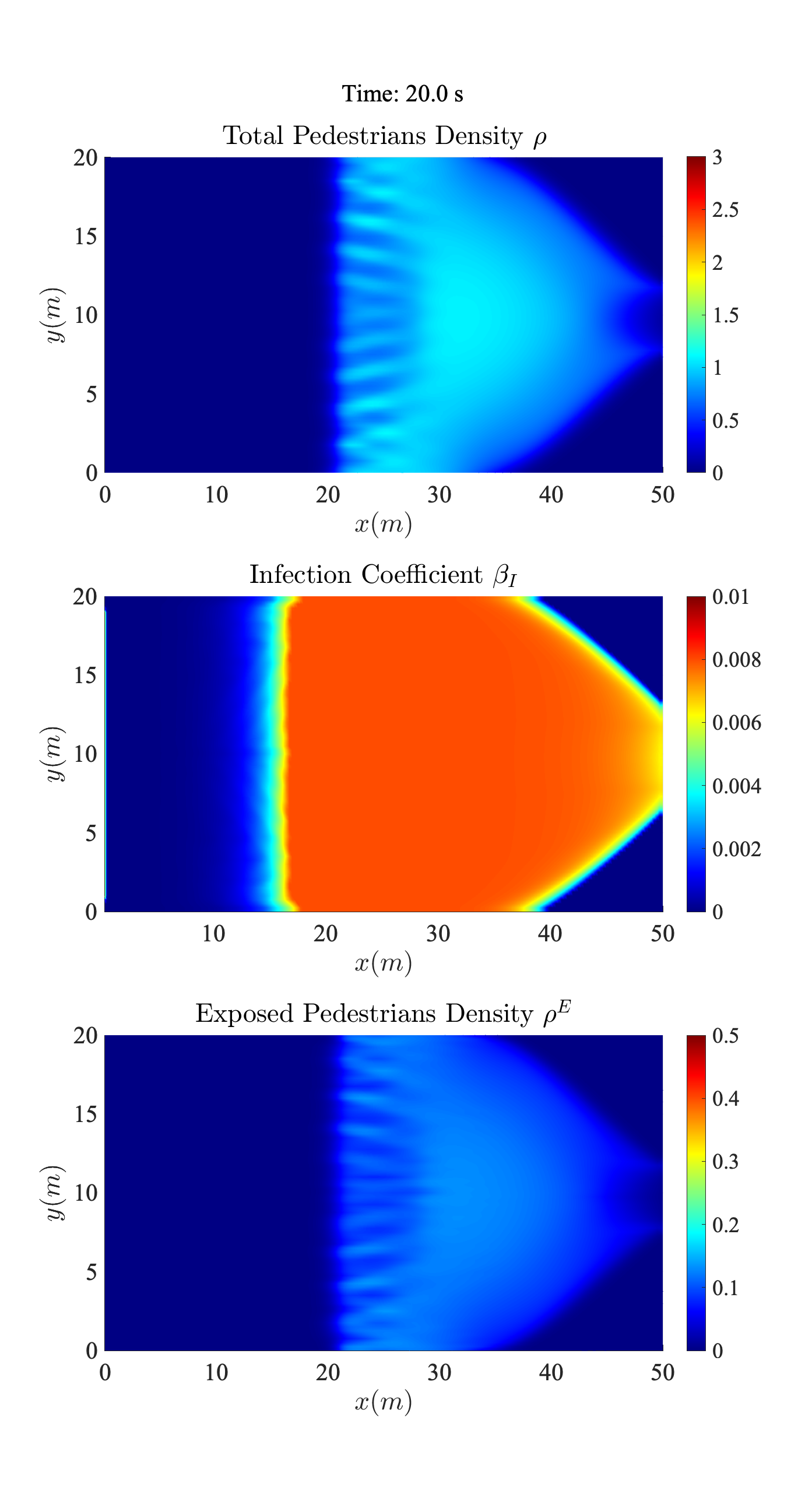} \hspace{-0.16in}
 \includegraphics[width=4.2cm,height=9cm]{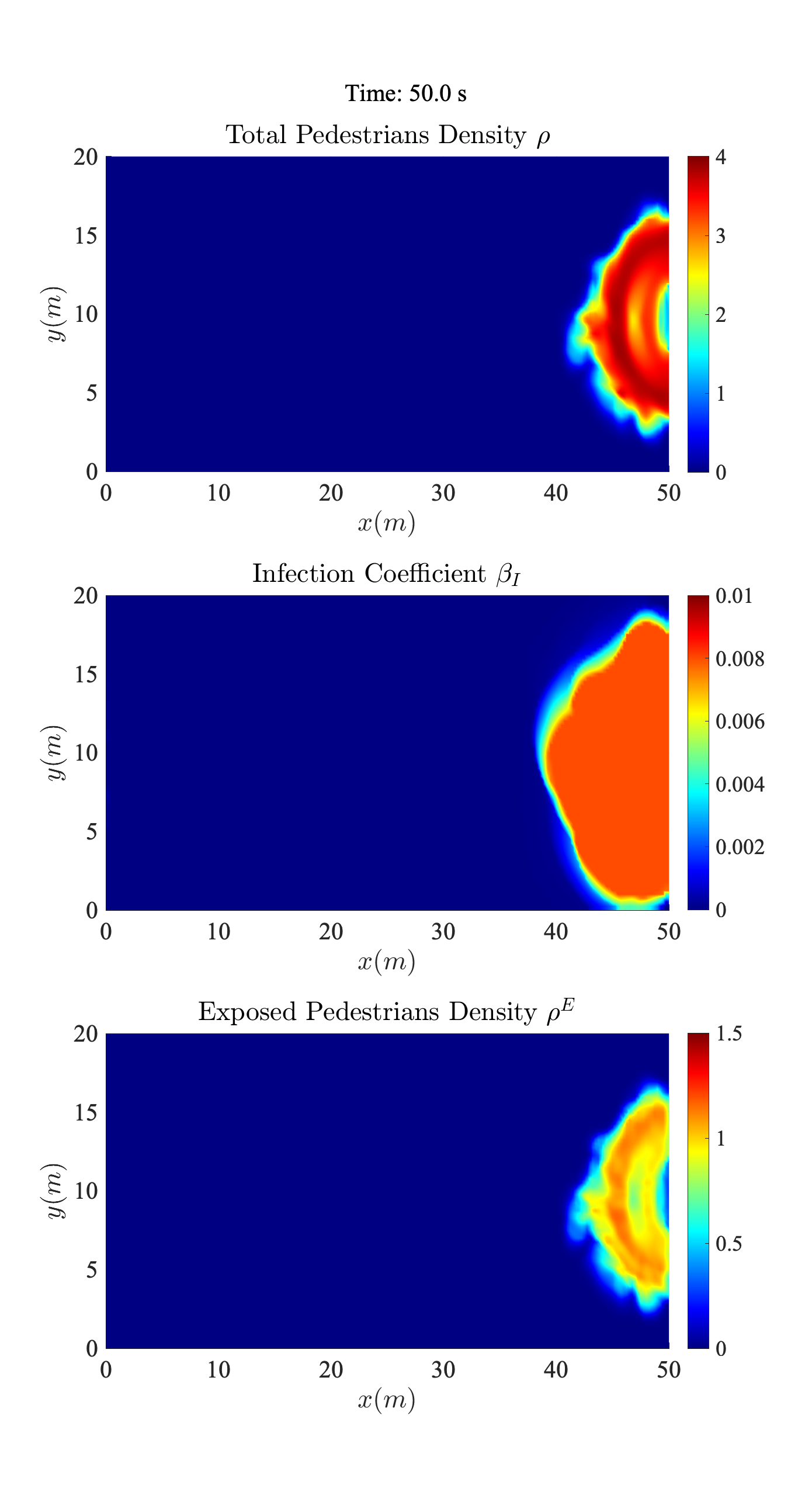}\hspace{-0.16in}
 \includegraphics[width=4.2cm,height=9cm]{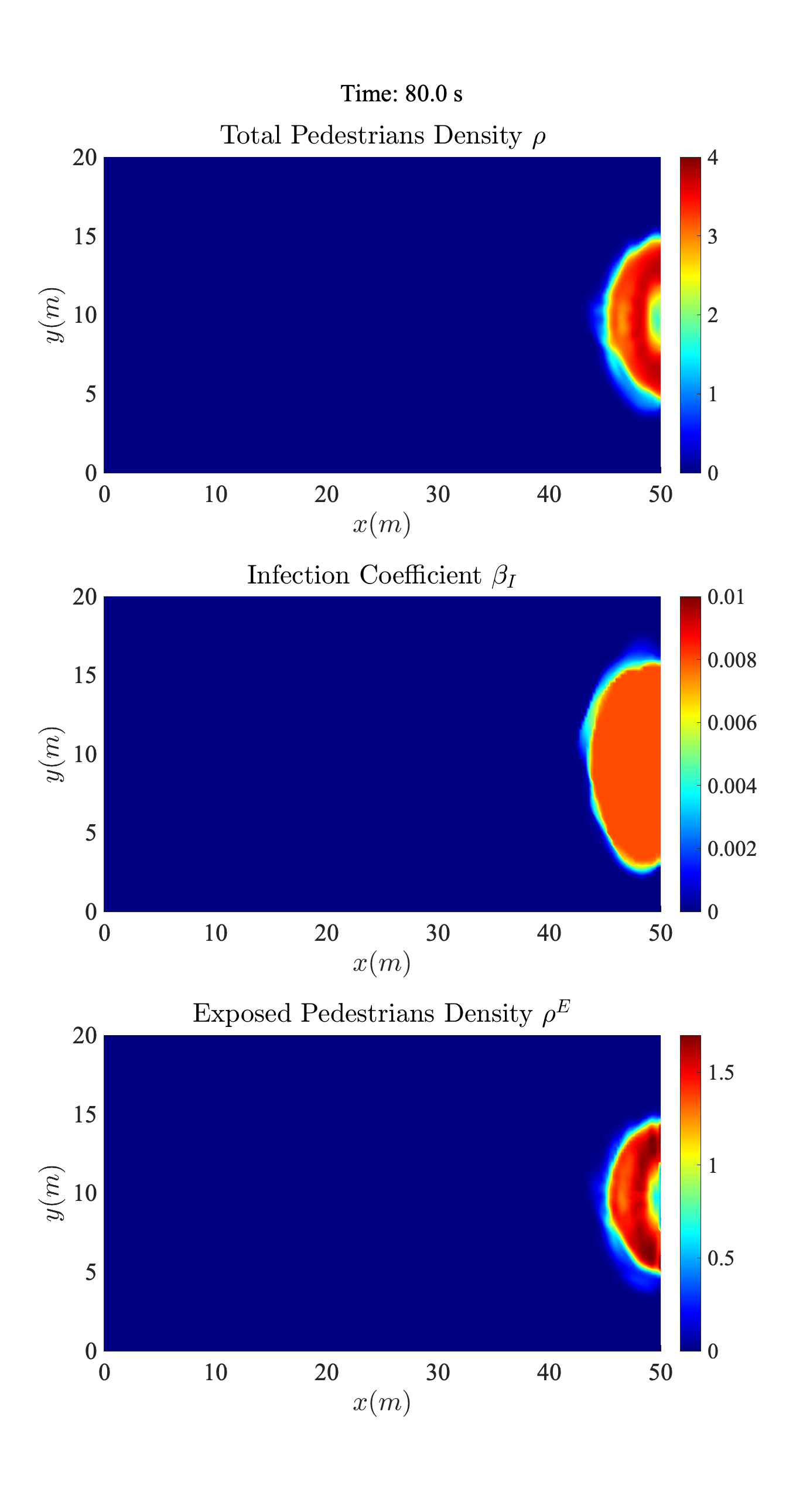}
\caption{Corridor with a $4\;m$ exit: Total density profiles (top), infection coefficient profile (middle), and exposed pedestrians' profile at different time instances with $u_{\max}=1.4$ $m/s $, $C_0=0.5$, and  with no ventilation.}
 \label{fig2d8}
 \end{figure}

\begin{figure}[h!]
\centering
 \includegraphics[width=4.2cm,height=9cm]{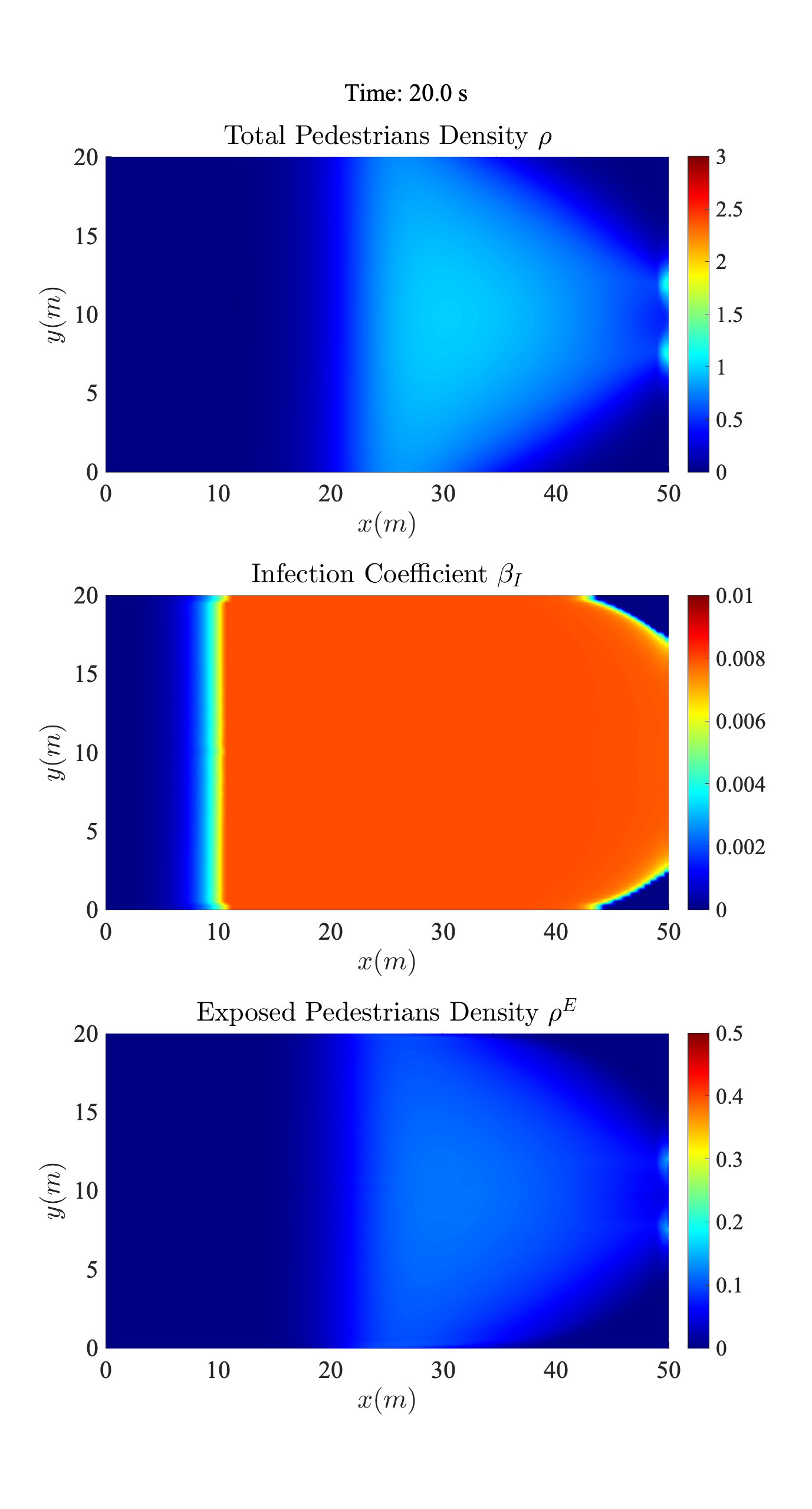} \hspace{-0.16in}
 \includegraphics[width=4.2cm,height=9cm]{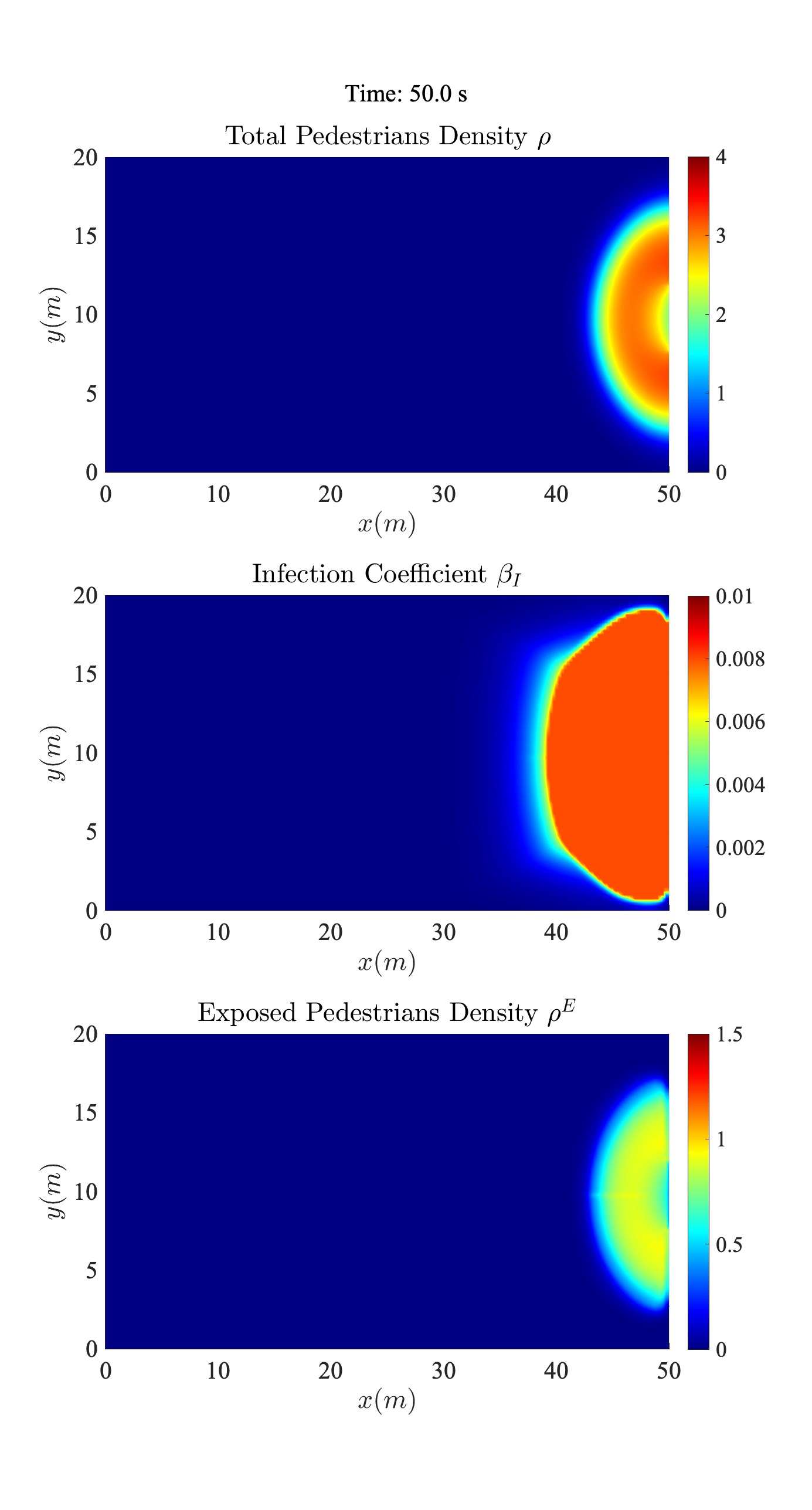}\hspace{-0.16in}
 \includegraphics[width=4.2cm,height=9cm]{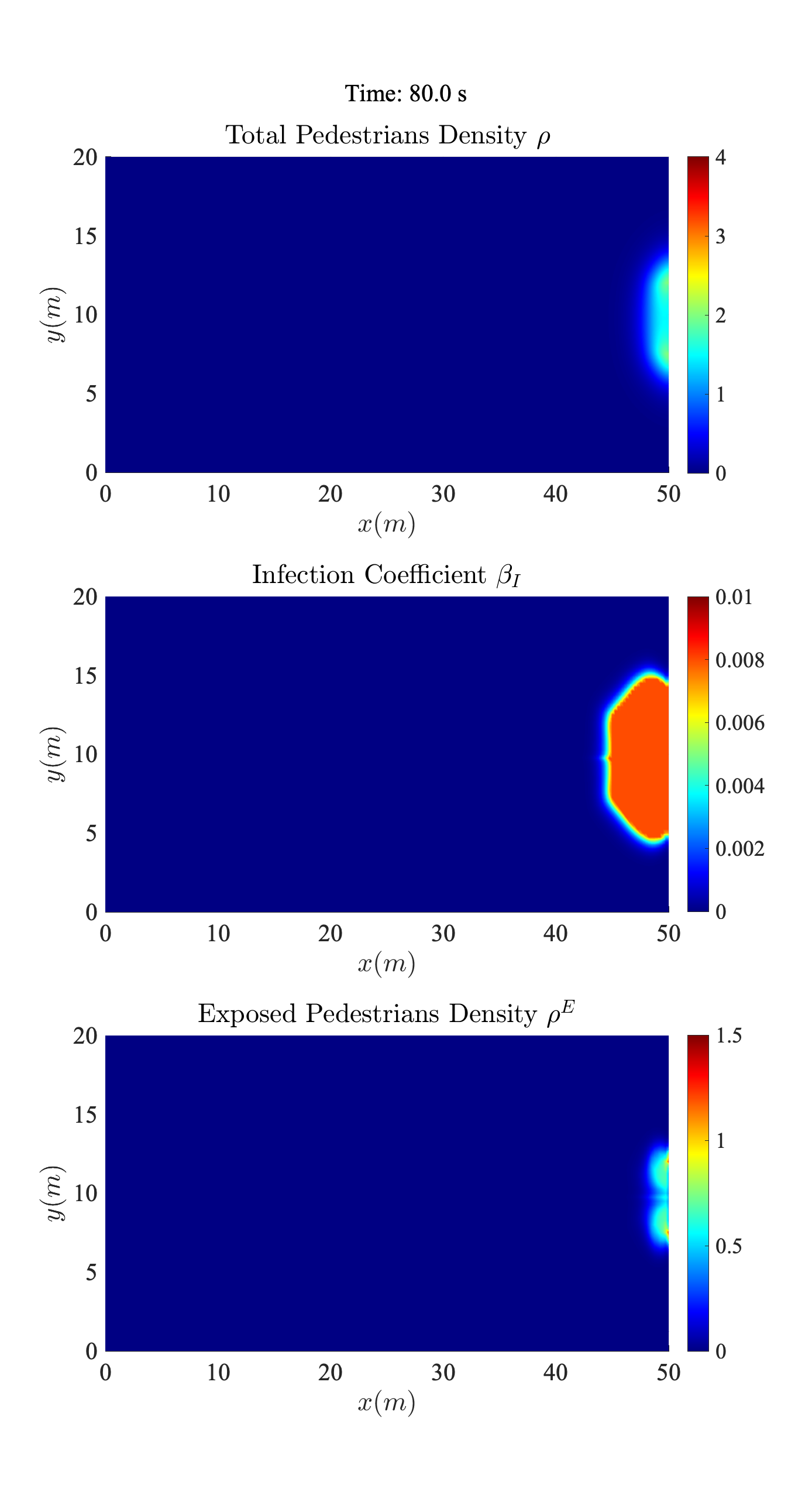}
\caption{Corridor with a $4\;m$ exit: Total density profiles (top), infection coefficient profile (middle), and exposed pedestrians' profile at different time instances with $u_{\max}=1.4$ $m/s $, $C_0=1.2$, and  with no ventilation.}
 \label{fig2d9}
 \end{figure}

The effect of imposing different values of $C_0$ is given in Figs \ref{fig2d8} and \ref{fig2d9} for the $4\;m$ exit, where total density profile, infection coefficient profile, and exposed pedestrians' density at different time instances are shown. Again, the effect of the internal pressure function that describes the repealing forces between individuals and prevents or not from overcrowding is evident. We can observe, also in this case, that for the smaller value of $C_0$ there is a significant decrease of the outflow of pedestrians through the exit, which essentially means that the clogging phenomenon persists more in time. On the other hand,  increasing $C_0$ leads to a less pronounced congestion effect and the outflow is smoother leading to smaller densities and faster evacuation times, and thus, to spreading mitigation.
 
\begin{figure}[h!]
   \centering
 \includegraphics[width=9cm,height=5cm]{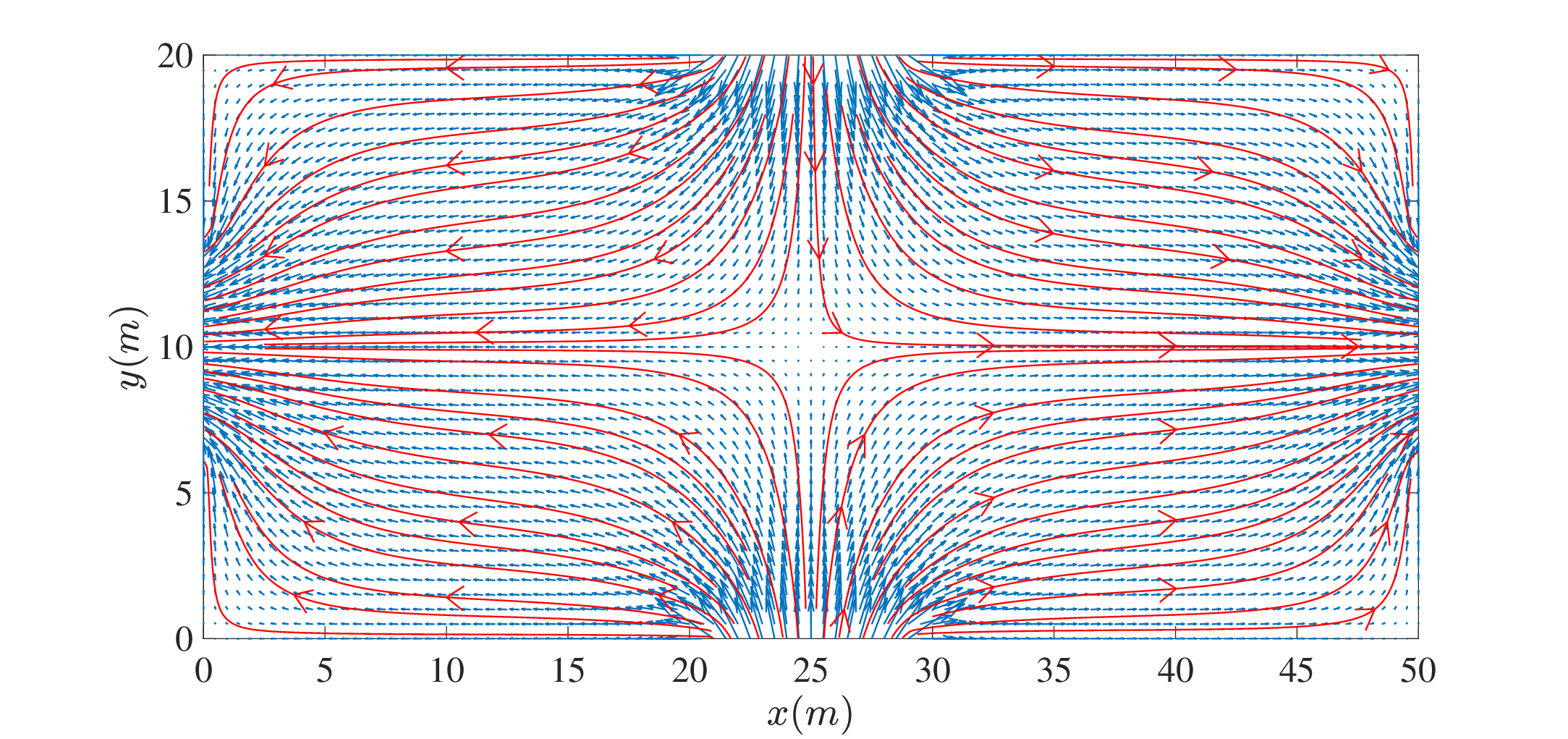} 
 \includegraphics[width=9cm,height=5cm]{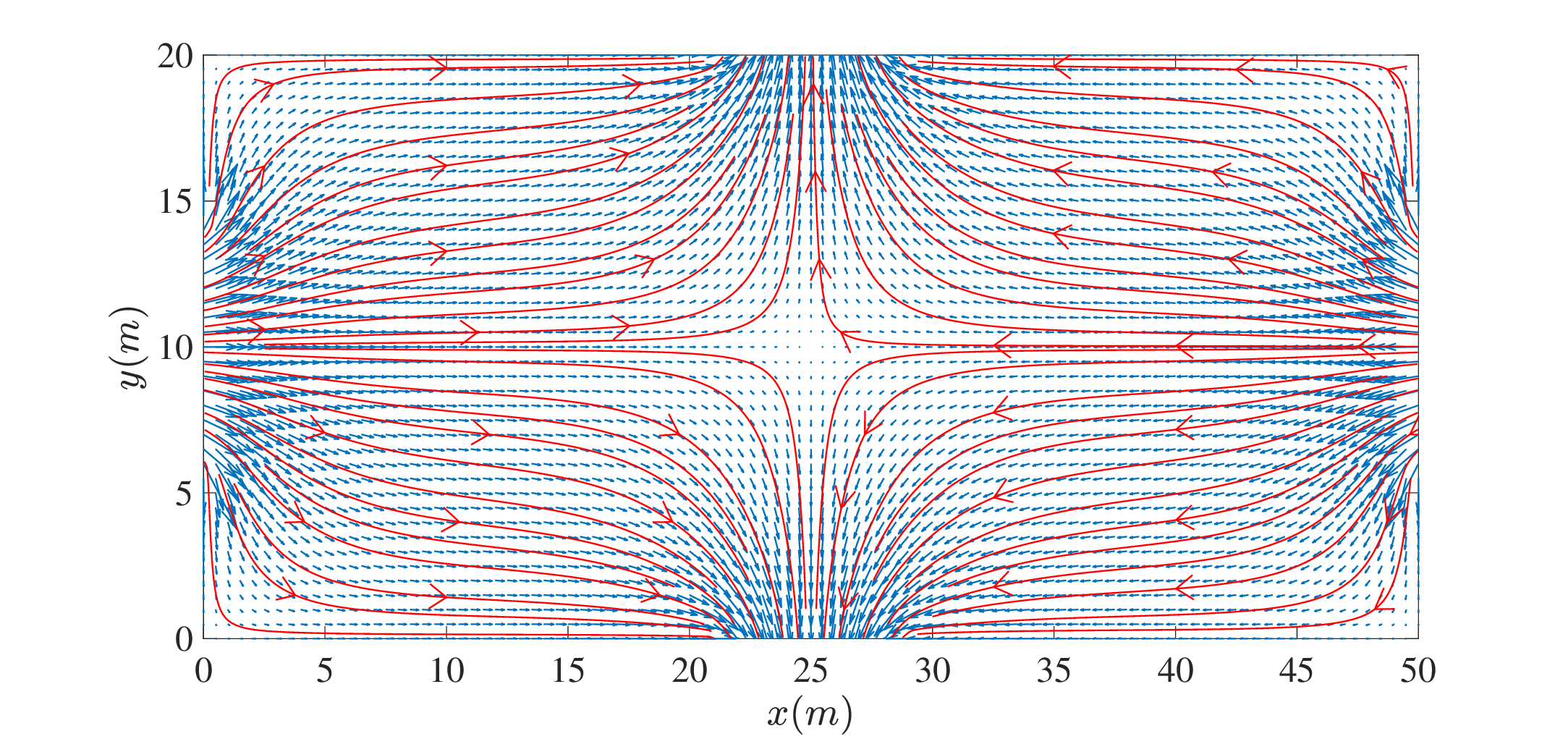}
\caption{Corridor with a $4\;m$ exit: Velocity field ${\bf U}_G$ vectors (in blue) and the corresponding streamlines (in red), with $u_{in}=10$ $ m/s$ for pattern I (top) and pattern II (bottom).}
 \label{fig12}
 \end{figure}

\begin{figure}[h!]
\centering
 \includegraphics[width=6.25cm,height=5.25cm]{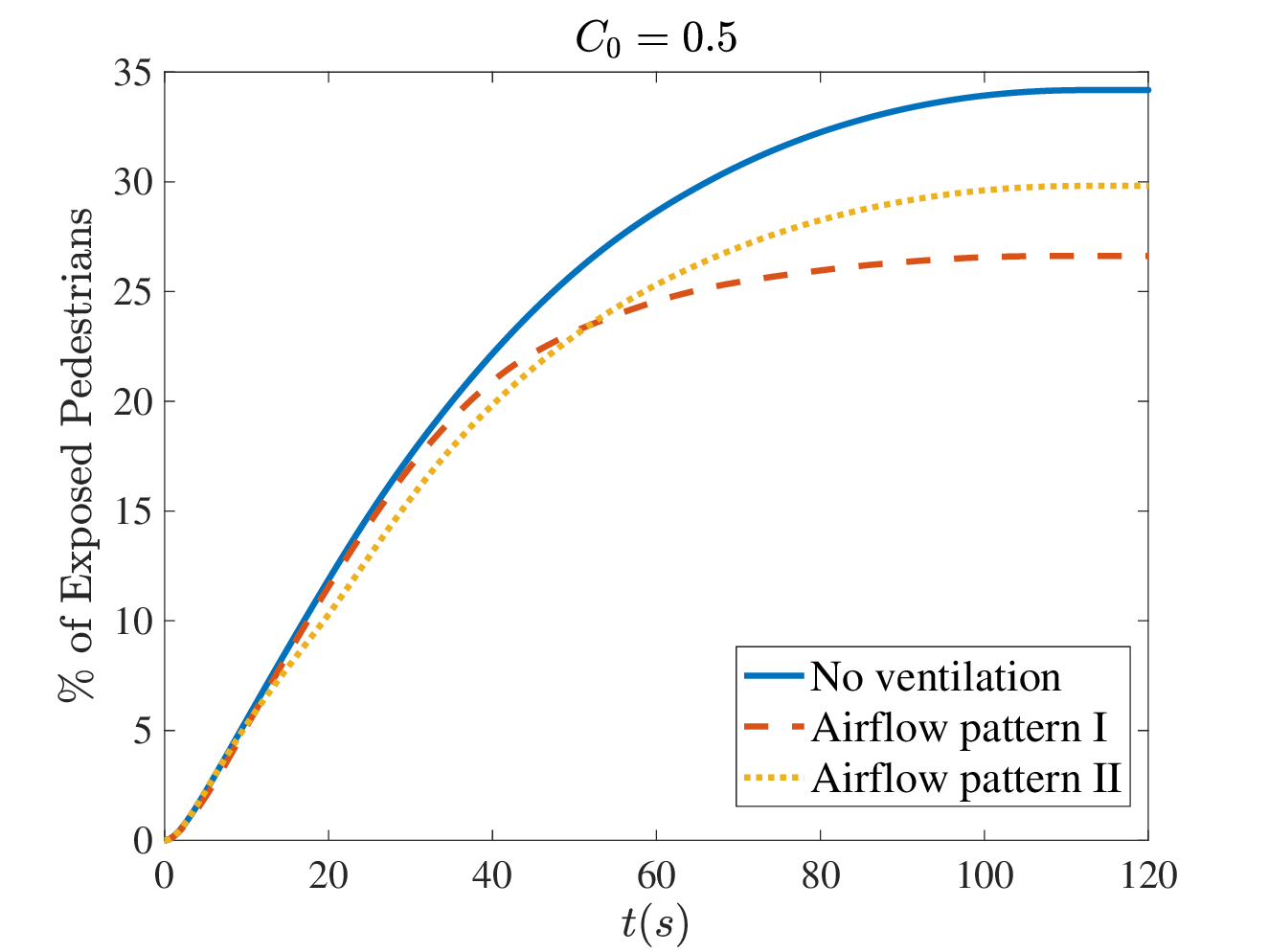} \hspace{-0.25in}
\includegraphics[width=6.25cm,height=5.25cm]{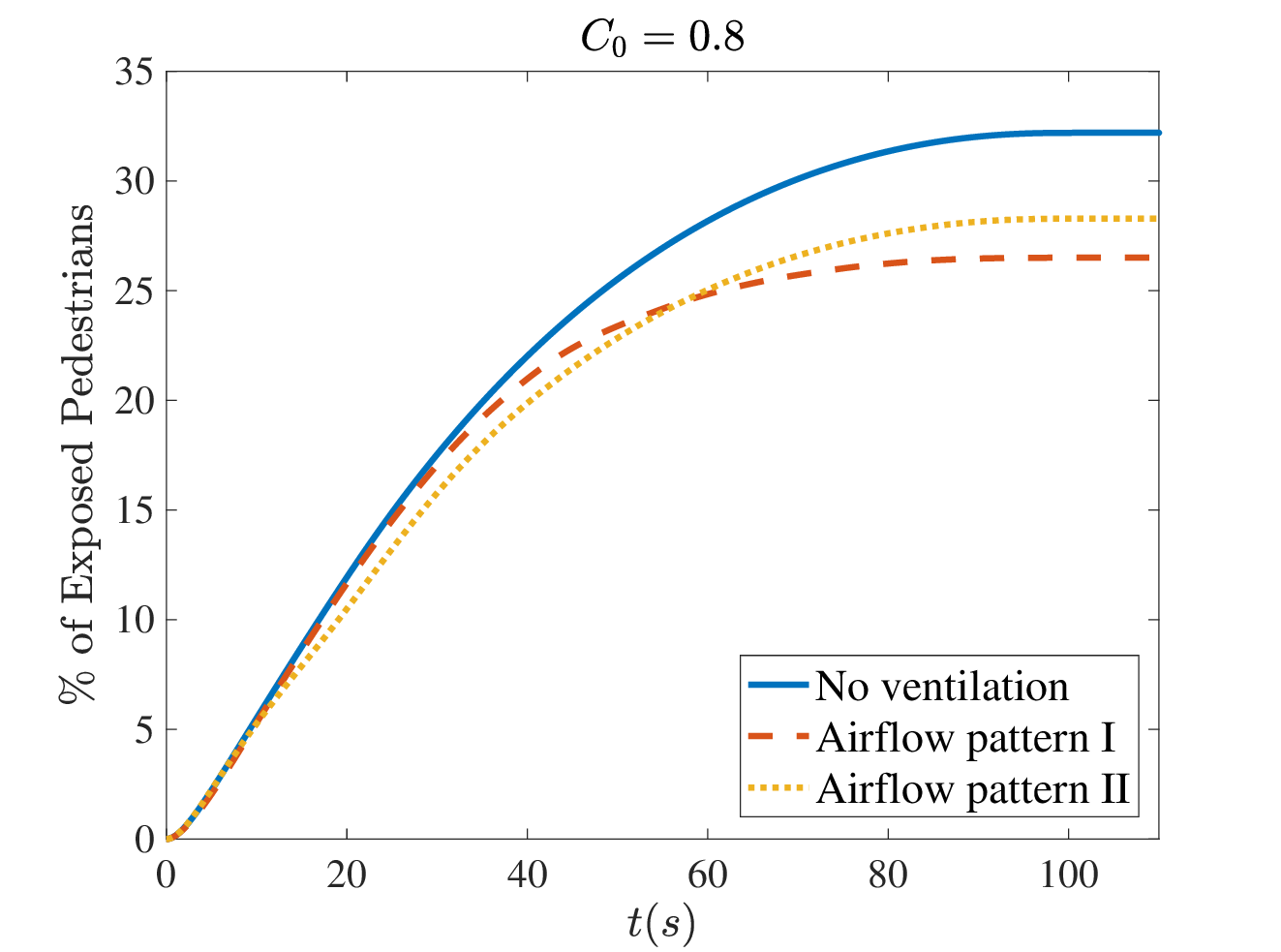} \\
\includegraphics[width=6.25cm,height=5.25cm]{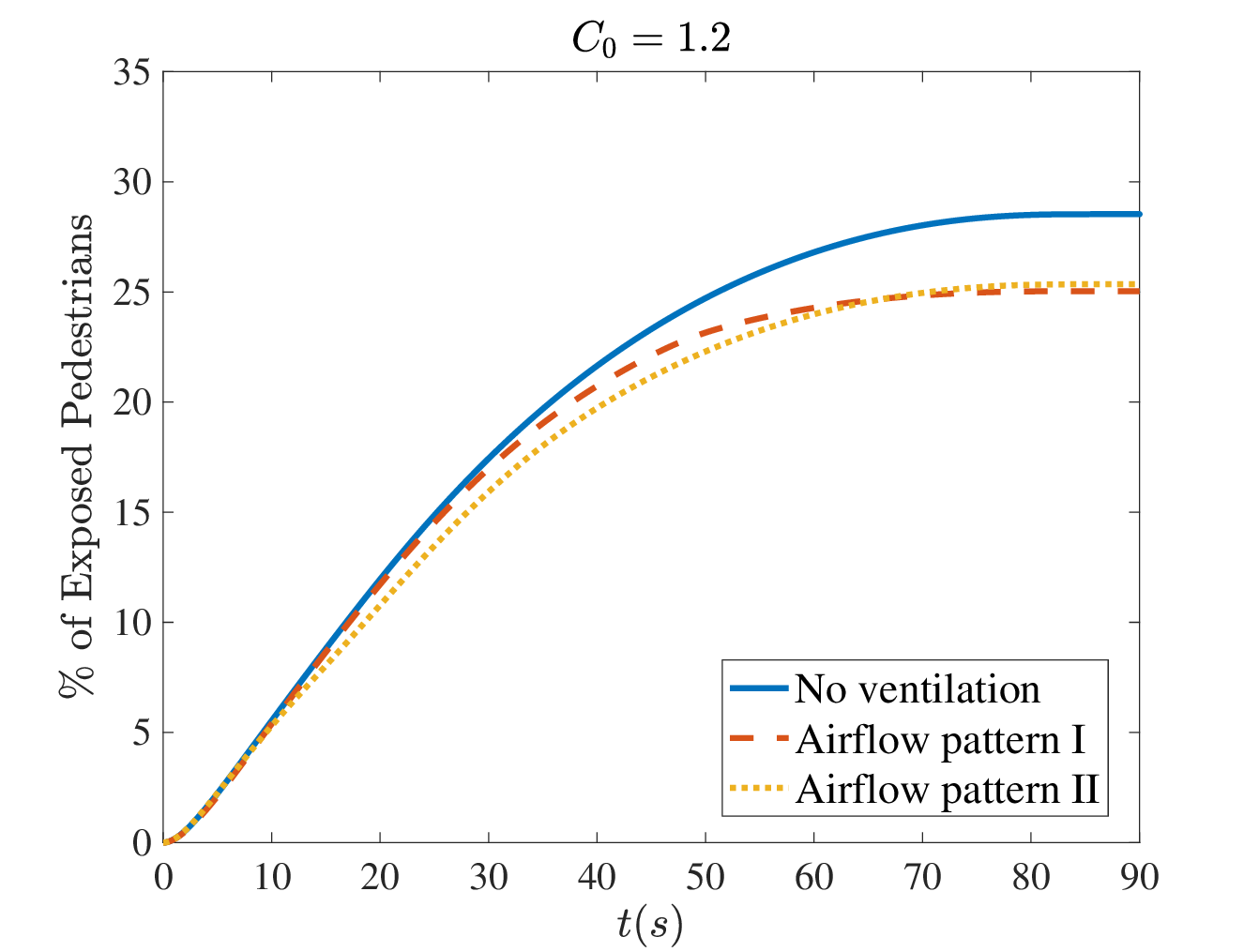} 
\caption{Corridor with a $4\;m$ exit: Percentage of exposed pedestrians  with  $u_{\max}=1.4$ $m/s $ for three different values of $C_0$.}
 \label{fig13}
 \end{figure}

 In the following simulations we focus our interest on the case of a  $4 m$ wide exit, which is the case that corresponds to the worst-case scenario, in terms of the predicted number of exposed pedestrians for all values of $C_0$.  Due to the larger scale of the facility, we impose  two inflow ventilation ducts at the middle of the top and bottom boundaries and two exhaust  ducts at the middle of the right and left boundaries, all being $8\;m$ wide, as shown in Fig. \ref{fig12} (top) and we call the obtained velocity field air-flow pattern I. Then we consider the reverse case called air-flow pattern II as shown in Fig. \ref{fig12} (bottom).  In both patterns, we consider the  ventilation speed to be $u_{in}=10$ $m/s$ and $u_{out}=-10$  $m/s$. The results obtained, for the two air-flow patterns, in terms of the percentage of the exposed pedestrians for each value of $C_0$ are presented in Fig. \ref{fig13} and compared with the no ventilation case. As  it can be observed for the case of $C_0=0.5$, both ventilation patterns reduce the amount of the predicted exposed individuals, with air-flow pattern I  resulting to an almost $10\%$ drop, as compared with the no ventilation case. A moderate drop of the expected exposed pedestrians can be observed in the case of $C_0=1.2$.
 
\begin{figure}[h!]
\centering
 \includegraphics[width=4.2cm,height=9cm]{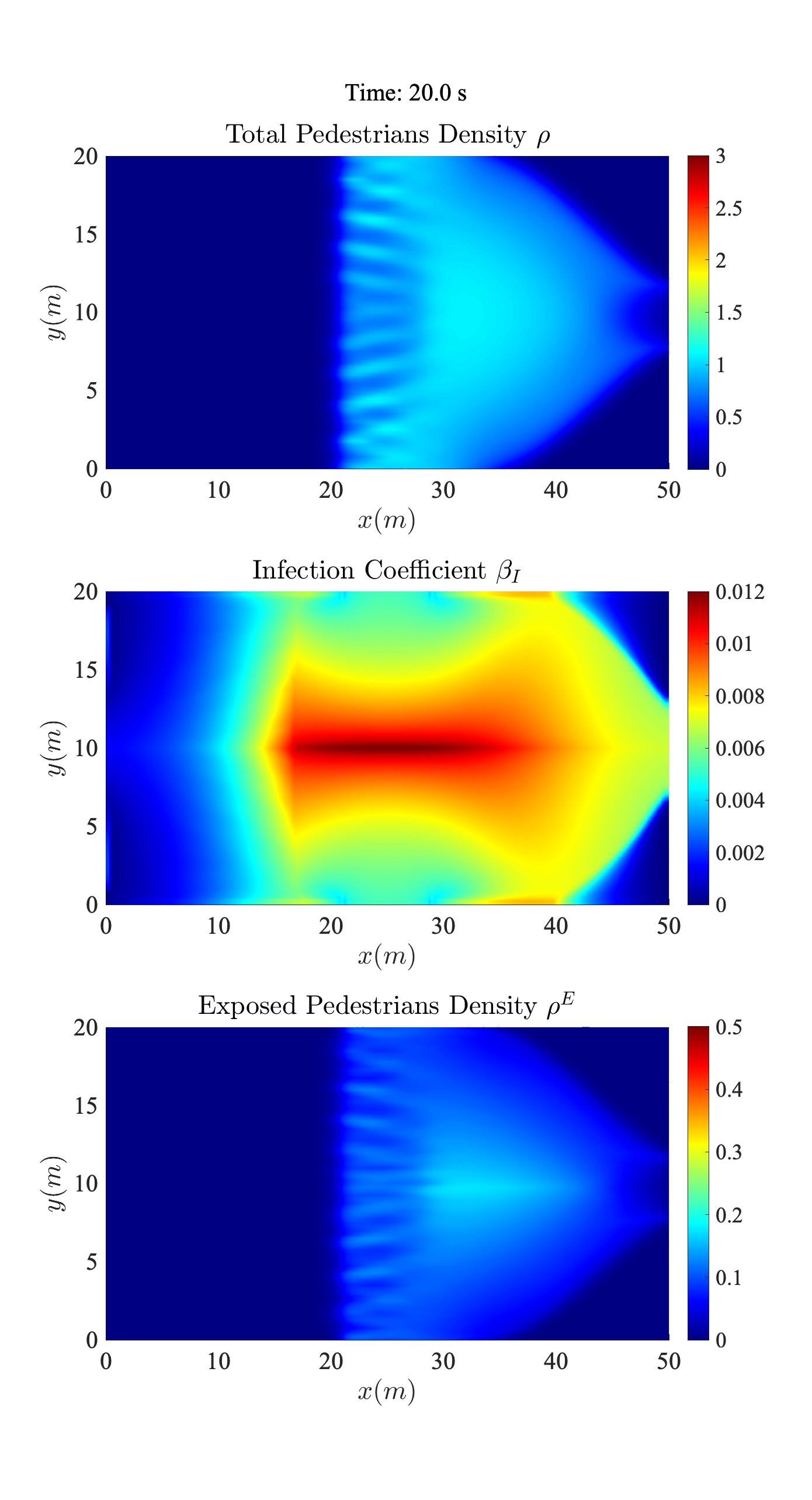} \hspace{-0.179in}
 \includegraphics[width=4.2cm,height=9cm]{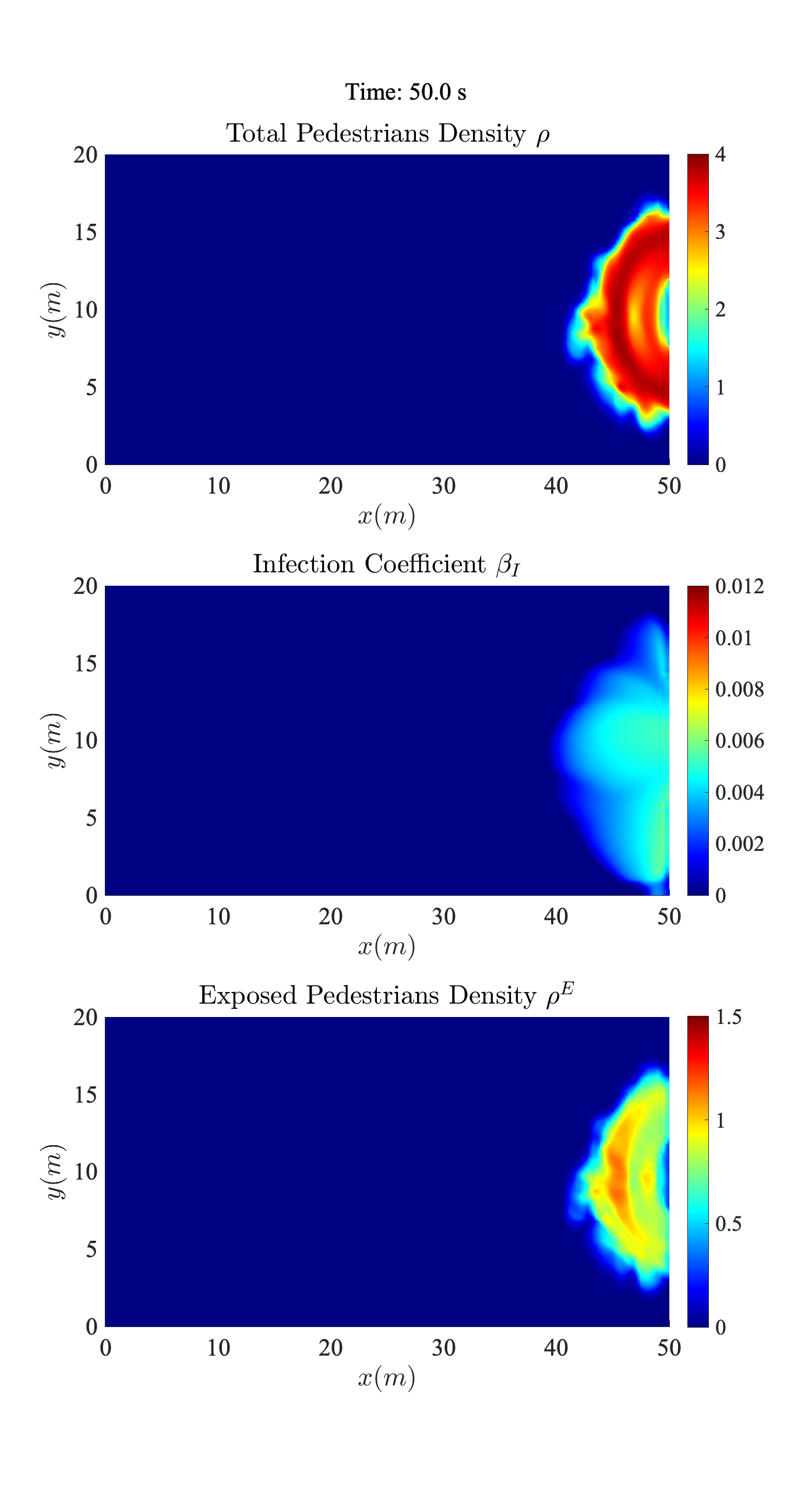}\hspace{-0.142in}
 \includegraphics[width=4.2cm,height=9cm]{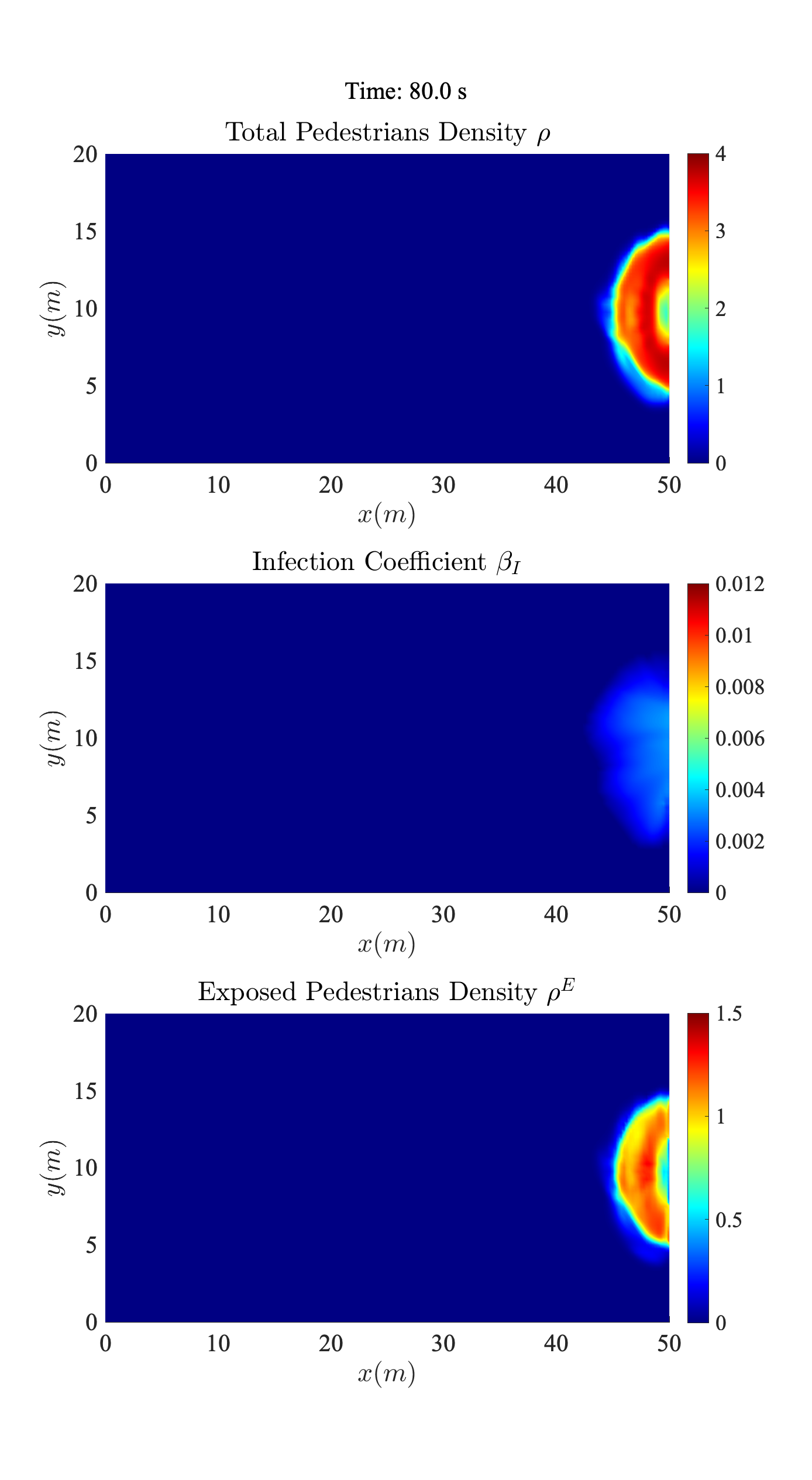}
\caption{Corridor with a $4\;m$ exit: Total density profiles (top), infection coefficient profile (middle), and exposed pedestrians' profile at different time instances with $u_{\max}=1.4$ $m/s $, $C_0=0.5$, and air-flow pattern I. }
 \label{fig2d10}
 \end{figure}

\begin{figure}[h!]
\centering
 \includegraphics[width=4.2cm,height=9cm]{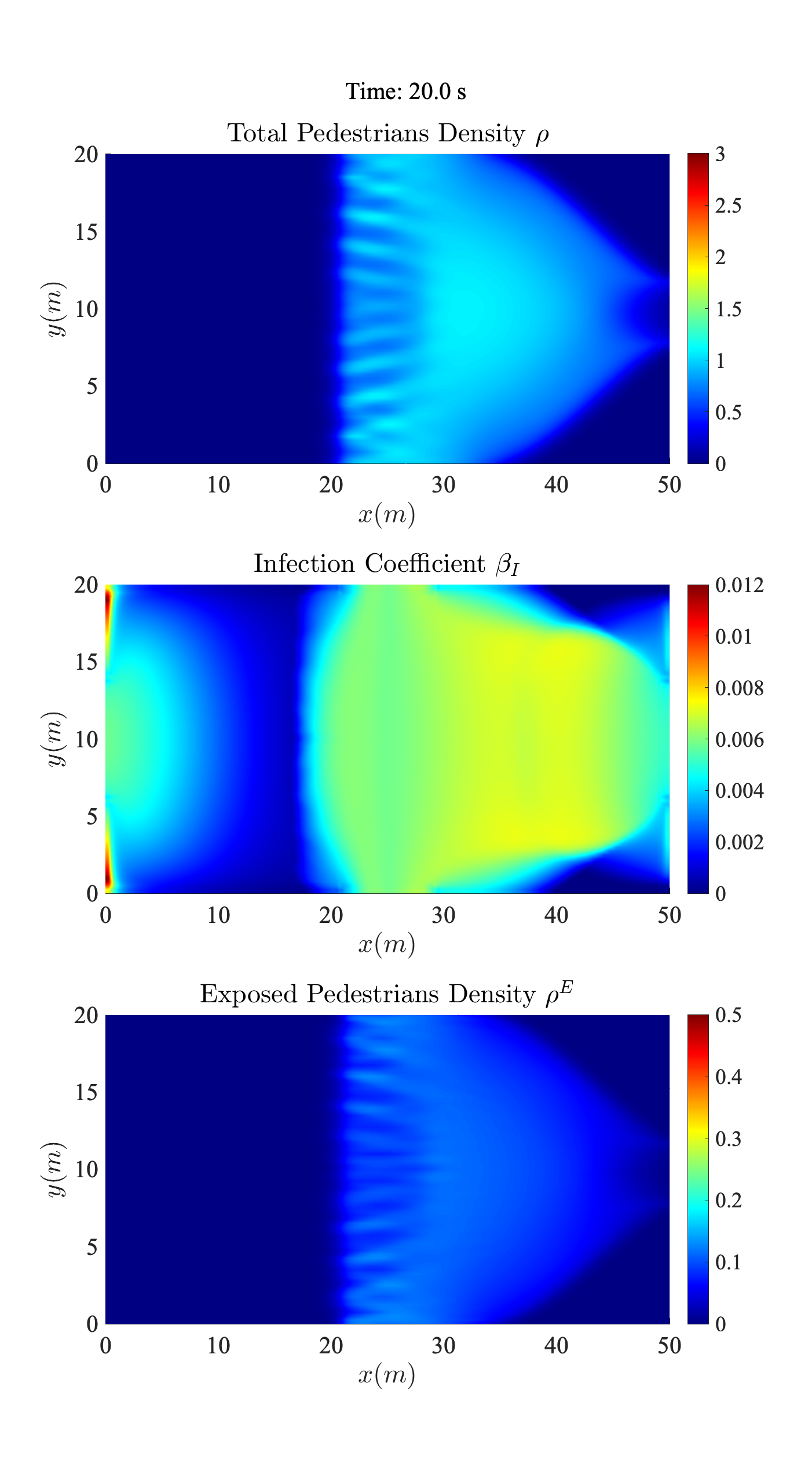} \hspace{-0.179in}
 \includegraphics[width=4.2cm,height=9cm]{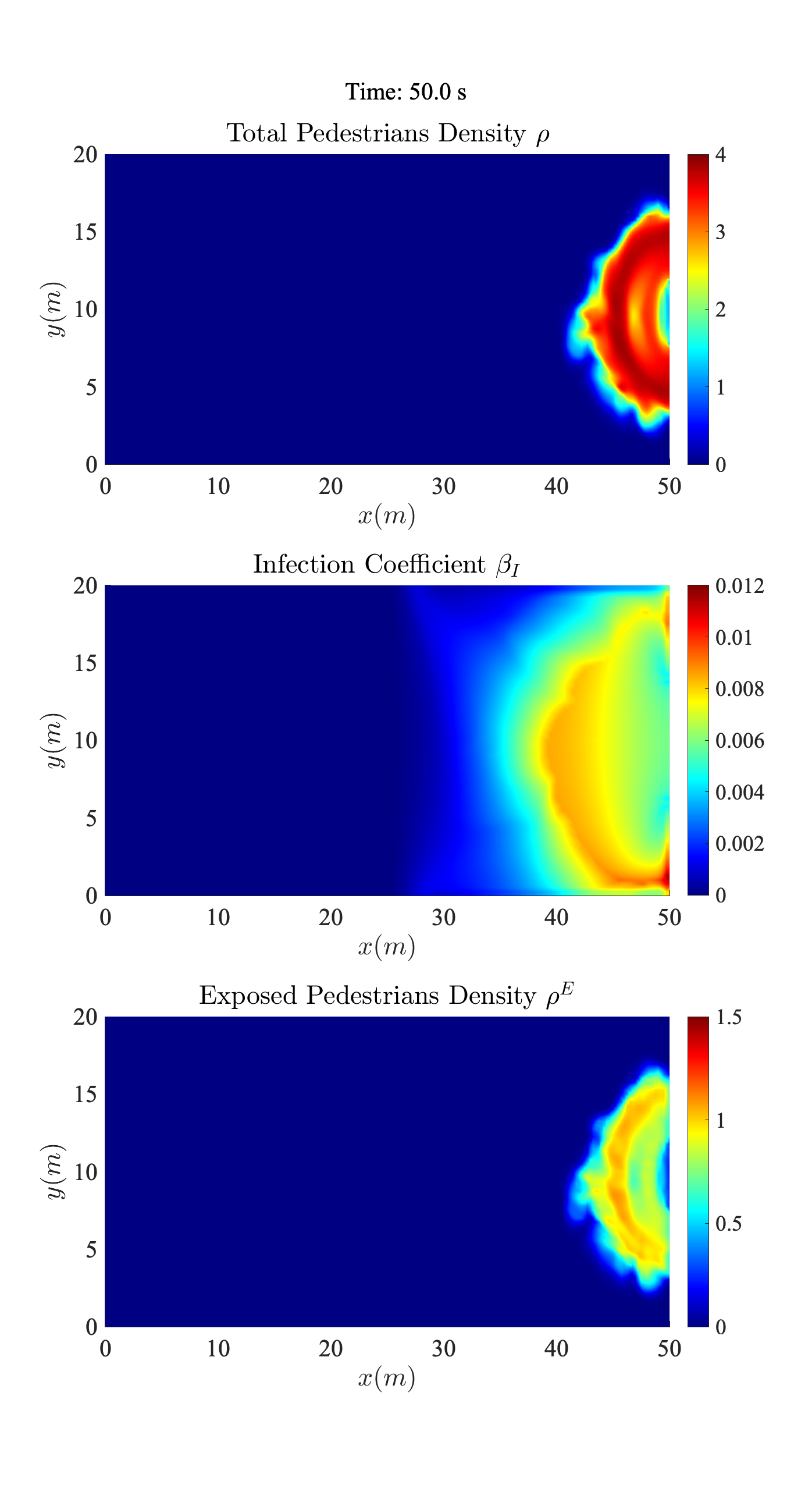}\hspace{-0.142in}
 \includegraphics[width=4.2cm,height=9cm]{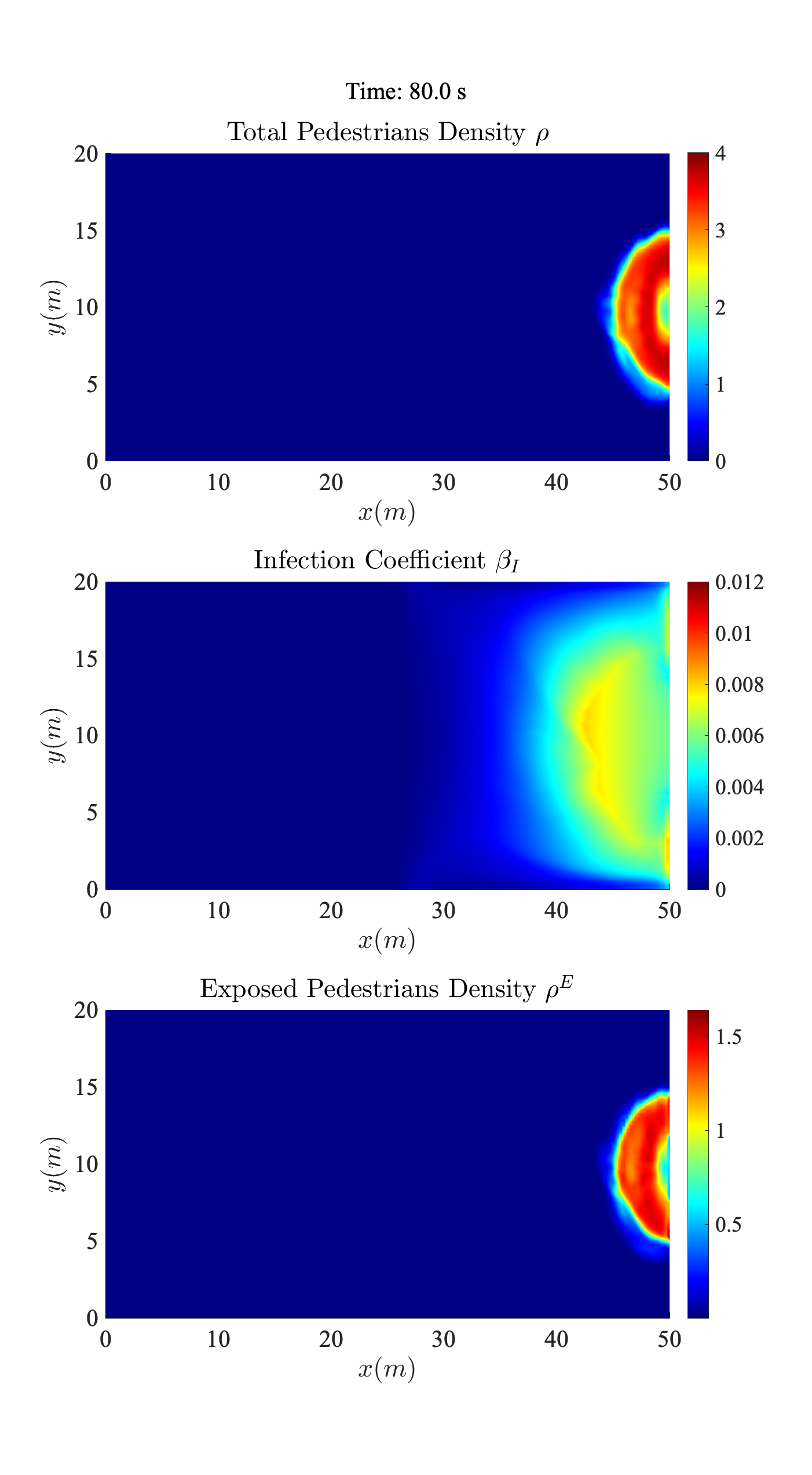}
\caption{Corridor with a $4\;m$ exit: Total density profiles (top), infection coefficient profile (middle), and exposed pedestrians' profile at different time instances with $u_{\max}=1.4$ $m/s $, $C_0=0.5$, and air-flow pattern II. }
 \label{fig2d11}
 \end{figure}

Referring now to Fig. \ref{fig2d10} and  Fig. \ref{fig2d11}   we observe that for such situations, where crowd occupies a large space for long time at the vicinity  of the exit, pattern~II leads to higher degree of spreading because infection coefficient is produced and  advected inside the domain, thus affecting a large portion of the crowd that is concentrated in a large area,  of about 10 meters, near the exit. In contrast, pattern I causes the infection coefficient to exit the domain more efficiently, or, in other words, pattern I cleans the air at a large area near the exit more efficiently, thus leading to smaller numbers of exposed individuals, due to imposing cleaner air near the exit. 
This is not in contradiction with the results of Sections 4.1 and 4.2, because the ``pocket" of clean air reported there, in the case of ventilation field against the pedestrians direction, which resembles more pattern II than pattern I, may be still evident, nevertheless that pocket  is concentrated at a very small area around the exit that does not have any benefit in the present case,  where the crowd occupies a much larger area around the exit.
 
 \section{Conclusions and Recommendations}
 
 In this paper, we performed several numerical investigations for studying the effect of employment of different, epidemics transport control measures, in scenarios in which pedestrians occupy and evacuate closed spaces. In particular, we performed tests varying the ventilation rate and direction, the maximum speed of pedestrians, and the average distances between pedestrians, as well as incorporating in the crowd masked or vaccinated individuals. We also studied the effect of increasing the size or number of the exits. The tests performed employing a macroscopic PDE model consisting of three main components, namely, a crowd flow dynamics component, an epidemics transport dynamics component, and an equation for computing the infection coefficient. 

Based on the results of the tests, we observed the following. Increasing the pressure coefficient $C_0$ leads, in general, to lower degree of spreading, as it leads to higher average distances between pedestrians, and thus, to a smoother crowd flow with less evident congestion/clogging effects at the exit (where the spreading is more profound). However, the effect of increasing $C_0$ may be less evident in situations where the capacity of the outflow at the exit is sufficiently large (e.g., due to multiple or large exits), as, in such scenarios, pedestrians may exit the domain keeping sufficiently high, average distances. Increasing the maximum speed coefficient $u_{\max}$ leads, in general, to lower degree of spreading, as it results in smaller evacuation times. However, very large values of $u_{\max}$ may result in higher spreading when the outflow capacity at the exits is not sufficiently large, as it may create more significant clogging/congestion effects. Incorporating masked/vaccinated individuals or increasing the ventilation rate magnitude reduces the number of total exposed pedestrians. Ventilation direction against pedestrians movement may be beneficial for spreading in scenarios of relatively small number of pedestrians, as a small pocket of clean air created at a small area around the exit may be beneficial. The reverse may, in general, be true, in cases of large crowds occupying a large area at the location of the exit during evacuation, as ventilation along the pedestrians flow may result in a larger area of clean air.
 
\section*{Acknowledgment}
Funded by the European Union (ERC, C-NORA, 101088147). Views and opinions expressed are however those of the authors only and do not necessarily reflect those of the European Union or the European Research Council Executive Agency. Neither the European Union nor the granting authority can be held responsible for them.

\bibliographystyle{plain}
\bibliography{Delis-crowd-contagion}

\begin{thebibliography}{10}

\bibitem{Agnelli2015}
J.P. Agnelli, F.~Colasuonno, and D.~Knopoff.
\newblock A kinetic theory approach to the dynamics of crowd evacuation from
  bounded domains.
\newblock {\em Mathematical Models and Methods in Applied Sciences}, 25(1):109
  – 129, 2015.

\bibitem{Agnelli2023}
Juan~Pablo Agnelli, Bruno Buffa, Damián Knopoff, and Germán Torres.
\newblock A spatial kinetic model of crowd evacuation dynamics with infectious
  disease contagion.
\newblock {\em Bulletin of Mathematical Biology}, 85(4), 2023.

\bibitem{Anderson}
R.~M. Anderson, H.~Heesterbeek, D.~Klinkenberg, and T.~D. Hollingsworth.
\newblock How will country-based mitigation measures influence the course of
  the covid-19 epidemic?
\newblock {\em Lancet}, 395:931--934, 2020.

\bibitem{Arino}
J.~Arino and P.~{Van den Driessche}.
\newblock A multi-city epidemic model.
\newblock {\em Mathematical Population Studies}, 10:175--193, 2003.

\bibitem{Ashare}
ASHRAE.
\newblock Ventilation for acceptable indoor air quality, 2015.

\bibitem{Bert}
G.~Bertaglia and L.~Pareschi.
\newblock Hyperbolic models for the spread of epidemics on networks: Kinetic
  description and numerical methods.
\newblock {\em ESAIM Math. Model. \& Numerical Analysis}, 55:381--407, 2021.

\bibitem{Buch2006}
S.~Buchmüller and U.~Weidmann.
\newblock Parameters of pedestrians, pedestrian traffic and walking facilities.
\newblock {\em IVT Schriftenreihe. Institut für Verkehrsplanung und
  Transportsysteme (IVT), ETH Zürich}, 132, 2006.

\bibitem{Delmastro}
M.~Delmastro and G.~Zamariola.
\newblock Depressive symptoms in response to covid-19 and lockdown: a
  cross-sectional study on the italian population.
\newblock {\em Nature Scientific Reports}, 10, 2020.
\newblock article no 22457.

\bibitem{Toro}
E.F.Toro.
\newblock {\em Riemann Solvers and Numerical Methods for Fluid Dynamics}.
\newblock Springer Berlin, Heidelberg, 2009.

\bibitem{Guan}
L.~Guan, C.~Prieur, L.~Zhang, C.~Prieur, D.~Georges, and P.~Bellemain.
\newblock Transport effect of covid-19 pandemic in france.
\newblock {\em Annual Reviews in Control}, 50:394--408, 2020.

\bibitem{Hag}
M.~Haghani.
\newblock Crowd dynamics research in the era of covid-19 pandemic: Challenges
  and opportunities.
\newblock {\em Safety Science}, 153, 2022.
\newblock paper no. 105818.

\bibitem{Hans18}
F.~S. Hanseler and S.~Hoogendoorn.
\newblock Optimal crowd management for congested metro stations.
\newblock In {\em In hEART 2018: 7th Symposium of the European Association for
  Research in Transportation}, 5-7 September, Athens, Greece, 2018.

\bibitem{Harten}
A.~Harten and J.~M. Hyman.
\newblock Self adjusting grid methods for one-dimensional hyperbolic
  conservation laws,.
\newblock {\em J. Comput. Phys.}, 50:253--269, 1983.

\bibitem{Hoss}
A.H. Hosseinloo, S.~Nabi, A.~Hosoi, and M.A. Dahleh.
\newblock Data-driven control of covid-19 in buildings: a
  reinforcement-learning approach.
\newblock {\em IEEE Trans. on Automation Science and Engineering}, 2023.

\bibitem{Hudg}
R.~L. Hughes.
\newblock A continuum theory for the flow of pedestrians.
\newblock {\em Transp. Res. Part B: Methodological}, 36(6):507, 2002.

\bibitem{Kah}
P.~Kachroo, S.A. Wadoo, S.~J. Al-nasur, and A.~Shende.
\newblock {\em Pedestrian Dynamics: Feedback Control of Crowd Evacuation}.
\newblock Springer, Berlin, 2008.

\bibitem{potential}
M.~Kaushik.
\newblock Potential flow theory.
\newblock In {\em Theoretical and Experimental Aerodynamics}. Springer,
  Singapore, 2019.

\bibitem{Kim}
D.~Kim and A.~Quaini.
\newblock Coupling kinetic theory approaches for pedestrian dynamics and
  disease contagion in a confined environment.
\newblock {\em Math. Models and Meth. in Applied Sciences}, 30:1893--1915,
  2020.

\bibitem{Leveque}
R.~J. Leveque.
\newblock {\em Finite Volume Methods for Hyperbolic Problems}.
\newblock Cambridge University Press, U.K., 2002.

\bibitem{Leveque2}
R.~J. Leveque.
\newblock {\em Finite Difference Methods for Ordinary and Partial Differential
  Equations: Steady-State and Time-Dependent Problems}.
\newblock SIAM, Philadelphia, 2007.

\bibitem{LW}
M.~J. Lighthill and G.B. Whitham.
\newblock On kinematic waves, $i$:flow movement in long rivers. $ii$:a theory
  of traffic on long crowded roods.
\newblock {\em Proc. Royal Soc. A}, A229:281--316, 1955.

\bibitem{Maity24}
Somnath Maity, S.~Sundar, and Jörg Kuhnert.
\newblock A high-resolution meshfree particle method for numerical
  investigation of second-order macroscopic pedestrian flow models.
\newblock {\em Applied Mathematical Modelling}, 131:205 – 232, 2024.

\bibitem{Niazi}
M.U.B Niazi, C.~{Canudas-de-Wit}, A.~Kibangou, and P.-A. Bliman.
\newblock Optimal control of urban human mobility for epidemic mitigation.
\newblock In {\em 2021 60th IEEE Conference on Decision and Control (CDC)},
  pages 6958--6963, TX, USA, 2021.

\bibitem{Payne}
H.~J. Payne.
\newblock Models of freeway traffic and control.
\newblock {\em Math. Models Publ. Sys. Simul. Council Proc.}, 28:51--61, 1971.

\bibitem{Qian}
H.~Qian and X.~Zheng.
\newblock Ventilation control for airborne transmission of human exhaled
  bio-aerosols in buildings.
\newblock {\em Journal of Thoracic Disease}, 10:S2295--S2304, 2018.

\bibitem{salam2}
P.~S.~A. Salam, W.~Bock, A.~Klar, and S.~Tiwari.
\newblock Disease contagion models coupled to crowd motion and mesh-free
  simulation.
\newblock {\em Math. Models and Meth. in Applied Sciences}, 31:1277--1295,
  2021.

\bibitem{salam1}
P.~S.~A. Salam, W.~Bock, A.~Klar, and S.~Tiwari.
\newblock Coupling pedestrian flow and disease contagion models.
\newblock In N.~Bellomo and L.~Gibelli, editors, {\em Crowd Dynamics, Volume 4:
  Analytics and Human Factors in Crowd Modeling}, pages 223--246. Springer
  International Publishing, Cham, 2023.

\bibitem{Satt}
L.~Sattenspiel and K.~Dietz.
\newblock A structured epidemic model incorporating geographic mobility among
  regions.
\newblock {\em Mathematical Biosciences}, 128:71--91, 1995.

\bibitem{Tizz}
M.~Tizzoni, P.~Bajardi, A.~Decuyper, G.~{Kon Kam King}, C.~M. Schneider,
  V.~Blondel, Z.~Smoreda, M.~C. Gonzalez, and V.~Colizza.
\newblock On the use of human mobility proxies for modeling epidemics.
\newblock {\em Plos Computational Biology}, 10, 2014.

\bibitem{Treiber}
M.~Treiber and A.~Kesting.
\newblock {\em Traffic flow dynamics: Data, Models and Simulation}.
\newblock Springer Berlin, Heidelberg, 2013.

\bibitem{Goatin1}
M.~Twarogowska, P.~Goatin, and R.~Duvigneau.
\newblock Comparative study of macroscopic pedestrian models.
\newblock {\em Transportation Research Procedia}, 2:477--485, 2014.

\bibitem{Goatin2}
M.~Twarogowska, P.~Goatin, and R.~Duvigneau.
\newblock Macroscopic modeling and simulations of room evacuation.
\newblock {\em Applied Mathematical Modelling}, 38(24):5781--5795, 2014.

\bibitem{Beek2024}
Arco van Beek, Yan Feng, Dorine~C. Duives, and Serge~P. Hoogendoorn.
\newblock Studying the impact of lighting on the pedestrian route choice using
  virtual reality.
\newblock {\em Safety Science}, 174, 2024.
\newblock Art. no. 106467.

\bibitem{Vyn}
E.~Vynnycky and R.~White.
\newblock {\em An Introduction to Infectious Disease Modelling}.
\newblock Oxford Univ. Press, 2010.

\bibitem{Wadoo10}
S.~A. Wadoo and P.~Kachroo.
\newblock Feedback control of crowd evacuation in one dimension.
\newblock {\em IEEE Transactions on Intelligent Transportation Systems},
  11:182--193, 2010.

\bibitem{Willem}
L.~Willem, F.~Verelst, J.~Bilcke, N.~Hens, and P.~Beutels.
\newblock Lessons from a decade of individual-based models for infectious
  disease transmission: a systematic review (2006-2015).
\newblock {\em BMC Infectious Diseases}, 17, 2017.
\newblock paper no. 612.

\bibitem{Zhao2}
B.~Zhao, Z.~Zhang, X.~Li, and D.~Huang.
\newblock Comparison of diffusion characteristics of aerosol particles in
  different ventilated rooms by numerical method.
\newblock In {\em ASHRAE Transactions}, volume 110 PART 1, page 88 – 95.
  2004.

\bibitem{Zhao}
H.~Zhao.
\newblock A fast sweeping method for eikonal equations.
\newblock {\em Mathematics of Computation}, 74(250):603--627, 2005.

\end{thebibliography}


 \section*{Appendix}
 
 The idea behind the MUSCL scheme is to replace the piecewise constant approximation of the conserved variables of the first-order  finite volume scheme by reconstructed states, derived from cell-averaged states. For each computational cell, slope limited, reconstructed left  $(L)$ and right $(R)$ states are obtained and used to calculate the numerical fluxes (namely the Roe and Rusanov ones in this work) at the cell interfaces. Using a piecewise linear approximation of the variables at each cell,  results in a scheme that is second-order accurate in space for smooth solutions. 
 This results in the following reconstructed states for each component $q$ of an unknown vector ${\bf Q}$ along the $x-$direction for the $x_{i+1/2,j}$ interface 
 \begin{eqnarray}
 q_{i+1/2,j}^{R} \!\!\! &=& \!\!\! q_{i+1,j}-{\frac {\phi \left(r_{i+1,j}^x\right)}{4}}\left[\left(1-\omega \right)\Delta q_{i+3/2,j}+\left(1+\omega \right)\Delta q_{i+1/2,j}\right], \\
q_{i+1/2,j}^{L} \!\!\! &=& \!\!\! q_{i,j}+{\frac {\phi \left(r_{i,j}^x\right)}{4}}\left[\left(1-\omega \right)\Delta q_{i-1/2,j}+\left(1+\omega \right)\Delta q_{i+1/2,j}\right],
 \end{eqnarray}
 and for the for $x_{i-1/2,j}$ interface 
 \begin{eqnarray}
 q_{i-1/2,j}^{R} \!\!\! &=& \!\!\! q_{i,j}-{\frac {\phi \left(r_{i,j}^x\right)}{4}}\left[\left(1-\omega \right)\Delta q_{i+1/2,j}+\left(1+\omega \right)\Delta q_{i-1/2,j}\right], \\
q_{i-1/2,j}^{L} \!\!\! &=& \!\!\! q_{i-1,j}+{\frac {\phi \left(r_{i-1,j}^x\right)}{4}}\left[\left(1-\omega \right)\Delta q_{i-3/2,j}+\left(1+\omega \right)\Delta q_{i-1/2,j}\right],
 \end{eqnarray}
 where $\Delta q_{i+1/2,j} = (q_{i+1,j} - q_{i,j}) $, $\Delta q_{i-1/2,j} = (q_{i,j} - q_{i-1,j}) $, and 
 \begin{equation}
 r_{i,j}^x = \frac{\Delta q_{i-1/2,j}}{\Delta q_{i+1/2,j} }.
 \end{equation}
 The parameter $\omega$ is allowed to lie in the interval $[-1,1]$ and for the results presented in this work we have set $\omega =0$.
 Finally, the function $\phi(r^x)$ is a limiter function that limits the slope of the piecewise approximations to ensure the solution is Total Variation Diminishing (TVD) thereby avoiding the spurious oscillations that would otherwise occur around discontinuities or shocks. The limiter is equal to zero when $r^x \le 0$  and is equal to unity when $ r^x = 1$. In this work the van Albaba  limiter has been utilized, which reads as
 \begin{equation}
 \phi(r) = \frac{r^2 + r}{1+r^2}.
 \end{equation}
 A similar reconstruction is performed along the $y-$direction to obtain the reconstructed values along the $y_{i,j\pm 1/2}$ interfaces.

\end{document}